\documentclass[article]{aa}  

\usepackage{graphicx}
\usepackage{epstopdf}
\usepackage{txfonts}
\usepackage{ulem} 
\usepackage{amsmath,amsfonts,amssymb}
\usepackage{color}

\usepackage{ulem}
\usepackage[dvipsnames]{xcolor}
\usepackage{enumitem}

\newcommand\gaia{\textit{Gaia}\xspace}
\newcommand\gdrtwo{\gaia~DR2\xspace}
\newcommand\gdrone{\gaia~DR1\xspace}

\begin{document} 

\title{ Deciphering the evolution of the Milky Way discs: \\ \gaia APOGEE \textit{Kepler} giant stars and the Besan\c con Galaxy Model }

\author{Lagarde N.\inst{1}\and Reyl\'e C.\inst{1} \and  Chiappini, C.\inst{2} \and Mor, R.\inst{3}\and Anders, F.\inst{3}  \and Figueras, F.\inst{3} \and Miglio, A.\inst{4,5,6} \and Romero-G\'omez, M.\inst{3}\and Antoja, T.\inst{3} \and Cabral, N.\inst{8} \and Salomon, J.-B.\inst{1,2,9}\and  Robin, A. C.\inst{1} \and Bienaym\'e, O.\inst{7} \and  Soubiran, C.\inst{10}  \and Cornu, D.\inst{11} \and Montillaud, J.\inst{1}}

   \institute{Institut UTINAM, CNRS UMR6213, Univ. Bourgogne Franche-Comt\'e, OSU THETA Franche-Comt\'e-Bourgogne, Observatoire de Besan\c con, BP 1615, 25010 Besan\c con Cedex, France. \\
   \email{nadege.lagarde@utinam.cnrs.fr}
   \and  Leibniz-Institut f\"ur Astrophysik Potsdam (AIP), Germany 
   \and Institut de Ci\`encies del Cosmos, Universitat de Barcelona (IEEC-UB), Mart\'i i Franqu\`es 1, 08028 Barcelona, Spain 
   \and School of Physics and Astronomy, University of Birmingham, Edgbaston, B15 2TT, UK 
   \and Dipartimento di Fisica e Astronomia, Universit\`{a} degli Studi di Bologna, Via Gobetti 93/2, I-40129 Bologna, Italy 
    \and INAF – Osservatorio di Astrofisica e Scienza dello Spazio di Bologna, Via Gobetti 93/3, I-40129 Bologna, Italy 
   \and Observatoire Astronomique de Strasbourg, Universit\'e de Strasbourg, CNRS, UMR 7550, 11 rue de l'Universit\'e, 67000 Strasbourg, France 
   \and LGL-TPE, UMR 5276, CNRS, Claude Bernard Lyon 1 University, ENS Lyon, Villeurbanne Cedex, France 
   \and Racah Institute of Physics, Hebrew University, Jerusalem 91904, Israel 
   \and  Laboratoire d'astrophysique de Bordeaux, Universit\'e Bordeaux, CNRS, B18N, all\'ee Geoffroy Saint-Hilaire, 33615 Pessac, France
   \and Observatoire de Paris, LERMA, CNRS, PSL Univ., F-75014, Paris, France 
} 
\date{Received ?/ Accepted ?}
\authorrunning{Lagarde, N.} \titlerunning{The \gaia \textit{Kepler} giant stars \& the Besan\c con Galaxy Model}
 
  \abstract 
   { Thanks to ongoing efforts to compute accurate stellar ages,  we are able to characterise stars in different regions of the Milky Way. The \gaia and \textit{Kepler} space-missions, along with ground-based spectroscopic surveys such as APOGEE, provide a unique way to study the chemo-kinematics relations as a function of age through the Galactic stellar populations and provide new constraints to Galactic evolution models.}
  {We investigate the properties of the double sequences of the Milky Way discs visible in the [$\alpha$/Fe] versus [Fe/H] diagram, which are usually associated to the chemical thin and thick discs at the solar circle. In the framework of Galactic formation and evolution, we discuss the complex relationships between age, metallicity, [$\alpha$/Fe], and the radial, azimuthal, and vertical components of the space velocities.}
   { We study stars with measured chemical and seismic properties from the APOGEE spectroscopic survey and the \textit{Kepler} satellite, respectively. In addition, astrometry from the \gaia  satellite is available for the majority of the sample. We separate the  [$\alpha$/Fe]-[Fe/H] diagram into three stellar populations: the thin disc, the high-$\alpha$ metal-poor thick disc, and the high-$\alpha$ metal-rich thick disc and characterise each of these in the age-chemo-kinematics parameter space. Because of the model-dependent nature of the ages inferred from asteroseismology, and because they depend on the quality of the input spectroscopic information, we compare results obtained from different APOGEE data releases (DR14 and DR16). We also use age determinations from two recent works in the literature. In addition, we use the Besan\c con stellar populations synthesis model to highlight selection biases and mechanisms (such as mergers and secular evolution) not included in the model.}
    {The thin disc exhibits a flat age--metallicity relation while [$\alpha$/Fe] increases with stellar age.
    We confirm no correlation between radial and vertical velocities with [Fe/H], [$\alpha$/Fe], and age for each stellar population. Considering both samples, V$_\varphi$ decreases with age for the thin disc, while V$_\varphi$ increases with age  for the high-$\alpha$ metal-poor thick disc. We show that this difference is not due to sample selection. Although the age distribution of the high-$\alpha$ metal-rich thick disc is very close to that of the high-$\alpha$ metal-poor thick disc between 7 and 14 Gyr, its kinematics seems to follow that of the thin disc. This feature, not predicted by the hypotheses included in the Besan\c{c}on Galaxy Model, suggests a different origin and history for this population. Finally, we show that there is a maximum dispersion of the vertical velocity, $\sigma_Z$, with age for the high-$\alpha$ metal-poor thick disc around 8 Gyr. The comparisons with the Besan\c{c}on Galaxy Model simulations suggest a more complex chemo-dynamical scheme to explain this feature, most likely including mergers and radial migration effects.}
   {}
     \keywords{Galaxy:stellar content, Galaxy:evolution, Galaxy:structure, Galaxy:kinematics and dynamics}
                  \maketitle
%
 
\section{Introduction}
\label{intro}

The study of stellar populations is necessary to understand how galaxies assembled and formed. The Milky Way is the only galaxy where it is possible to resolve individual stars and disentangle their chemical and dynamical properties. Galactic archaeology \citep{Freeman02, Matteucci12} uses present-day abundances as relics to follow the history of the Milky Way. It relies on the assumption that the history of our Galaxy is encoded in the chemical abundances of stars and in their kinematics, providing crucial insights into star formation, assembly (e.g. merger, accretion or outflow), and the dynamical history of our Galaxy. Studying stars at different ages has proven to be the best way to investigate the evolution of chemical elements during the earlier stages of the Milky Way, allowing the Galactic evolution to be reconstructed.\\

Accurate age determinations are decisive in avoiding misinterpretations of the formation and evolution of the Milky Way  \citep[e.g. ][]{Chiappini14,Minchev19}. Stellar age determination is very challenging because age is not a directly observable quantity, and our knowledge of stellar ages is dependent on stellar evolution models \citep[e.g. ][]{Lagarde17}. Traditionally, in Galactic studies the metallicity and $\alpha$-abundances of individual stars are used as proxies for their age \citep[e.g.][]{Tinsley79, Ryan96, Bovy12a, Ting13}. However, these proxies are  limited by the significant scatter of abundances in any given age bin \citep[e.g. ][]{Minchev17, Mackereth17, Anders18}. Similarly, the [C/N] ratio has been used to determine stellar ages of red-giant field stars \citep[e.g. ][]{Martig15,Masseron15}. However, these studies do not take into account the effects of mixing occurring in the stellar interiors, stellar input physics, and possible changes of these relations at different evolutionary stages. \citet{Lagarde17} underlined the importance of taking these aspects of stellar evolution into account, especially the impact of transport processes occurring in red-giant stars in the determination of ages for Galactic archaeology studies. The oldest method commonly adopted  to derive stellar ages is isochrone fitting \citep[e.g.][]{Jorgensen05, Yi01}. Recently, asteroseismology paved the way to a better understanding of stellar interiors, providing detailed insight into stellar properties such as mass, radius, evolutionary state \citep[e.g.][]{Stello08,Mosser12b, Bedding11, Vrard16}, and rotational profile \citep[e.g.][]{Mosser12a,Beck12,Gehan18}, as well as into the properties of helium ionisation regions \citep{Miglio10}. The CoRoT \citep{Baglin06}, \textit{Kepler} \citep{Borucki10},  K2 \citep{Howell14}, and TESS \citep{TESS} space missions offer a unique opportunity to obtain some fundamental properties by observation of mixed modes in red giants \citep[e.g.][]{ChMi13}. The masses of red-giant stars can be directly related to stellar interior physics and stellar evolution allowing one to determine ages \citep[e.g.][]{Casagrande16, Anders17a, Anders17b,Silva18, Valentini19, Rendle19,Miglio21} without being limited to surface properties, and with higher accuracy than the determinations from isochrones \citep{Lebreton14a,Lebreton14b}. Seismic data collected for a large number of red-giant stars belonging to the Galactic-disc populations, coupled with other types of observations, are of crucial importance in constraining stellar and Galactic physics \citep[e.g. ][]{Miglio13,Miglio17}. \\

Observations of the Milky Way reveal the existence of a thick disc, as observed photometrically in external disc galaxies \citep{Tsikoudi79,Burstein79,Dalcanton02,Yoachim06,Comeron15}. Recent spectroscopic surveys have given rise to a new paradigm where the separation of different Galactic components is based on chemical properties, by opposition to a kinematic or geometric definition. 
The distribution of stars in the [$\alpha$/Fe] versus [Fe/H] plane has been shown to be bi-modal and has been used to separate the thin- and (chemical) thick-disc populations using low-resolution spectroscopic surveys, such as SEGUE \citep{Lee11b,Yanny09} and RAVE \citep{Boeche13,Steinmetz20a, Steinmetz20b}, or high-resolution surveys, such as HARPS \citep{Adibekyan13},  Gaia-ESO survey \citep{RecioBlanco14}, GALAH survey \citep{Duong18}, and APOGEE survey \citep{Hayden15,Queiroz20}. \\
 
All these surveys are highly complementary to the \gaia satellite astrometry. The second \gdrtwo data release \citep{Brown18} provides data of superior quality for five astrometric parameters (parallaxes, proper motion and position) for more than 1.3 billion stars in the whole Milky Way, allowing a detailed Galactic map to be constructed. This enormous data set has allowed the community to discover crucial events in the evolution of the Milky Way, such as a collision with another galaxy \citep{Helmi18,Belokurov18}, helping us to better understand halo and thick-disc populations \citep{Haywood18, Sahlholdt19,DiMatteo19, Gallart19, Mackereth19b,Deason19,Mackereth20,Naidu20}. \\

Although the coupling of these different kinds of observations is already common in the literature, many questions related to the evolution of the Milky Way are unanswered because of a lack of accurate stellar ages. A massive undertaking is underway 
to provide more accurate ages from asteroseismology. In particular, the metal-poor regime is being investigated \citep[e.g.][]{Valentini19} as well as stars that were born in situ \citep{Chaplin20} and stars accreted from Gaia-Euceladus \citep{Montalban21}, and the effects of stellar models on this derivation are being  quantified \citep{Miglio21}.
Here we take advantage of the large APOKASC catalogue \citep[APOGEE+\textit{Kepler}][]{Pinsonneault18} which provides the spectroscopic and asteroseismic properties of 6676 evolved stars, as well as their age estimates. APOKASC is a combination of spectroscopic properties from the Apache Point Observatory Galactic Evolution Experiment \citep[APOGEE,][]{APOGEE} project during the fourth epoch of the Sloan Digital Sky Survey (SDSS-IV) and asteroseismic information from the \textit{Kepler} mission launched in 2009. 
We also consider the recent stellar age determinations of a subsample of \textit{Kepler} giants with APOGEE DR14 by \citet{Miglio21}. We have added distances from StarHorse \citep{Queiroz20}, the authors of which used parallaxes from \gdrtwo and complementary spectroscopic and photometric information in a bayesian approach, and then computed kinematic properties using these distances. \\

The present study is motivated by two main goals. First, we discuss the main chrono-chemo-kinematics relations to highlight key constraints to Milky Way evolution. We use stellar ages derived from asteroseismology with two different methods. Comparing these results to recent studies, we derive robust correlations between age, metallicity, and kinematics for stars in the three considered populations. 
The second goal is to highlight the differences between observations and the prescriptions from Galactic theory using a stellar population synthesis model. The Besan\c con Galaxy model (hereafter BGM) is a state-of-the-art stellar population synthesis model \citep{Robin03,Lagarde17} intended to reproduce the main basic properties of the Milky Way stellar populations at large scale and for the full sky. This powerful tool allows us to compute mock catalogues 
to statistically compare them with any type of large survey data. Observational features not well reproduced by the mock catalogue can reveal physical processes not yet included in the model and thus improve our knowledge of Galactic evolution. \\

This paper is structured as follows. In Section \ref{data} we present the observed samples considered in this study. In Section \ref{simulations}, we simulate samples with the Besan\c con galaxy model, and discuss the possible biases on the main studied parameters. In Sect. \ref{chemicalpop}, we show the separation of stars in the [$\alpha$/Fe]-[Fe/H] diagram into three populations \citep [see e.g.][]{Adibekyan13,Anders18, Guiglion19} and discuss their age dispersion and their age--chemical properties relation. The age-chemo-kinematics relations of the stellar populations are studied in Section~\ref{Kinematics}. A discussion and conclusions are presented in Section~\ref{conclu}.

\section{Data}
\label{data}

In this study, we use observational information provided by spectroscopy, asteroseismology, astrometry, and photometry. In this first section we present the data used in the present work.
\subsection{Asteroseismic sample}

Accurate stellar ages deduced from asteroseismic measurements are available for all stars in our sample. These determinations constitute the most accurate stellar ages \citep{ChMi13,Lebreton14b}. Although the stellar mass determinations are directly deduced from seismic diagnostic, we must bear in mind that stellar ages are not independent of stellar evolution models, and are therefore also not independent of the underlying stellar physics. In order to better assess our results and to take into account these possible variations, we study two recently published age catalogues.

\subsubsection{Seismic sample based on \citet{Pinsonneault18}}

As a first sample, we consider the APOKASC catalogue published by \citet{Pinsonneault14,Pinsonneault18}, and we take into account the following selection including criteria published in \citet{Pinsonneault14,Pinsonneault18}: 

\begin{itemize}
\item Apparent magnitude range of 7$<$H$<$11; 
\item Stars with an effective temperature ($\rm{T}_{\text{eff}}$) lower 
than 5300~K to select only giant stars; 
\item Stars within the observed domains of the large separation ($\Delta\nu$) and the frequency of maximum power ($\nu_{\text{max}}$) ($0.4<\Delta\nu<20~\mu Hz$ ; $2<\nu_{max}<250~\mu Hz$);
\item Binaries identified by the radial velocity variation are rejected using VSCATTER$< $1~km s$^{-1}$ 
\item We adopt `ASPCAPFLAG'\footnote{The `ASPCAPFLAG' is used in the APOGEE catalogue to flag potential issues in the data analysis process of the stellar spectra in ASPCAP pipeline. We used ASPCAPFLAG=0.}  to select the best well-defined spectroscopic determinations \citep{Holtzman18};
\item The most accurate seismic determinations only, using the information in the \textit{Notes column} indicated in \citet{Pinsonneault18}.
\end{itemize}

In the APOKASC catalogue \citep{Pinsonneault18},  ages are determined using a modified version of the grid-based modelling as implemented in the BeSPP code \citep{Serenelli13} in which the input data are stellar mass, surface gravity, [Fe/H], and [$\alpha$/Fe]. Empirical calibrations are used to rescale and combine asteroseismic results from different pipelines \citep[more details in][]{Pinsonneault18}. The catalogue contains 5426 giant stars.

\subsubsection{Seismic sample based on \citet{Miglio21}}
\label{M21sample}

\citet[][hereafter  M21]{Miglio21} studied $\sim$5400 \textit{Kepler} giant stars using the PARAM  code \citep{Rodrigues17}, exploring the uncertainties on mass and age estimates. M21 focussed on stars with robust age estimates by removing low-mass clump stars that are expected to be more affected by mass loss. In addition, they also restricted their red giant branch (RGB) sample to stars with a radius of less than 11R$_\odot$. This allows contamination of early-AGB stars to be avoided, and removes stars with low $\nu_{max}$. 
The authors explore three different sets of stellar models in order to investigate the sensitivity of the age determinations (see more details in the article).
Finally, they highlight 3297 giants stars with robust and accurate stellar mass and age. In the present study, we use this subsample with stellar ages derived using stellar evolution models including the effects of microscopic diffusion. We focus on stars in common with APOKASC and with stellar ages lower than 14 Gyr, reducing our sample to 2814 stars.  \\

\subsubsection{Age comparison}

     \begin{figure}
  \centering
     \includegraphics[width=0.75\hsize,clip=true,trim= 0cm 0cm 0cm 0cm]{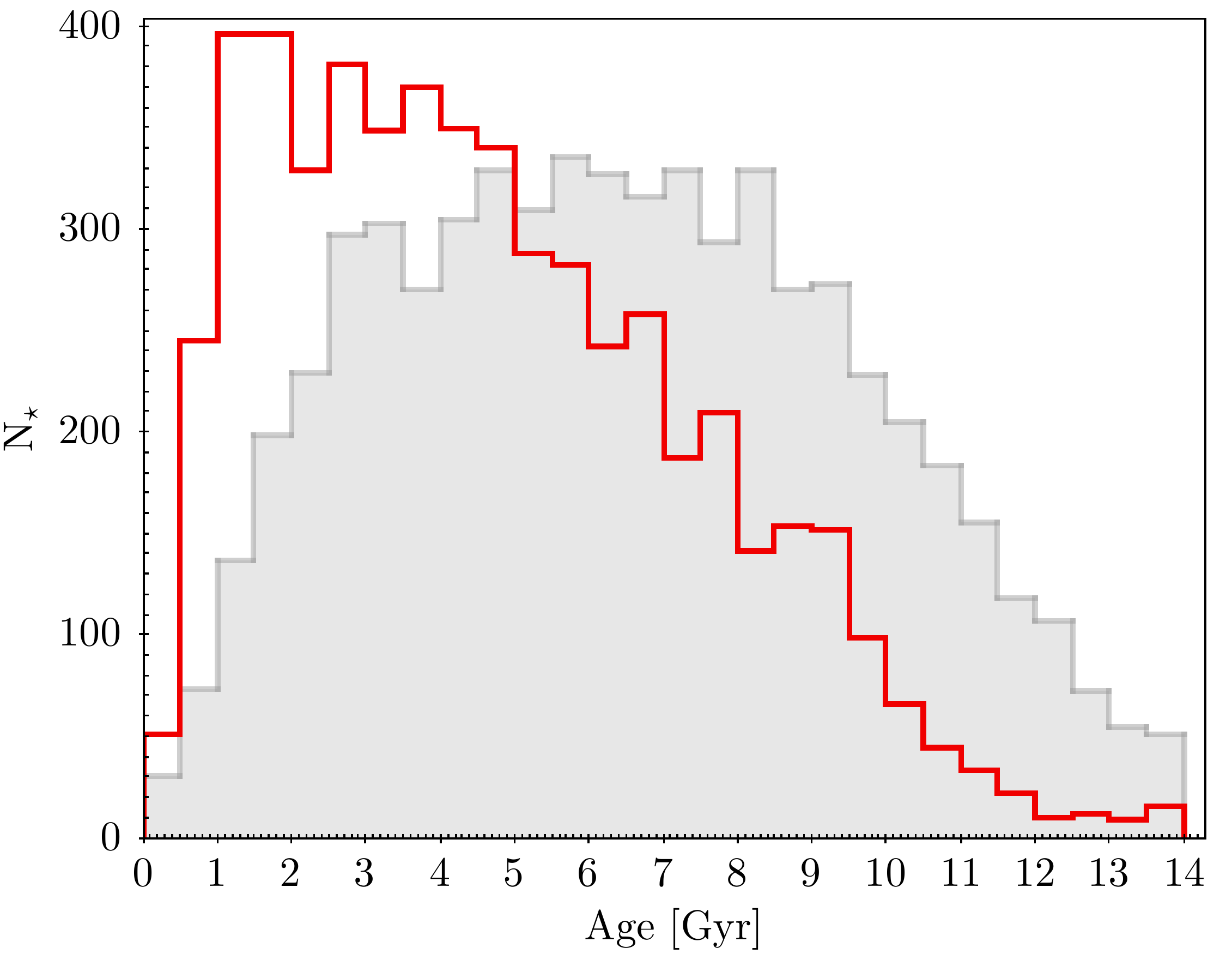} 
     \includegraphics[width=0.75\hsize]{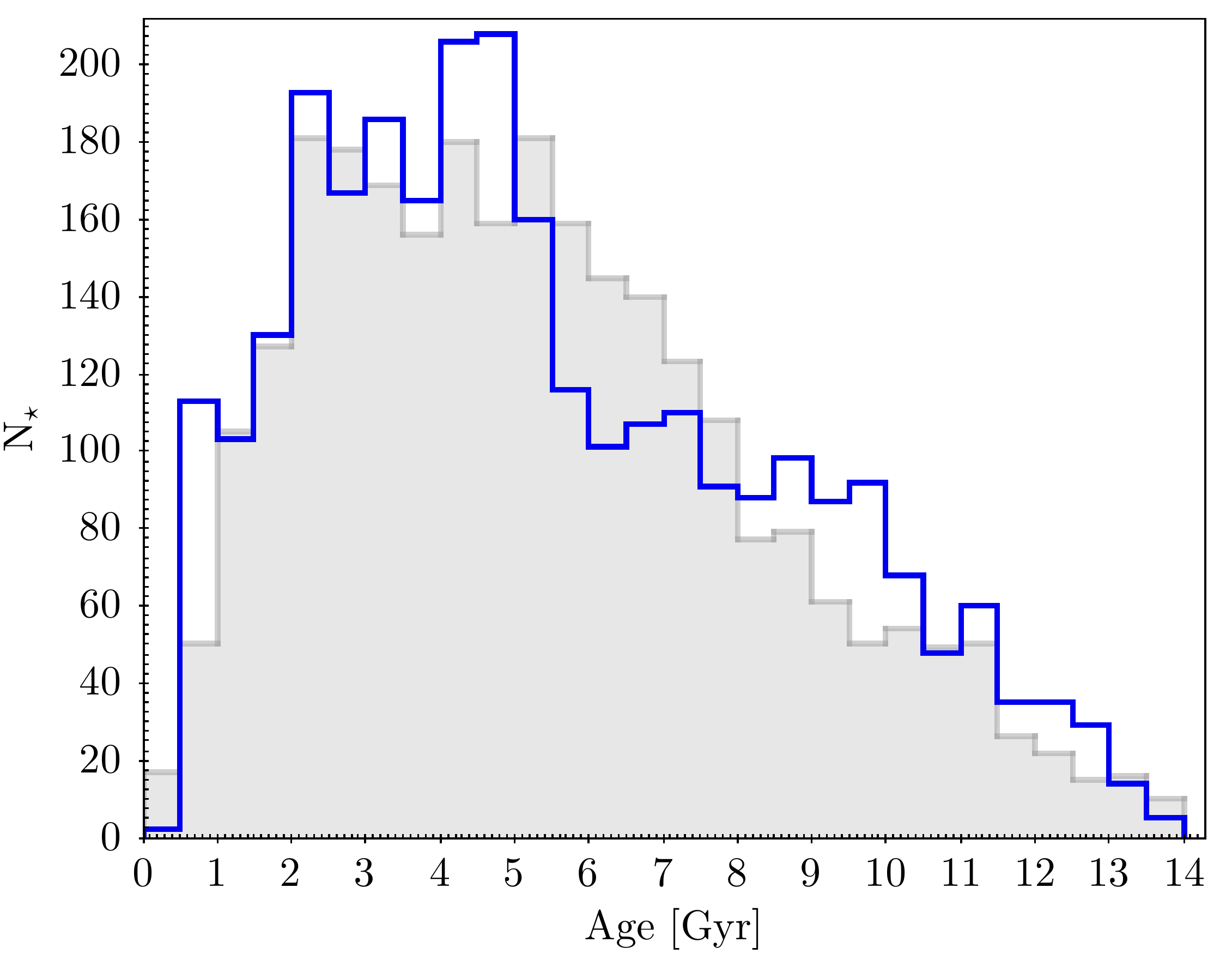} 
   \caption{ Age distribution from the APOKASC catalogue (top panel) and the M21 catalogue (bottom panel) compared to the corresponding BGM simulation (grey shaded regions). BGM simulations are presented in Sect.\ref{simulations}.}
   \label{distrib_age_model}
 \end{figure}

Figure \ref{distrib_age_model} shows the age distributions from the APOKASC and M21 samples. The two samples show a broad age distribution, peaking at different ages (between 1 and 2 Gyr for the APOKASC sample and 4.5-5.5 Gyr for the M21 sample). The Besan\c con simulations appear to be more compatible with the M21 distribution (see discussion in Section \ref{agedisp}). In Sect.\ref{simulations}, the biases induced by the two selections are discussed.  

    \begin{figure}
  \centering
    \includegraphics[width=0.85\hsize]{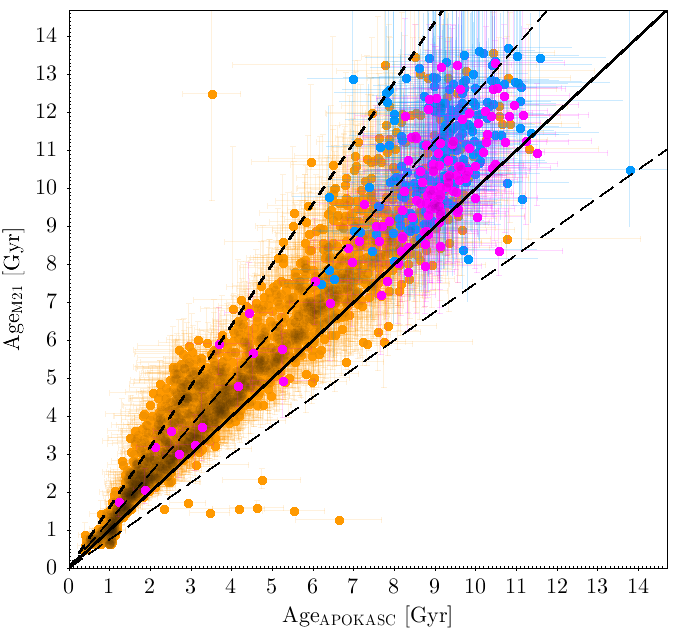}
   \caption{Comparison of the stellar ages derived by APOKASC and by M21 {for stars in common}. The solid, long-dashed, and dashed lines represent 1:1, $\pm$25\%, and 60\% in the relative differences of ages, respectively. The thin-disc, high-$\alpha$ metal-rich, and high-$\alpha$ metal-poor  thick disc stars are represented by orange, magenta, and blue circles, respectively. The separation into three stellar components is discussed in Sect.\ref{chemicalpop}.}
   \label{compa_age}
 \end{figure}
 
Figure \ref{compa_age} compares both stellar age determinations for stars in common in the two samples (2814 stars), showing that the ages obtained in M21 are systematically higher than those of APOKASC.  
The age determinations agree within 25\% for 70\% of the high-$\alpha$ metal-rich stars and for 85\% of the high-$\alpha$ metal-poor stars, whereas the discrepancy is higher for young disc stars. 
This results from the choice of stellar models to compute the stellar ages, and also the physics included in these stellar evolution models (as demonstrated in M21).

 \subsection{Chemical properties }
  
\begin{figure}
   \centering
  \includegraphics[width=\hsize,clip=true,trim= 0cm 0cm 2cm 3cm]{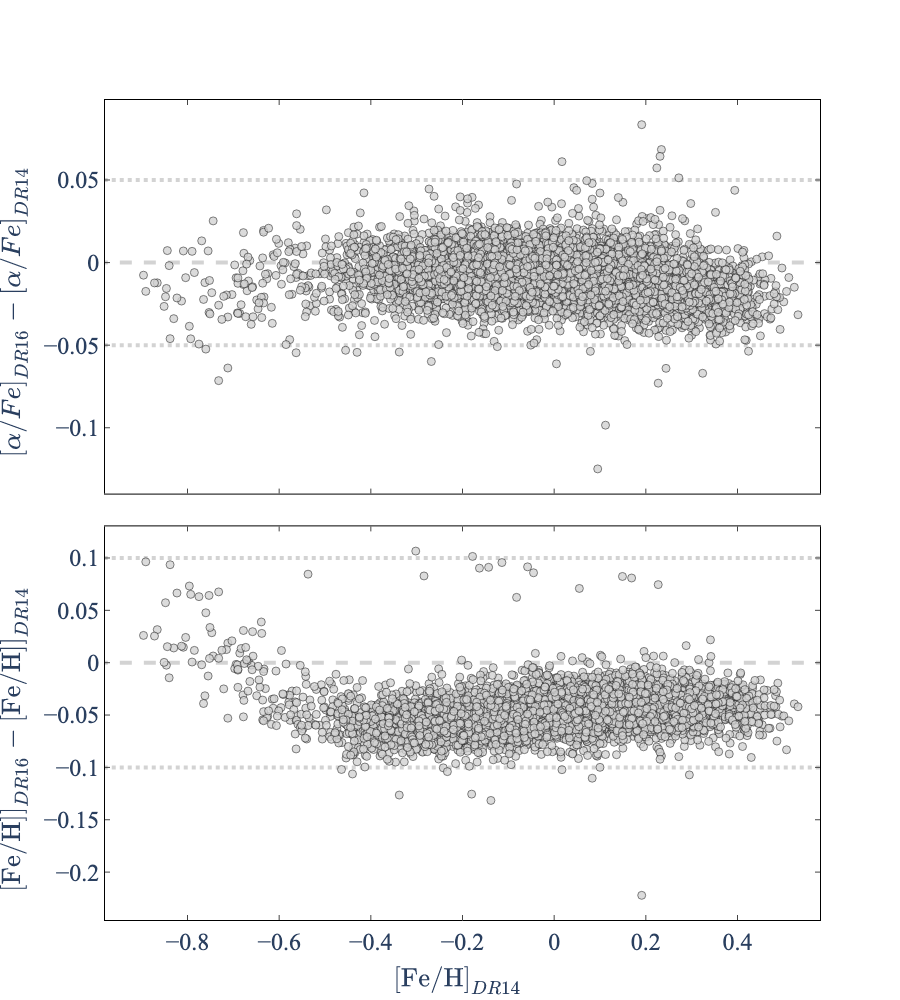}
    \caption{Difference between the metallicity and [$\alpha$/Fe] (bottom and top panel, respectively) derived in the DR16 and APOKASC catalogues (APOGEE DR14) as a function of metallicity for our sample.}
   \label{diffDR14DR16}
 \end{figure}

 We use the spectroscopic parameters from APOKASC which are taken from APOGEE DR14. \citet{APOGEEDR16} published the sixteenth data release of APOGEE. For this data release, entirely new synthetic grids were created based on MARCS stellar atmospheres \citep{Gustafsson08}. The metallicity and [$\alpha$/Fe] abundances differ from the DR14 to the DR16, implying a difference in [Fe/H] and in [$\alpha$/Fe] of less than 0.1 dex. These differences are illustrated in Fig. \ref{diffDR14DR16}. To investigate all possible parameters that could change the conclusions of the present work, we performed the same study (in parallel) using also the [Fe/H] and [$\alpha$/Fe] abundances derived from DR16. We investigate the effect of these changes between DR14 and DR16 on the relations between kinematics and metallicity or [$\alpha$/Fe] in Section~\ref{Kinematics}. For the discussion, we used DR14 because stellar ages are derived using the spectroscopy from DR14. 

\subsection{Distances and velocities}

      \begin{figure}
   \centering
\includegraphics[width=0.99\hsize,clip=true,trim= 0cm 0cm 2cm 1cm]{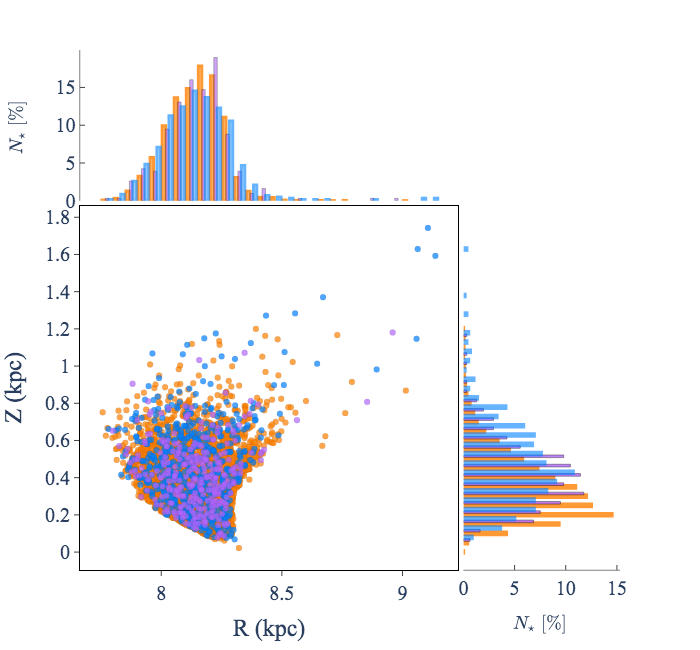}
   \caption{Vertical distance from the Galactic plane (Z) as a function of galactocentric distance (R). We also show the respective normalised distributions of R and Z. The thin-disc, high-$\alpha$ metal-rich, and high-$\alpha$ metal-poor thick-disc populations are shown in orange, purple, and blue, respectively. This separation into three stellar components is discussed in Sect.\ref{chemicalpop}.}
   \label{ZRfig}
 \end{figure}
 
In order to characterise Galactic stellar populations in velocity space, we need the distances of stars. Using the \gdrtwo catalogue, \cite{Katz2018} quantified the distance bias introduced when using the inverse of the parallax and found that the cut at 20\% relative uncertainty in parallax leads to unbiased distances out to about 1.5 kpc,  with overestimates of the order of 17\% at 3 kpc. Alternatively, Bayesian methods could be used to derive \gdrtwo\ distances from parallaxes \citep{BailerJones15,BailerJones18,Anders19, Queiroz20}. As our sample extends beyond 1.5 kpc, but within 4 kpc, we adopted the StarHorse distances. StarHorse is a Bayesian spectrophometric code that derives distances from a combination of the ASPCAP stellar parameters with Gaia DR2 and photometric surveys, thus achieving high precision \citep[][]{Queiroz18, Queiroz20}. The use of additional information coming from APOGEE spectra and complementary photometry makes StarHorse distances even more precise, especially beyond 1-2 kpc \citep[for a detailed description see][]{Anders19, Queiroz20}. 
The resulting sample contains 5149 and 2660 stars for APOKASC and M21 sample, respectively (see Table \ref{composition_pops}). The Galactic velocities (V$_R$, V$_\varphi$, V$_Z$) and coordinates (R, $\varphi$, Z) were computed using the distances computed by \citet{Queiroz20} and following the same method developed by \citet{Katz2018}. We adopted the distance of the Sun to the Galactic centre R$_\odot$= 8.34kpc, the circular velocity at the solar radius of V$_c$ = 240 km s$^{-1}$ from \citet{Reid14}, and the height of the Sun given by \citet{Chen01} of 27 pc.  Here, V$_R$ is oriented outwards the Galactic centre, V$_\varphi$ following the direction of Galactic rotation, and V$_Z$ towards the North Galactic pole. We assume the peculiar velocity of the Sun with respect to the local standard of rest taken from \citet{Schonrich10}, (U$_\odot$, V$_\odot$,W$_\odot$) = (11.10, 12.24, 7.25) km s$^{-1}$. Figure \ref{ZRfig} shows the location of our APOKASC sample in vertical distance $Z$ to the Galactic plane and galactocentric distance $R$.

\section{Forward modelling using the Besan\c con Galaxy Model}
\label{simulations}

As discussed by \citet{Pinsonneault14,Pinsonneault18}, the selection function of the APOKASC sample is not straightforward. Indeed it  requires very good knowledge of the target selection for both \textit{Kepler} and APOGEE. It should be borne in mind that the \textit{Kepler} mission is designed to detect exoplanets and not to study the properties of stars in different stellar populations. 
As the selections made by APOKASC and M21 are different, the induced biases have to be investigated before drawing any conclusions from the characteristics of the stellar populations in the sample, such as age or metallicity distributions. Galactic stellar population synthesis models  can be used to provide a forward modelling that allows  mock catalogue simulations to be made where selection bias on observable parameters can be accurately reproduced, allowing us to identify possible biases in the inferred distributions. This method was used to study red giants observed by CoRoT and APOGEE by \citet{Anders16, Anders17a, Anders17b}. We use the comprehensive description of Galactic stellar populations provided by BGM \citep[e.g.][]{Lagarde17} to produce a realistic data simulation. 

\subsection{Model description}

BGM is based on a scenario for Galaxy formation and evolution that reflects our present understanding of the Milky Way. Four stellar populations are considered: a thin disc, a thick disc, a bar, and a halo, with each stellar population having a specific density distribution. The simulation was computed using the revised scheme of BGM \citep{Czekaj14} where the stellar content of each population is modelled through an initial mass function (IMF) and a star formation history (SFH), and follows evolutionary tracks \citep[revised in][]{Lagarde17, Lagarde19}. The resulting astrophysical parameters are used to compute their observational properties using atmosphere models and assuming a 3D extinction map. A Galactic dynamical model is used to compute radial velocities and proper motions, as described by \citet{Robin17}. As this study focuses on the thin and thick discs, the main ingredients for both populations are described below: 

\begin{itemize}
\item {The IMFs} for both stellar populations are taken from the analysis of the Tycho-2 data \citep{Mor18}. \\
\item {The SFH} of the thin disc is from \citet{Mor18}, while for the thick disc the SFH is modelled assuming a two-episode formation \citep{Robin14}, describing the young and the old thick discs, and using a Gaussian age distribution from 8 to 12 Gyr and from 10 to 13 Gyr, respectively. \\
\item The iron abundance [Fe/H] metallicity and the metallicity dispersion for the thin disc are estimated assuming {the age--metallicity relation} deduced  from \citet{Haywood06} \citep[for more details see][]{Czekaj14}: 
\begin{equation}
[Fe/H]=-0.016\times age +0.01,
\end{equation}
\begin{equation}
\sigma_{[Fe/H]}=0.010\times age +0.1, 
\end{equation}

with the stellar age given in gigayears, 
and a radial metallicity gradient of -0.07 dex kpc$^{-1}$ limited to Galactocentric radii of between 5 and 12 kpc.

Considering the thick disc, a mean metallicity is assumed for the young thick disc and the old thick disc (-0.5 and -0.8 dex, respectively) with a dispersion of 0.3 dex.\\
\item The adopted [$\alpha$/Fe] versus [Fe/H] relations follow the trend observed in the DR14 of APOGEE for both stellar populations.
Namely, for $[Fe/H]\leq0.1$, 
\begin{equation}
      [\alpha/Fe] =
     \left\{
     \begin{array}{rl}
     -0.121\times\text{[Fe/H]}+0.0259\\
      \text{for\ the\ thin\ disc\ stars}, \\
     0.320-\text{exp}(1.19375\times\text{[Fe/H]}-1.6) \\
    \text{ for the thick disc stars}.
     \end{array}
     \right.
     \end{equation}

For [Fe/H]$>$0.1, [$\alpha$/Fe] is assumed to be solar. To these relations an intrinsic Gaussian dispersion of 0.02 dex is added.\\
\item The adopted {velocity dispersions} as a function of age have been constrained from the RAVE survey \citep[DR4, ][]{Kordopatis13a} and proper motions from the TGAS part of the \textit{Gaia} DR1 \citep{Gaia16} by \citet[][see their Table 4]{Robin17}.  \\

\item The rotation curve is given by \citet{Sofue15}. Furthermore, the asymmetric drift is also taken into account and comes from \citet{Robin17}. The dynamical statistical equilibrium is ensured by assuming the St\"{a}ckel approximation of the potential from \citet{Bienayme15, Bienayme18}. \end{itemize}

\begin{figure}
   \centering
    \includegraphics[width=0.75\hsize,clip=true,trim= 0cm 0cm 0cm 0cm]{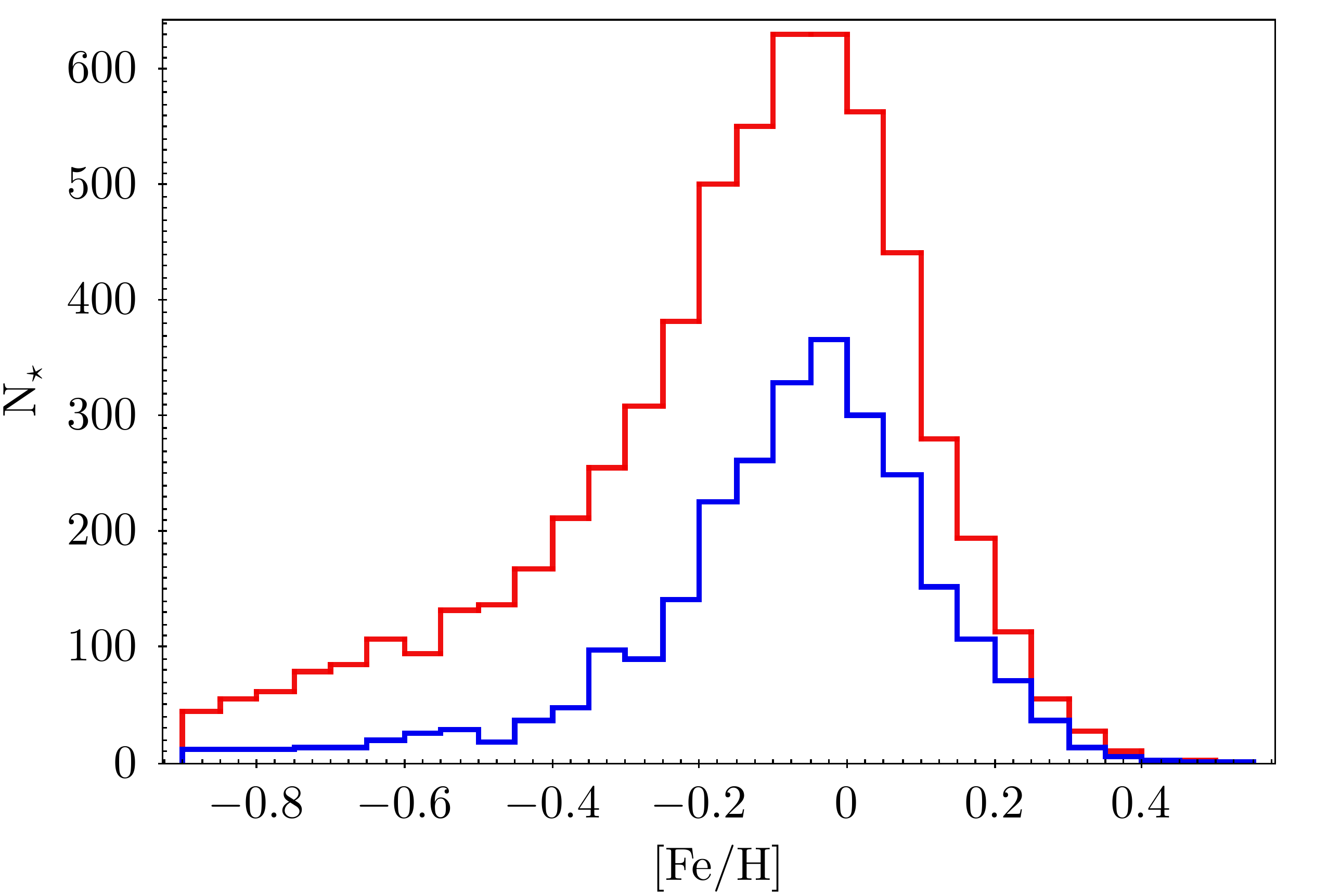} 
    \includegraphics[width=0.75\hsize,clip=true,trim= 0cm 0cm 0cm 0cm]{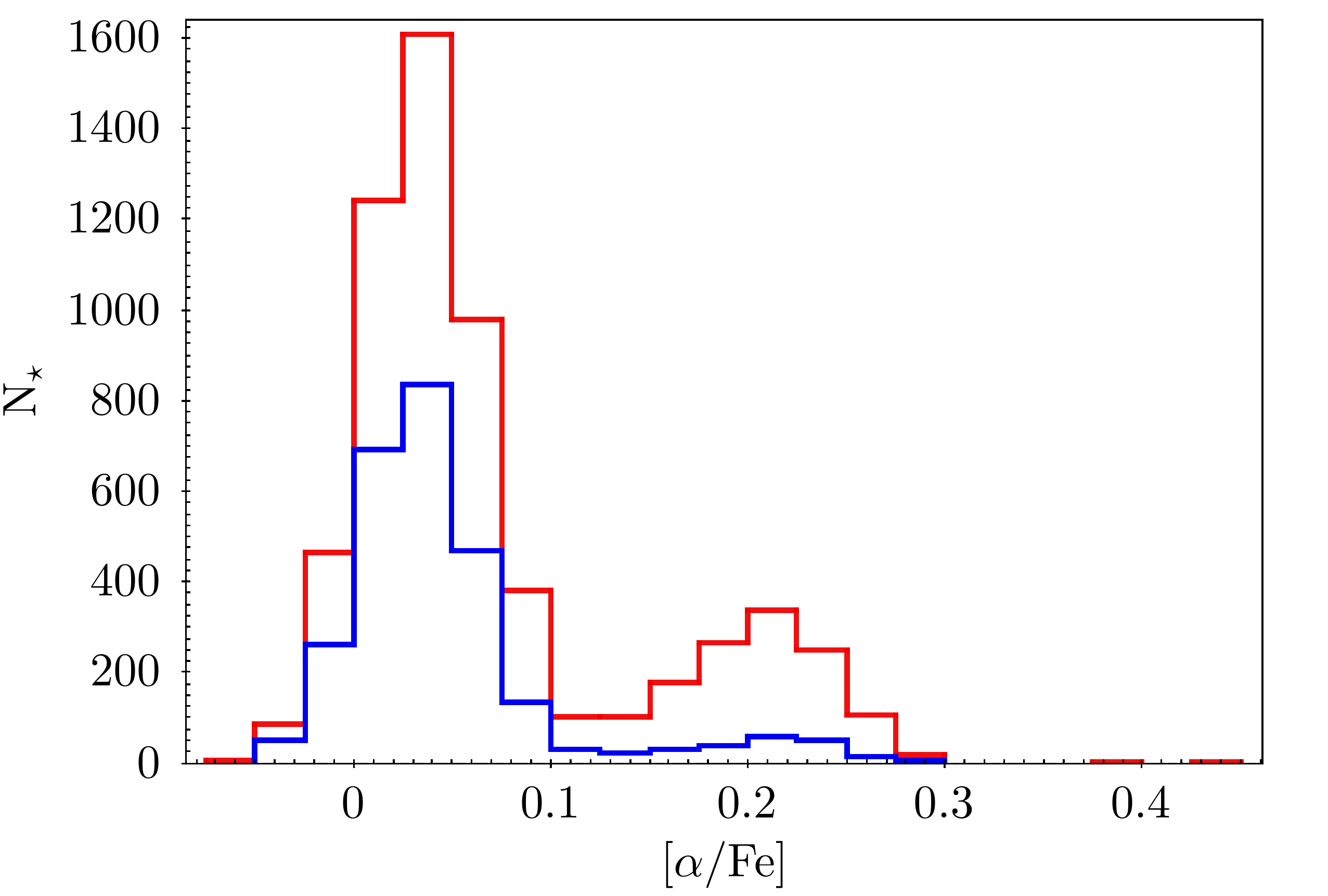}  
    \includegraphics[width=0.75\hsize,clip=true,trim= 0cm 0cm 0cm 0cm]{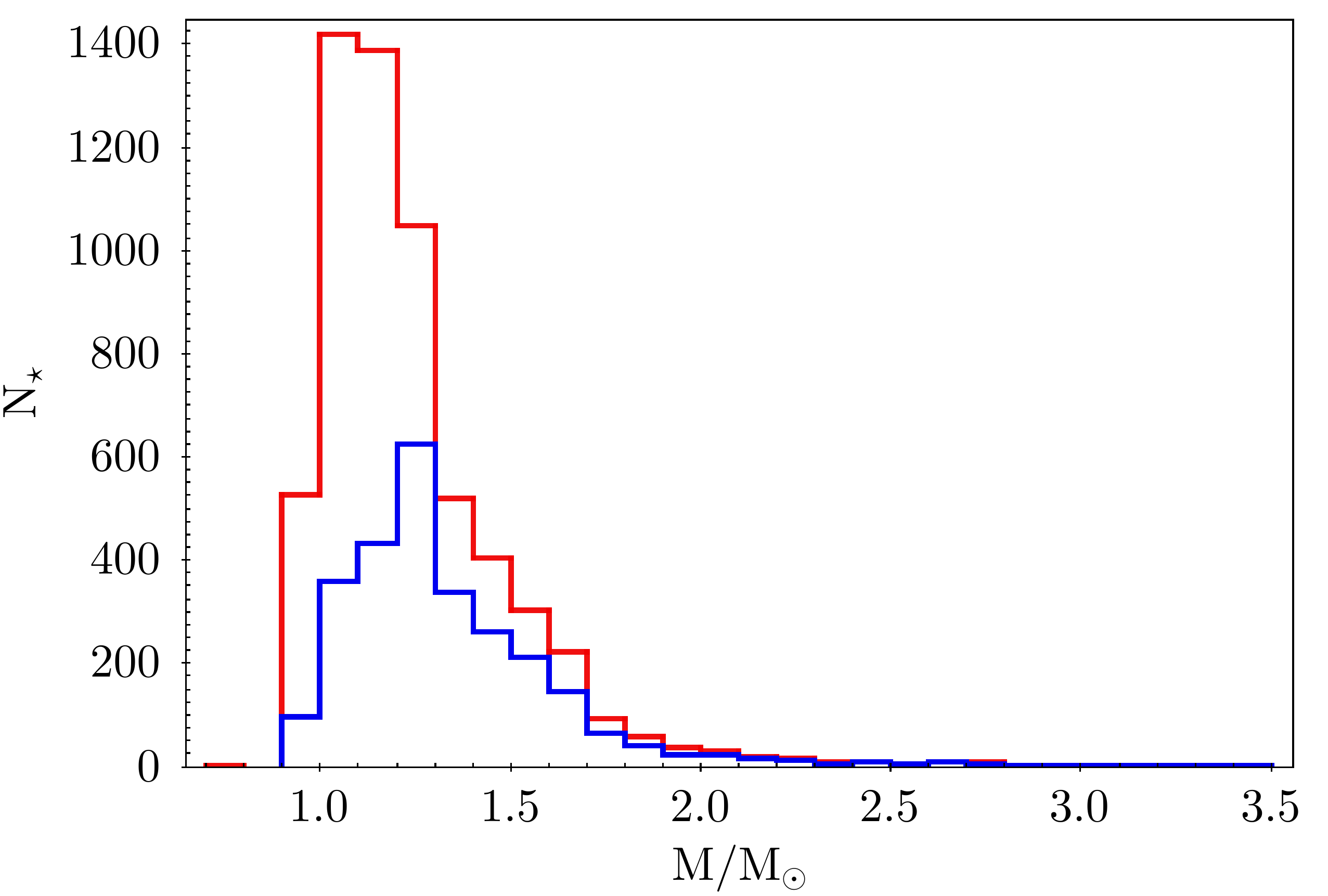} 
    \includegraphics[width=0.75\hsize,clip=true,trim= 0cm 0cm 0cm 0cm]{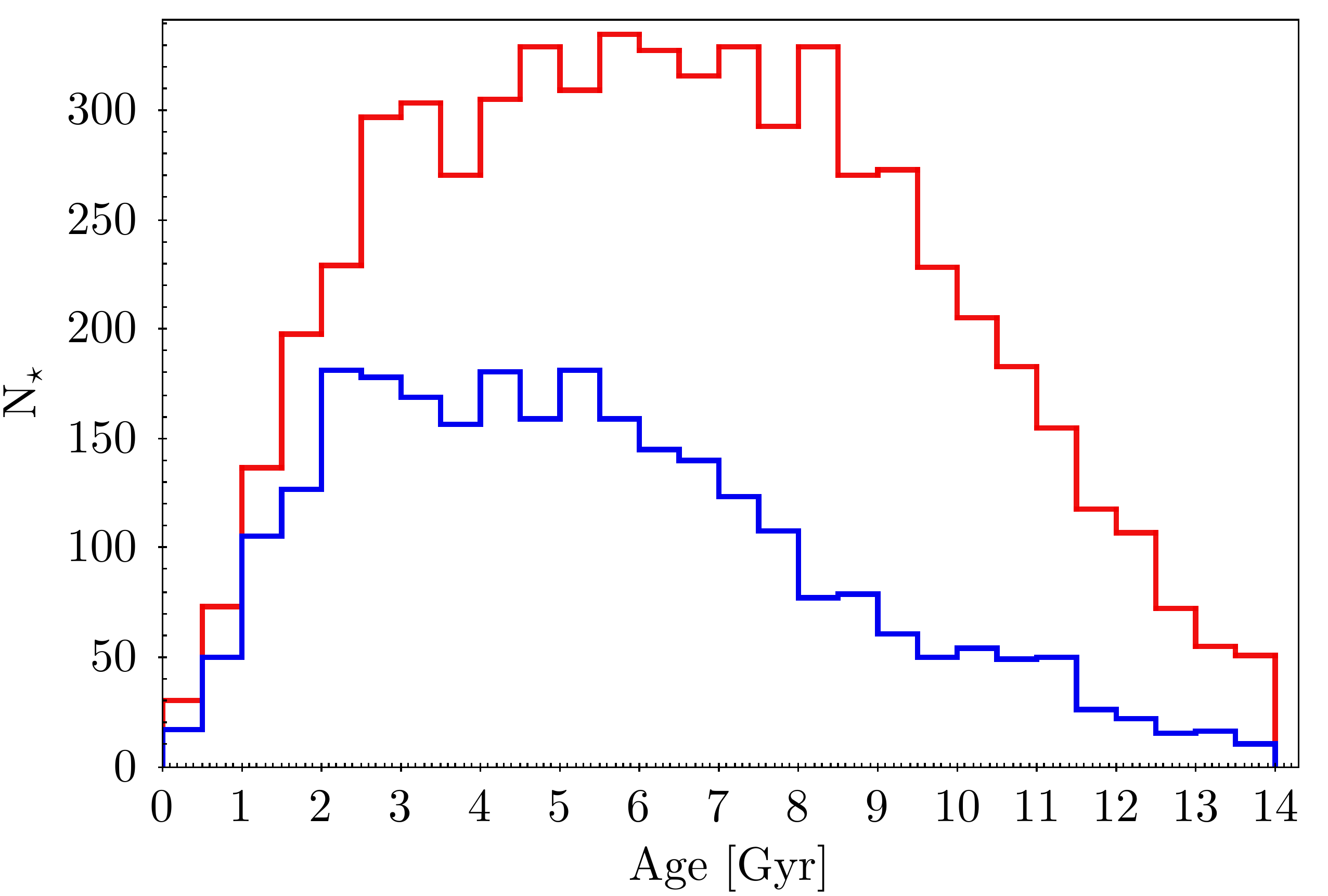}     
    \caption{Distributions of [Fe/H], [$\alpha$/Fe], stellar mass, and age from the mock catalogues of APOKASC (red solid lines) and M21  (blue solid lines) samples, computed with the BGM. The age distribution takes into account the typical error on the seismic age determination, and follows a Gaussian distribution with a dispersion of 25\%. 
    }
    \label{compa_simu}
 \end{figure} 

\subsection{Simulating samples}

\subsubsection{APOKASC sample}
\label{APOKASCsim}

To compare the distributions before and after the observational selections and to reveal whether or not these cuts change the overall distributions, we ran a simulation of the \textit{Kepler} field.  We then applied the same selections as for the data, as described in Section \ref{data}. 
We compare our BGM simulation with the APOKASC data  including the $H$ magnitude and T$_{\rm{eff}}$ selections using a Kolmogorov-Smirnov (KS) test of the magnitude distributions in different bands ($H,J,K$). As the sample is not complete at higher and lower values of $H$ magnitude, the KS test returns a low p-value, indicating that the synthetic stellar sample and the observed one do not come from the same parent distribution. For that reason, we prefer to restrict our observed sample to 7.4$\leq H$<10.8 in this study. Using this selection in the $H$ magnitude, we obtain a KS test that shows better agreement (for $H$ band p-value=0.999, $K$ band p-value=0.97, and $J$ band p-value=0.99), indicating a better representative observed sample. \\

In summary, we apply the following selection to both the BGM simulation and the observations: we select H in the range 7.4$\leq H$<10.8; T$_{\rm{eff}}<5300$~K;
$0.4<\Delta\nu<20~\mu Hz$, and $2<\nu_{max}<250~\mu Hz$. Thanks to recent improvements, BGM can simulate seismic properties such as $\Delta\nu$, $\nu_{max}$ or the period spacing of g-modes ($\Delta\Pi_{(\ell=1)}$) \citep[see][for more details]{Lagarde17}. Also, only single stars are selected. 

\subsubsection{Simulating the Miglio et al. 2020 sample}
\label{ M21 sim}

To obtain the mock catalogue of M21, we have to take into account their additional selections. We refine the previous selection done for the APOKASC catalogue using two additional criteria (see discussion in Sect.\ref{M21sample}), namely the mass of clump stars, M$_{Clump}\geq$1.2M$_\odot$, and
 the radius of RGB stars, R$_{RGB}$<11R$_\odot$.

\subsection{Comparison of both simulations}

Figure \ref{compa_simu} presents the distributions of stellar properties  using the mock catalogues computed with the BGM, including the selection discussed in Sect. \ref{APOKASCsim} and Sect.\ref{ M21 sim} for the APOKASC and M21 samples, respectively. Additional selections done by M21 favour lower [$\alpha$/Fe] and solar-metallicity stars, which leads to a larger proportion of thin disc stars ($\sim$91\%) being selected from the M21 mock catalogue  than from the APOKASC mock catalogue ($\sim$77\%). Thus,  M21 selections reject older low-mass stars (see lower panels of Fig. \ref{compa_simu}), and are thus biased towards younger ages compared to the APOKASC mock catalogue. 
Figure \ref{distrib_age_model} presents a comparison between the simulated and observed age distributions. Better agreement is seen between the BGM simulations and the observations from M21. This is discussed in detail in Sect. \ref{agedisp}. In the following sections, we use these two simulations to interpret the relationships seen between observed stellar properties and age.


\begin{figure}
   \centering
    \includegraphics[width=0.96\hsize,clip=true,trim= 0cm 0cm 0cm 2.5cm]{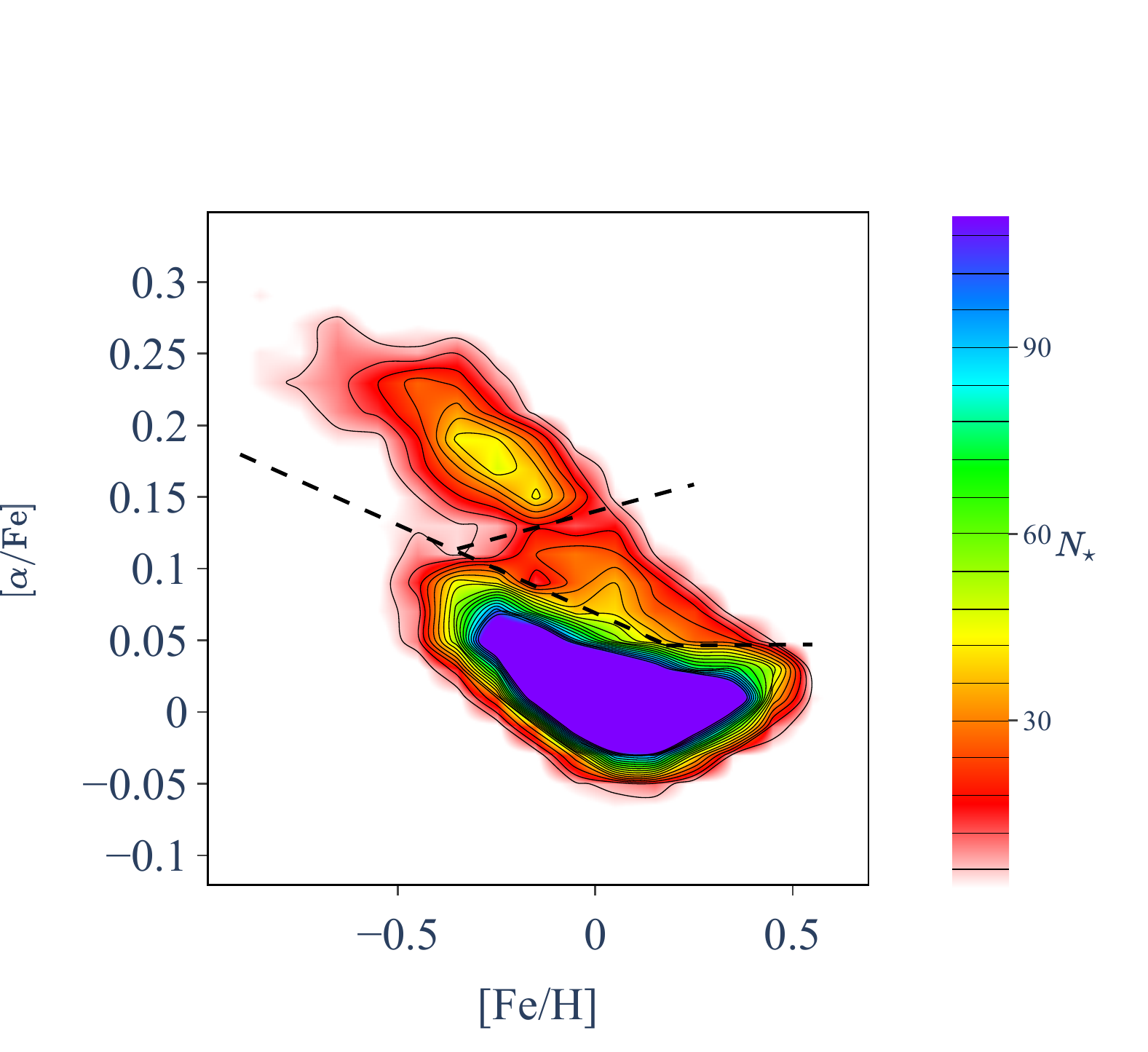}
    \includegraphics[width=0.96\hsize,clip=true,trim= 0cm 0cm 0cm 2.5cm]{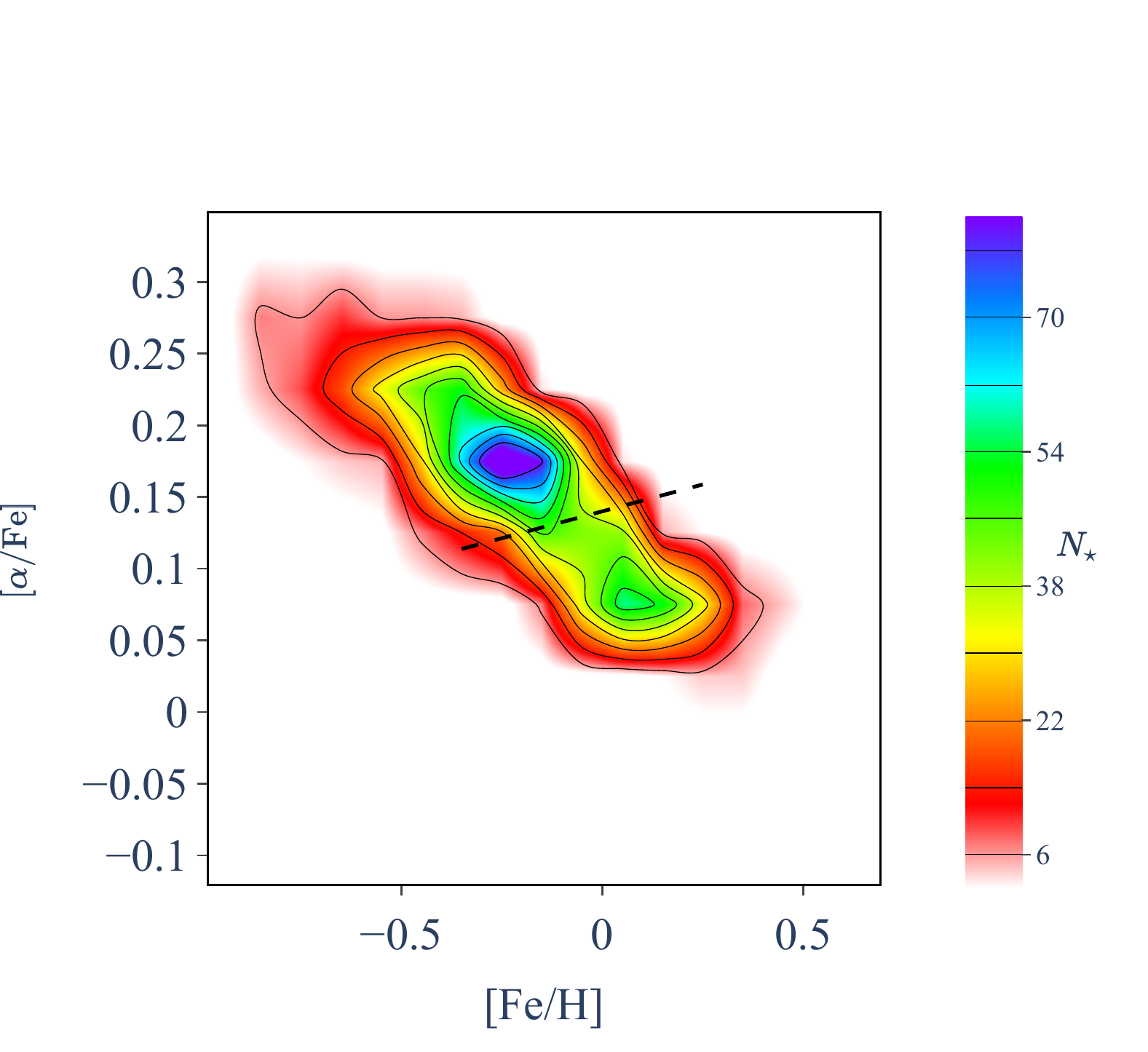}
    \includegraphics[width=0.85\hsize,clip=true,trim= 0cm 0cm 1cm 5.5cm]{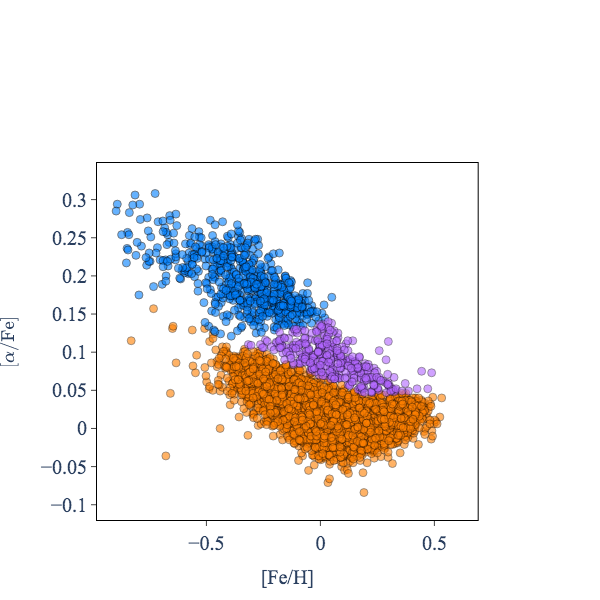}
        \caption{Distribution of stars in the [$\alpha$/Fe] vs. [Fe/H] plane. Chemical properties are taken from APOKASC. The dashed lines show the limits used to separate the three disc populations: h$\alpha$mp, h$\alpha$mr thick discs, and thin discs (top and middle panels). The middle panel excludes the thin disc stars to focus on the thick disc stellar populations. In the bottom panel, the sample is separated into three populations: Thin disc, h$\alpha$mr, and h$\alpha$mp thick-disc populations are represented with orange, purple, and blue dots, respectively.}
   \label{contourfig}
 \end{figure}

\section{Age--chemical properties of stellar populations}
\label{chemicalpop}

The existence of the Galactic thin--thick disc dichotomy was originally discussed by \citet{Yoshi82} and \citet{GiRe83}, leading studies of the structural and chemo-dynamical transition between the two disc populations. Using data from a low-resolution spectroscopic survey, \citet{Lee11b} studied the thin--thick disc transition with a robust statistical approach, inside and outside the solar neighbourhood. Using a sample of G-type stars with spectroscopic measurements from the SEGUE survey, \citet{Yanny09} showed that the distribution of stars in the [$\alpha$/Fe] versus [Fe/H] plane is bimodal and used it to separate the thin- and thick-disc populations, although no clear gap was observed between the two populations. Interpretations of the SEGUE data vary in the literature, from authors that claim no thin--thick disc distinction \citep{Bovy12b} to those allowing the existence of a distinct Galactic thick disc. Selection effects have been evoked to explain the difference between \citet{Lee11b} and  \citet{Bovy12b}. \citet{Boeche13} studied giant stars observed by the RAVE survey and identified three stellar populations that could be associated with the Galactic thin disc, a dissipative component corresponding to a thick disc as well as a component including halo stars. Their [$\alpha$/Fe] versus [Fe/H] plane does not show a clear separation between thin and thick discs. Using high-resolution spectroscopic observations, studies of the solar neighbourhood underlined the existence of a gap between the thin and thick discs \citep{Fuhrmann04, Reddy06,Bensby07}. Later studies confirmed this separation using observations with higher spectral resolution, such as the HARPS sample of F, G, and K stars in the solar neighbourhood by \citet{Adibekyan13}, using the GIRAFFE spectra of FGK-type stars collected in the Gaia-ESO survey \citep{RecioBlanco14}, using GALAH survey data \citep{Duong18}, using APOGEE Data Release 10 \citep{Anders14} and APOGEE Data Release 12 \citep{Hayden15}, and more recently, extending all the way into the innermost regions of the Galaxy by using  the APOGEE Data Release 16 \citep{Queiroz20}.

\subsection{Chemical separation of the Galactic disc populations in the sample}

The APOKASC catalogue offers the opportunity to characterise the distribution of disc stars in the [$\alpha$/Fe] versus [Fe/H] plane. Figure \ref{contourfig} (top panel) shows the density distribution of our sample in this plane. The well-known decreasing evolution of $\alpha$-abundances with metallicity \citep{Pagel87, MatBro91} is present within the expected values for disc stars. The bimodal density distribution with the $\alpha$-rich, metal-poor thick-disc population, and the $\alpha$-poor, metal-rich thin disc population is clearly visible. Whereas most previous studies attribute the whole upper sequence to the thick disc, isolating the `common' thick disc (middle panel of Fig.\ref{contourfig}), two density peaks appear in the  [$\alpha$/Fe] versus [Fe/H] plane: one centred at ([Fe/H],[$\alpha$/Fe]) $=$ ($-$0.25,0.18) and a second centred at ([Fe/H],[$\alpha$/Fe]) $=$ (0.05,0.075)\footnote{The bi-modality observed in the [Fe/H]-[$\alpha$/Fe] and two density peaks in the thick disc are also visible using the APOGEE-DR16. }. These two groups were already underlined by \citet{Adibekyan13} while studying HARPS FGK dwarfs. In addition, \citet{Anders18}, using the dimensionality reduction algorithm t-SNE method,  also identified a metal-rich $\alpha$-rich population \citep[see also, ][]{Guiglion19}.  
In what follows, we consider these as two different populations, namely the high-$\alpha$ metal-poor and high-$\alpha$ metal-rich thick-disc populations (hereafter, h$\alpha$mp, and h$\alpha$mr thick-disc populations, respectively), in order to investigate their properties separately. We defined the limits to separate the populations following isodensity contours, which are shown by dashed lines in Fig. \ref{contourfig}. The resulting separation in the [$\alpha$/Fe] versus [Fe/H] plane is shown in the bottom panel of the Fig. \ref{contourfig}. Table \ref{composition_pops} presents the proportion of stars in different stellar populations for both samples. Although the following equations depend on the spectroscopic data used, we define the three stellar populations as follows: 

\begin{itemize}
\item \textbf{Thin disc stars:} \\
if [Fe/H]$\ge$0.18 dex, $[\alpha/Fe]< 0.002\times[Fe/H]+0.046$, and\\ 
if [Fe/H]$\le$0.18 dex, $[\alpha/Fe]< $-$0.123\times[Fe/H]+0.069.$ 
\item \textbf{h$\alpha$mp thick disc stars:} \\
if [Fe/H]$\ge$$-$0.35dex, $[\alpha/Fe]\ge0.075\times[Fe/H]+0.14$, and\\
if [Fe/H]<$-$0.35 dex, $[\alpha/Fe]\ge$-$0.123\times[Fe/H]+0.069.$
\item \textbf{h$\alpha$mr thick disc stars:} \\
if [Fe/H]$\ge$0.18 dex, $[\alpha/Fe]\ge 0.002\times[Fe/H]+0.046$, and\\
if $-$0.35<[Fe/H]$\le$0.18 dex, \\
$-0.123\times[Fe/H]+0.069<[\alpha/Fe]< 0.075\times[Fe/H]+0.14.$
\end{itemize}

Using the simulations from the BGM, we estimate that the proportion of halo stars in the h$\alpha$mp population is negligible (i.e. three halo stars in the h$\alpha$mp population, which is ~0.22\% of the h$\alpha$mp populations). 
 

\begin{table}
    \centering
     \caption{Proportion of stars in stellar populations as defined in Fig.~\ref{contourfig} from the APOKASC catalogue and using crossed-match with  M21. }
         \begin{tabular}{ c | c c | c c }
                                           & \multicolumn{2}{c|}{APOKASC} &  \multicolumn{2}{c}{ M21 } \\
        Stellar populations &   All & \gaia & All & \gaia \\
         \hline
        Thin disc &  4491 &  4265 & 2479 & 2345\\
        h$\alpha$mr thick disc  & 323   & 306 & 130 & 122 \\
        h$\alpha$mp thick disc& 612  & 578 & 205 & 193\\
        \hline
        \hline
        Total &5426  &  5149 & 2814 & 2660 \\
    \end{tabular}
    \tablefoot{The column named \gaia gives the number of stars for which we computed the kinematics. }
    \label{composition_pops}
\end{table}

\subsection{Age distributions of the three chemically defined populations}
 \label{agedisp}

\begin{figure*}
  \centering
      \hspace{1cm}\textbf{thin disc} \hspace{4cm} \textbf{h$\alpha$mr thick disc}\hspace{4cm} \textbf{h$\alpha$mp thick disc} \hspace{0.5cm} \par
      \vspace{0.2cm}
    \includegraphics[width=0.33\hsize,clip=true,trim= 0cm 0cm 0cm 0cm]{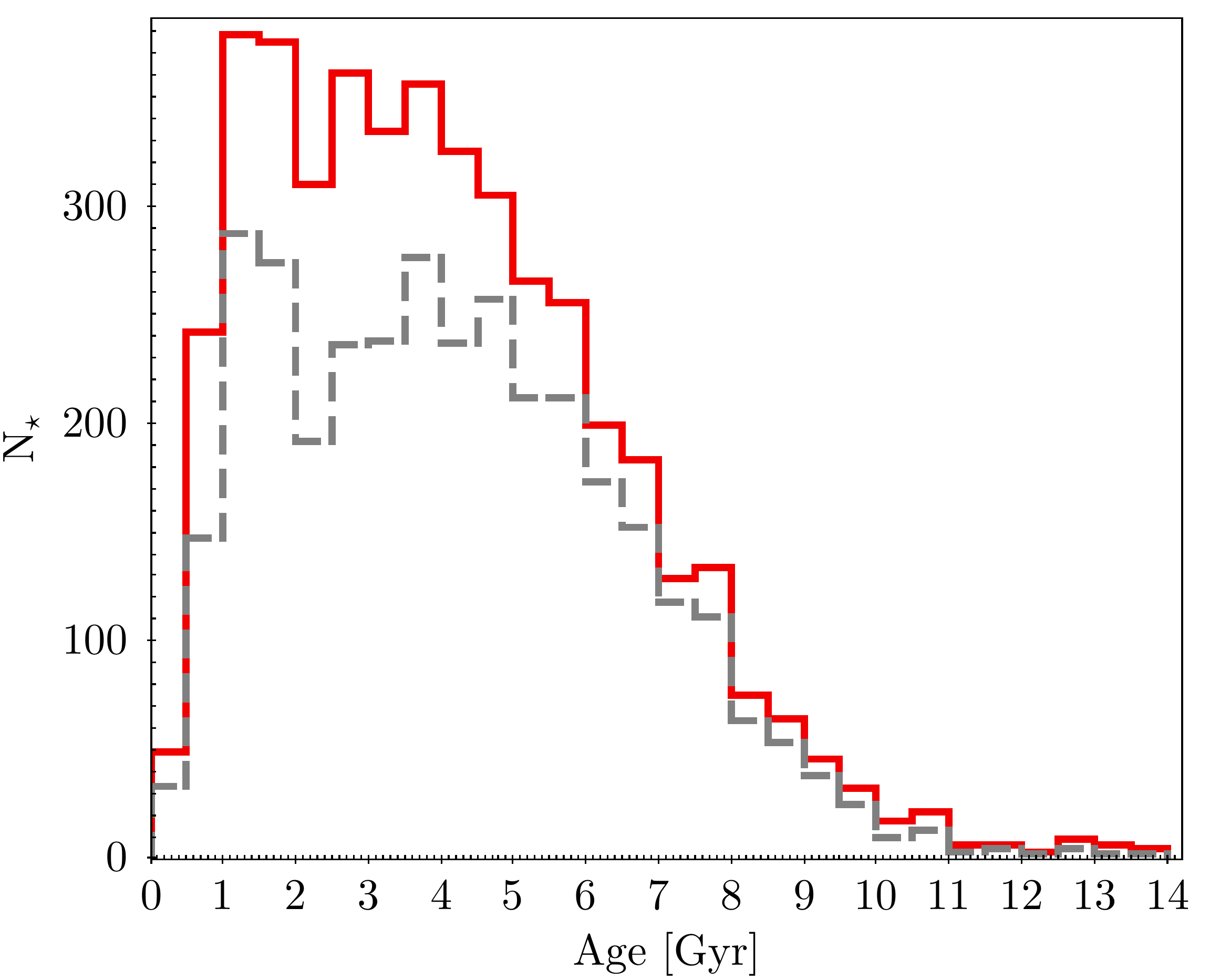}
    \includegraphics[width=0.33\hsize,clip=true,trim= 0cm 0cm 0cm 0cm]{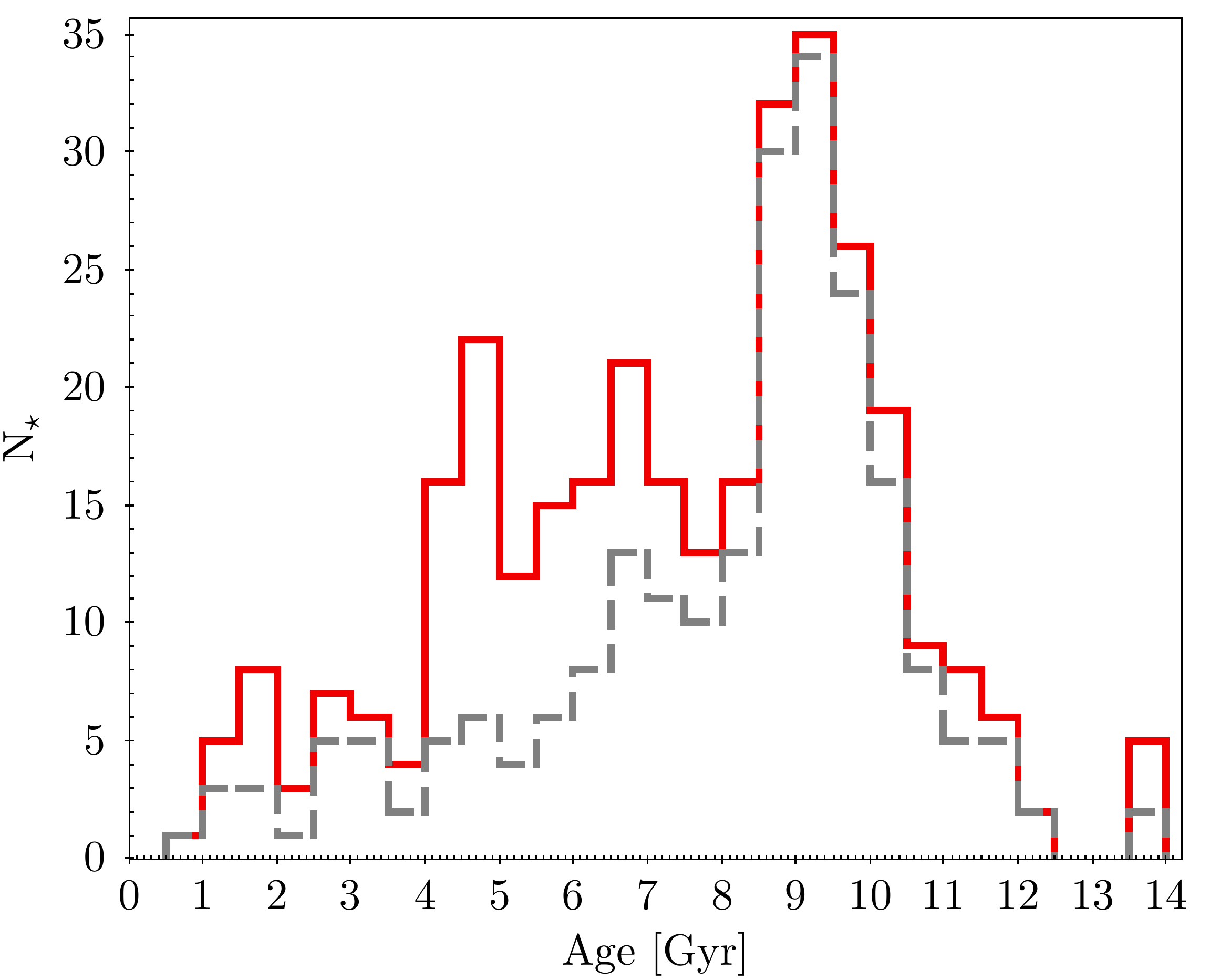}
    \includegraphics[width=0.33\hsize,clip=true,trim= 0cm 0cm 0cm 0cm]{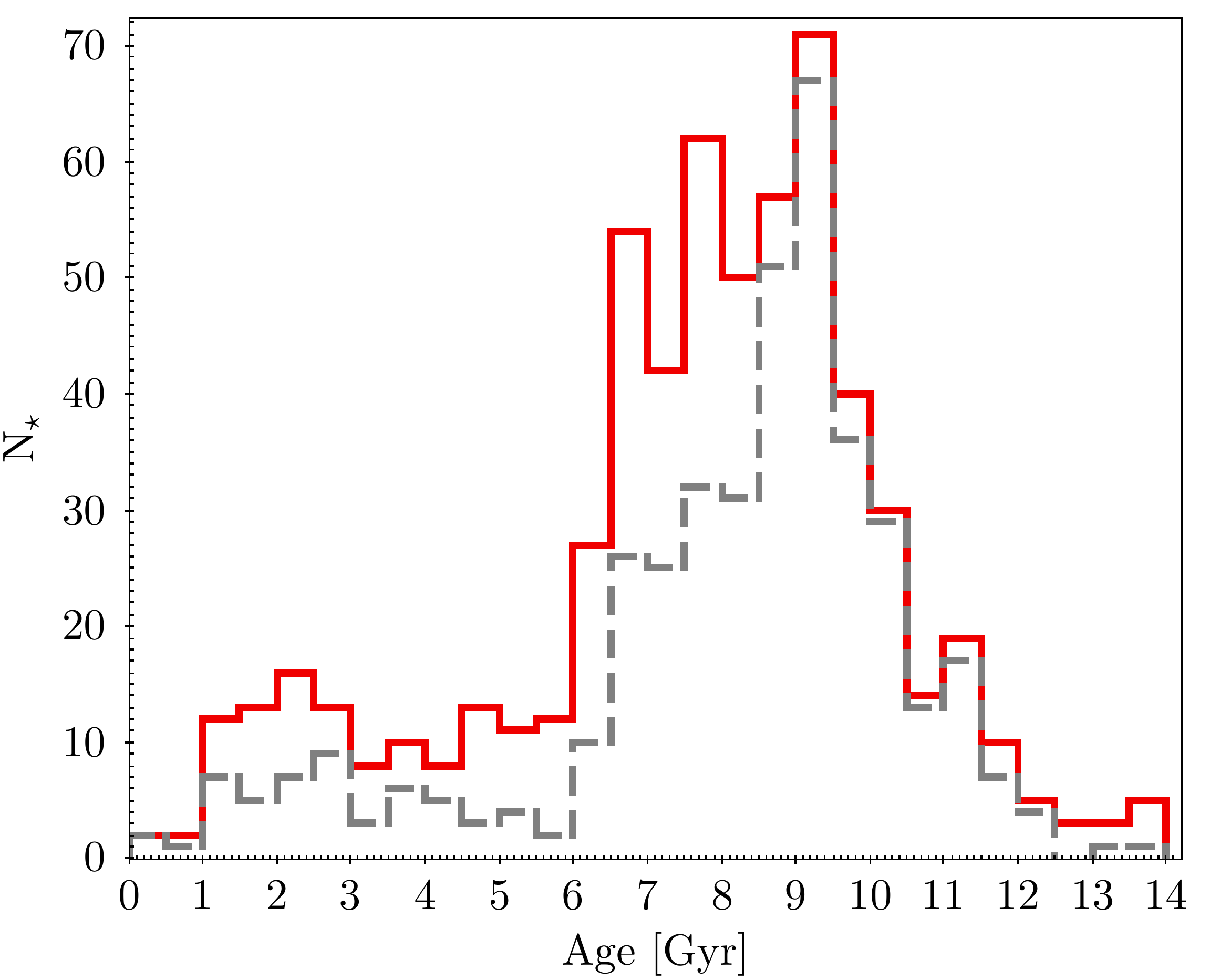}

    \includegraphics[width=0.33\hsize]{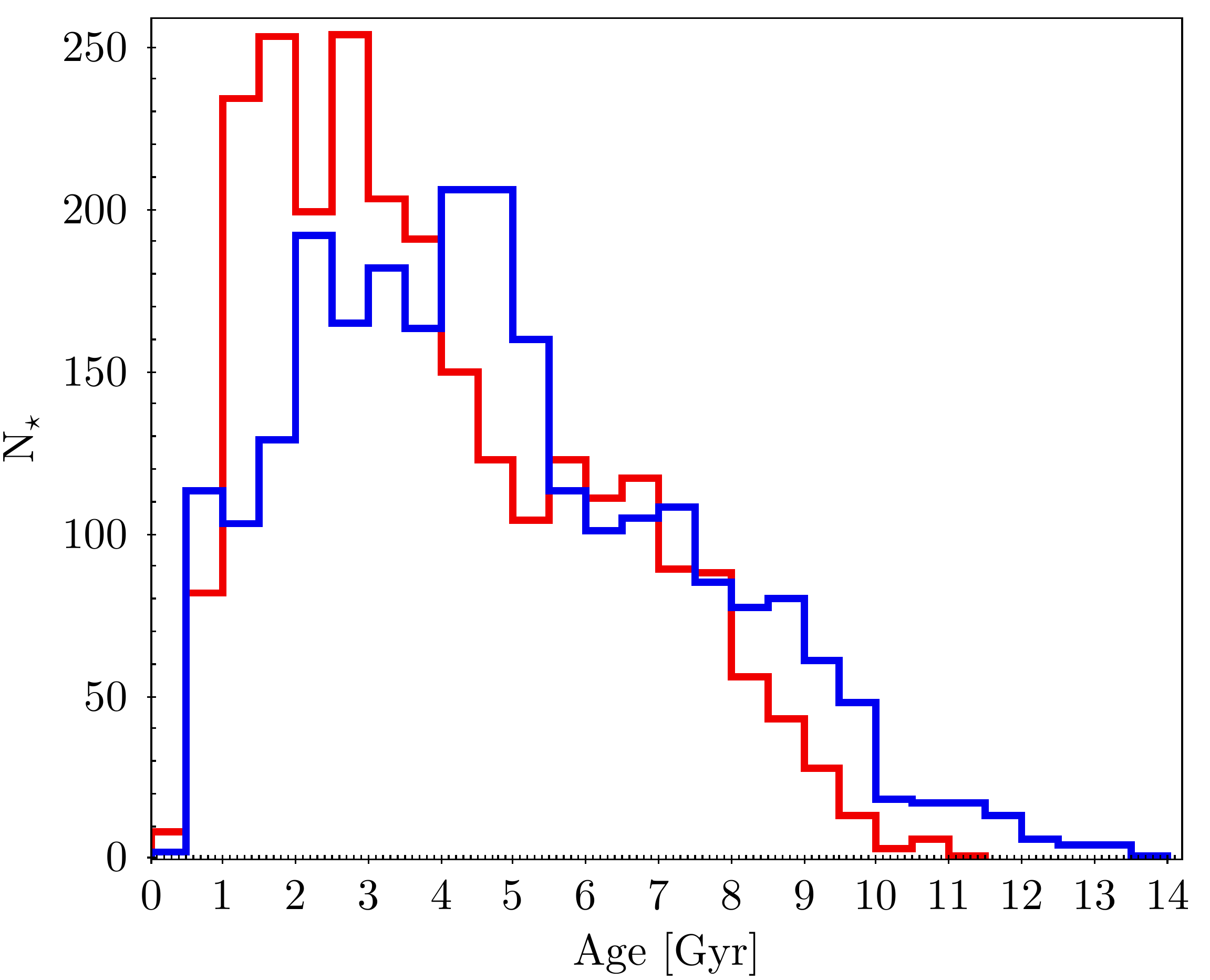}
      \includegraphics[width=0.33\hsize]{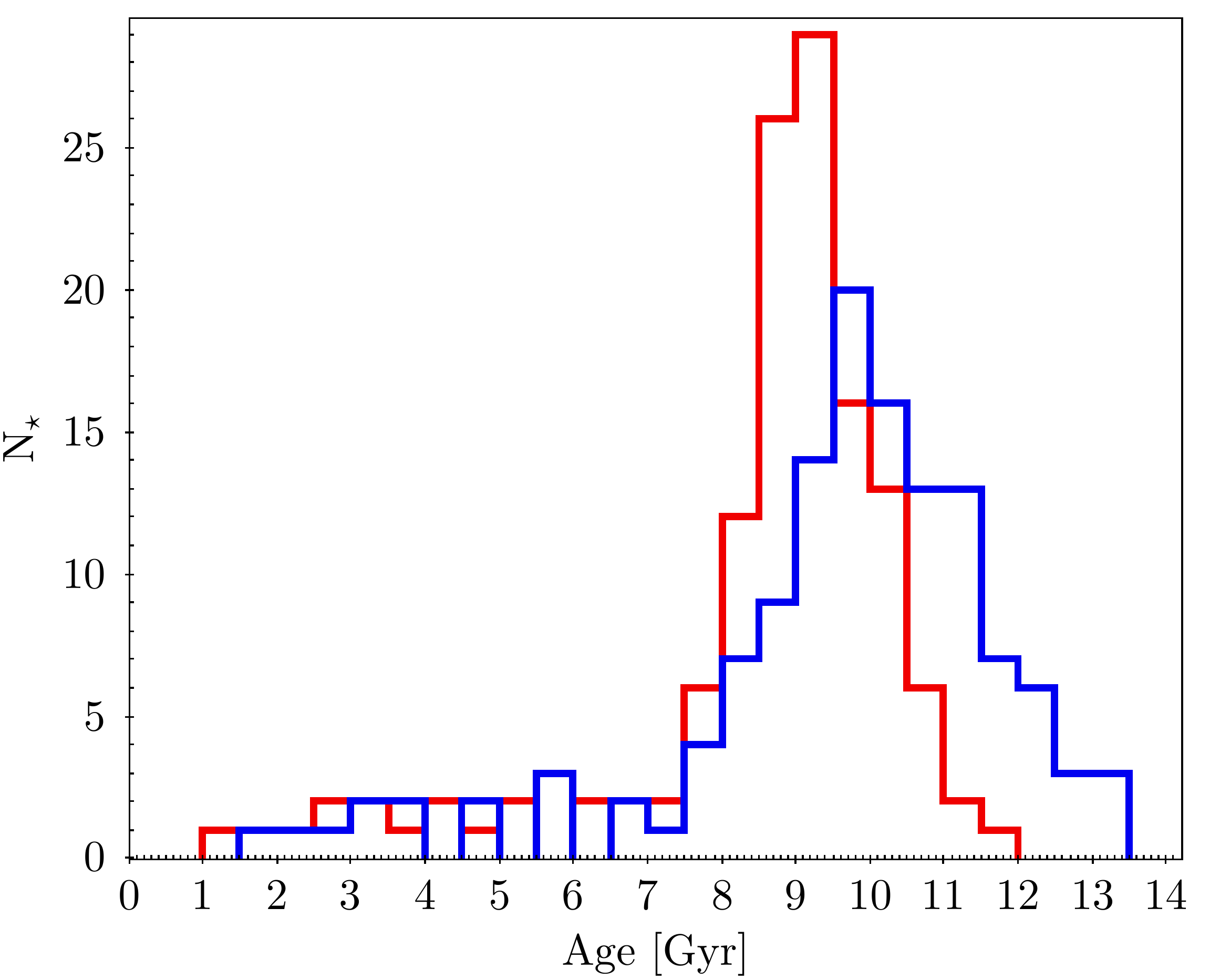}
    \includegraphics[width=0.33\hsize]{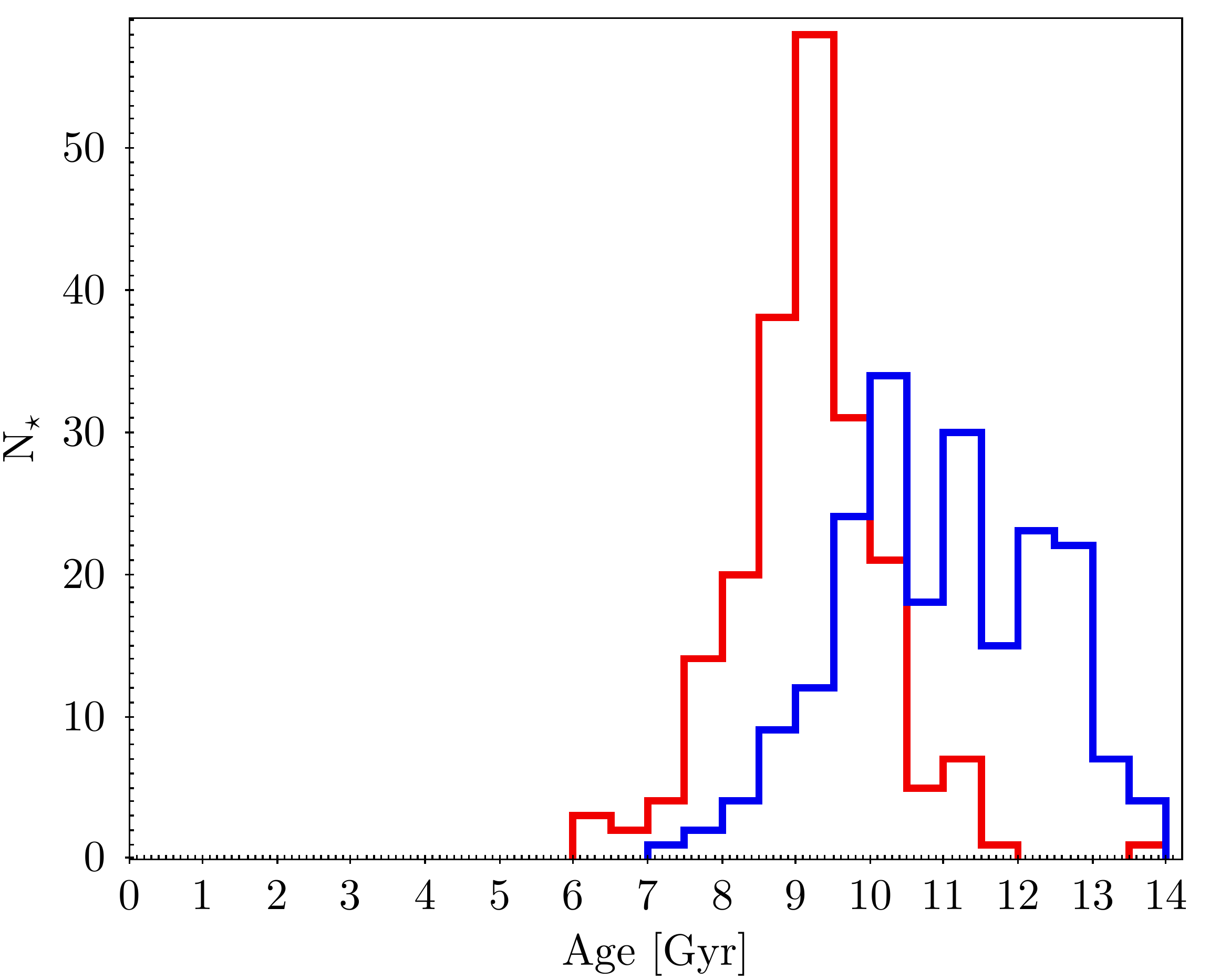}  

   \caption{Age distributions for the thin-, h$\alpha$mr, and h$\alpha$mp thick-disc populations, from left to right panels, respectively. \textit{Top panels:} Age distributions from the APOKASC catalogue (red histogram) and with a relative error of less than 25\% (grey dotted line). \textit{Bottom panels:} 
 Age distributions from M21 (blue solid line) and from APOKASC (with a relative error less than 25\%) for the subsample of stars in common (red solid line).  } 
   \label{distrib_age}
 \end{figure*}

While in Figure~\ref{distrib_age_model} we show the age distributions from the BGM mock catalogues compared to the samples without distinguishing between different groups in the [$\alpha$/Fe] versus [Fe/H] diagram, Fig.~\ref{distrib_age} shows the age distributions for the chemically defined thin-, h$\alpha$mr, and h$\alpha$mp thick-disc. These figures present stellar age distributions from APOKASC and from  M21. Because they represent the present-day distribution of existing giant stars only, the aim of such figures is not to infer the star formation history (SFH) but to provide information in order to better understand it.  However, overall the BGM is able to reproduce the age distribution of the M21 sample, overlooking statistical fluctuations, showing that the star formation history from \citet{Mor18} is able to reproduce the M21 sample.\\ 
Moreover, the differences seen between the M21 and APOKASC age distributions in Figs. \ref{distrib_age_model} and \ref{distrib_age} show that the interpretation of these plots is not trivial.  For example, the APOKASC age distribution shows a peak around 1-2 Gyr that is not seen in the M21 distribution. As discussed by \citet{Casagrande16} and \citet{MaCo17}, this peak may originate from selection biases where stars in the secondary clump are overrepresented (at ~1 Gyr) (see also discussion in \citet{Miglio21}). Perturbing the age of the simulations by the typical uncertainties observed in seismology ($\sim$25\%), this peak might be flattened in the BGM age distributions (see Fig.\ref{distrib_age_model}). Moreover, the M21 age distribution shows a crest between 2 and 5.5 Gyr that is compatible with the BGM simulations. Although \citet{Miglio21} showed that low ages are strongly dependent on hydrodynamical processes included in stellar evolution models during the main sequence (such as overshoot and rotation-induced mixing), several previous studies pointed out an increase in star formation in the period between ~2 and 3 Gyr which is probably reflected in the M21 distribution. More data are needed to statistically confirm this feature. 
 
\citet{Vergely02} showed a peak of star formation 1.6 Gyr ago using a sample from the Hipparcos catalogue \citep{Perryman12}. Also using Hipparcos data, but with a different approach, \citet{Cignoni06} found maximum star formation activity to be about 3 Gyr ago in the solar neighbourhood. These results are in agreement with the SFH found in the larger volume by \citet{Bernard17} using \gdrone \citep{Prusti16} and Tycho-Gaia Astrometric Solution (TGAS) parallaxes \citep{Lindegren16}. \citet{Mor19} found an imprint of a star formation burst 2 to 3 Gyr ago in the thin disc using the \gdrtwo\  data.  The reason for such enhanced star formation could be a local perturbation or a recent merger.
\citet{Ruiz20} proposed several enhanced formation episodes  occurred 5.7, 1.9, and 1 Gyr ago which could be induced by Sagittarius dwarf galaxy passages. In addition, using the Sloan Digital Sky Survey, Gaia, and LAMOST data, \citet{Donlon20} proposed to explain the star formation enhancement around 2-3 Gyr ago with the Virgo Radial Merger (VRM). These authors proposed that a VRM progenitor dwarf galaxy passed through the Galactic centre 2.7$\pm$0.2 Gyr ago.  Our results, despite being preliminary and awaiting larger asteroseismology samples, confer the advantage that they were derived through a direct method and using the most accurate individual age measurements available at present. \\

The age distributions in the APOKASC and M21 samples of the thin disc show a steep increase from 10 to 5 Gyr (see Fig.~\ref{distrib_age}). Figure \ref{distrib_age} shows that the h$\alpha$mp thick disc is dominated by objects older than 8 Gyr (in particular when restricting to ages of  higher precision; see the dotted line in the upper right panel).
The median age of this population is different in the APOKASC sample ($\sim$8.8 Gyr for better ages) and M21 sample ($\sim$10.94 Gyr). These age distributions are in agreement with the age of the thick disc stars previously discussed by many authors \citep[e.g.][]{Fuhrmann11,Haywood13,DelgadoMena17,Anders18,Nissen20,Grieves18,Silva18,Miglio21} who used different datasets and methods of age determinations. The h$\alpha$mp thick disc age distribution is particularly extended in the APOKASC data.  Approximately 80\% of the h$\alpha$mp stars are found to be older than 6 Gyr. The sample of younger h$\alpha$mp stars needs to be increased for this result to be confirmed.\\

Regarding the h$\alpha$mr thick-disc population (Figure \ref{distrib_age}, middle panels), its age distribution ranges from 1 to 14 Gyr with the majority of stars having an age of 7 to 14 Gyr. This is very similar to the h$\alpha$mp thick disc population in both samples (KS tests done between 7 and 14 Gyr give p-values of $\sim$0.99 and 0.92 for APOKASC and M21, respectively), with a higher contribution in the younger tail (Fig.~\ref{distrib_age}). 

\subsection{Age relations with metallicity and [$\alpha$/Fe]}
 \label{agerelations}
 
Taking into account the three stellar populations defined in the previous section, and stellar ages derived in the APOKASC and M21 catalogues, we investigate the age relations with metallicity and $\alpha$-abundances by comparison to BGM simulations. \\

\begin{figure*}
  \centering
      \hspace{0.8cm}\textbf{APOKASC} \hspace{2.5cm} \textbf{M21}\hspace{4cm} \textbf{APOKASC} \hspace{2.5cm} \textbf{M21}\hspace{1cm}\par
      \vspace{0.2cm}
     \includegraphics[width=0.48\hsize,clip=true,trim= 0cm 0cm 2cm 3cm]{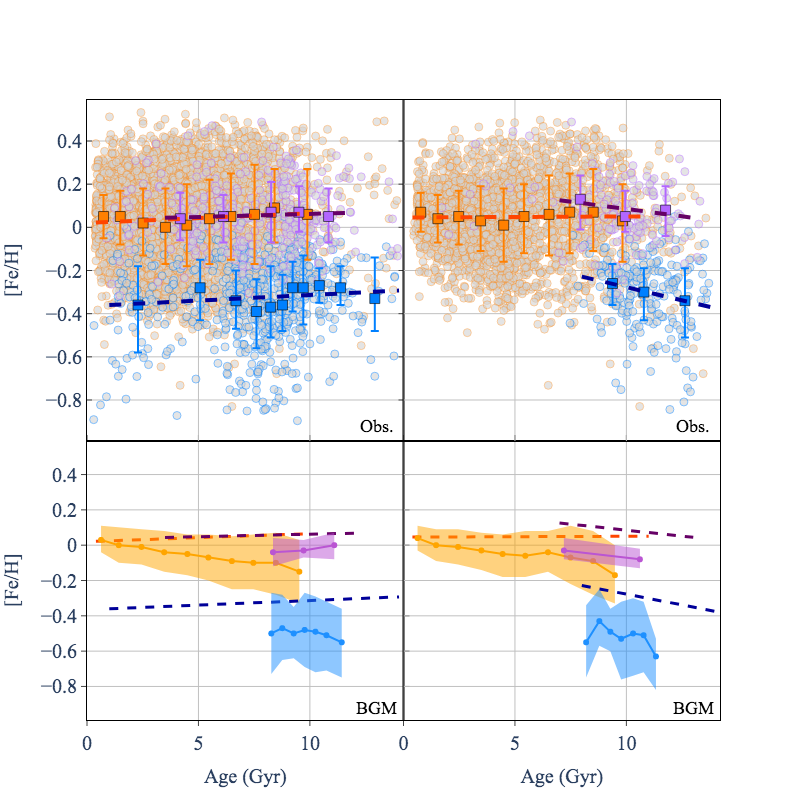}
   \includegraphics[width=0.48\hsize,clip=true,trim= 0cm 0cm 2cm 3cm]{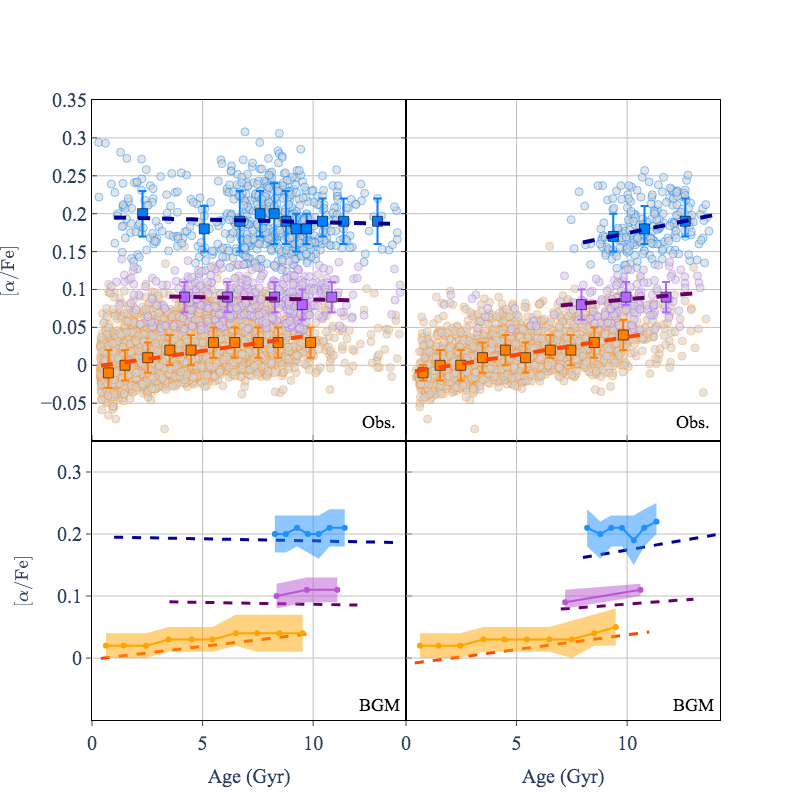}
   \caption{[Fe/H] (left) and [$\alpha$/Fe] (right) as a function of stellar age in our sample taking into account stellar ages from APOKASC and  M21. All stars from both samples are shown (coloured dots). The mean value per age bin is shown (square) for the thin-disc (orange), h$\alpha$mr, and h$\alpha$mp thick-disc (magenta and blue respectively) populations. The mean value was computed between the 5\% and 95\% quantile of the sample to avoid extreme values. The 1$\sigma$ error bars are shown. The dashed lines are linear fits. \textit{Bottom panels:} BGM predictions corresponding to observations and represented by solid lines. The $\pm$1$\sigma$ error bars are shown with shadow regions. Linear fits done for observations are recalled with dashed lines. }
   \label{alphafeh_age}
\end{figure*}

Simple Galactic evolution models predict chemical enrichment of the Galaxy as a function of time because successive generations of stars release more metal-rich material into the Galaxy. However, as clearly discussed in the literature, other processes are a play, such as radial migration, and this contributes to the observed scatter in the age--metallicity relation \citep[see][and references therein]{Minchev13, Minchev14a, SchBin09}. First, a large spread (about 1 dex) in metallicity was found at a given age \citep[see discussion in]{Minchev13, Minchev14b}. Indeed, the relation was reported to be flat at young ages, with a steep decrease in metallicity that occurs between 6 and 9 Gyr, depending on the study \citep[e.g.][and M21]{Haywood13,Bensby2014,Bergemann2014,Feuillet2016,Feuillet2018}.\\

The [Fe/H] versus age distribution of both samples (Fig.~\ref{alphafeh_age}, left panel) also shows a large spread of about 0.8 dex at all ages. The age--metallicity relations of the three populations are flat for the APOKASC sample. For the M21 sample, the age--metallicity relation of the thin disc population is also flat, while that of the two thick disc populations shows decreasing age with increasing metallicity.
On the other hand, the age--metallicity relations are more or less identical for the thin- and h$\alpha$mr thick-disc populations, with a mean metallicity close to solar (see Table \ref{meanvel}). The mean metallicity of the h$\alpha$mp thick-disc population is 0.3 dex smaller. As discussed in Fig.\ref{distrib_age}, the APOKASC and M21 samples show different age distributions for the h$\alpha$mp thick disc population\footnote{KS test p-value are $\sim$0.42}, leading to a different age--metallicity relation.
Figure~\ref{alphafeh_age} (right panel) shows the [$\alpha$/Fe] versus age distribution in both samples. Whereas the thin-disc population shows a slight increase with age, with a gradient of 0.0042$\pm$ 0.00078 dex Gyr$^{-1}$ for the APOKASC sample, the relations are flat for both h$\alpha$mr and h$\alpha$mp thick disc populations. Looking at the M21 sample, the gradient of thin disc population is in agreement with that of the APOKASC sample (0.0047$\pm$ 0.00053 dex Gyr$^{-1}$). The h$\alpha$mp and h$\alpha$mr thick discs show a positive gradient of [$\alpha$/Fe] with stellar age (0.006$\pm$ 0.00047 and 0.0026$\pm$ 0.00014 dex Gyr$^{-1}$, respectively).We also confirm that the [$\alpha$/Fe] versus age relation is tighter than the metallicity versus age relation, in particular for the thin-disc population (dispersion of about 0.1 dex). However, we are not able to come to any firm conclusions as to the larger scatter with increasing age, as seen by \cite{Bergemann2014} in the Gaia-ESO Survey, and contrarily to \cite{Haywood13,Bensby2014,Feuillet2018} who found that the relation tightens at ages of greater than 9 Gyr. We also show that our APOKASC sample contains $\alpha$-enriched young stars, such as those discussed by \citet[][and M21]{Martig15, Chiappini15, Jofre16, Silva18}. 

The two independent age determinations are in agreement concerning the thin-disc and h$\alpha$mr thick-disc properties, with a flat relation between [Fe/H] and age and an increase in [$\alpha$/Fe] with age. Concerning the h$\alpha$mp thick disc population, the situation appears to be different. While $\alpha$-elements are produced mainly by Type II SNe (which have a massive and short-lived stellar progenitor), Type Ia SNe are the main contributors to the iron in the disc of our Milky Way. The [$\alpha$/Fe] ratio is expected to decrease during Galactic evolution while [Fe/H] increases. This is what we observe for the M21 sample (see Fig.~\ref{alphafeh_age}). The authors of M21 highlighted $\alpha$-enriched young stars as `over-massive or rejuvenated stars' and remove them from their sample (see Sect.5.2 of M21 for more details) wheras they are kept in the APOKASC sample. Larger samples of h$\alpha$mp thick-disc stars with different age derivations are necessary in order to come to any firm conclusions on the relation between [$\alpha$/Fe] and age. 

The lower panels of Fig.~\ref{alphafeh_age} show the metallicity and [$\alpha$/Fe] versus age relations computed with the BGM for the selections made in APOKASC and M21 studies. The relations adopted in the BGM are compatible with observations within a systematic `offset'. The main difference occurs for the thin disc. The BGM predicts a negative gradient following that of \citet{Haywood06} which is not observed with our two samples.
These could be revised in the BGM in light of the present study. 

     \begin{figure*}
  \centering
    \includegraphics[width=0.33\hsize,clip=true,trim= 0cm 0cm 0cm 2cm]{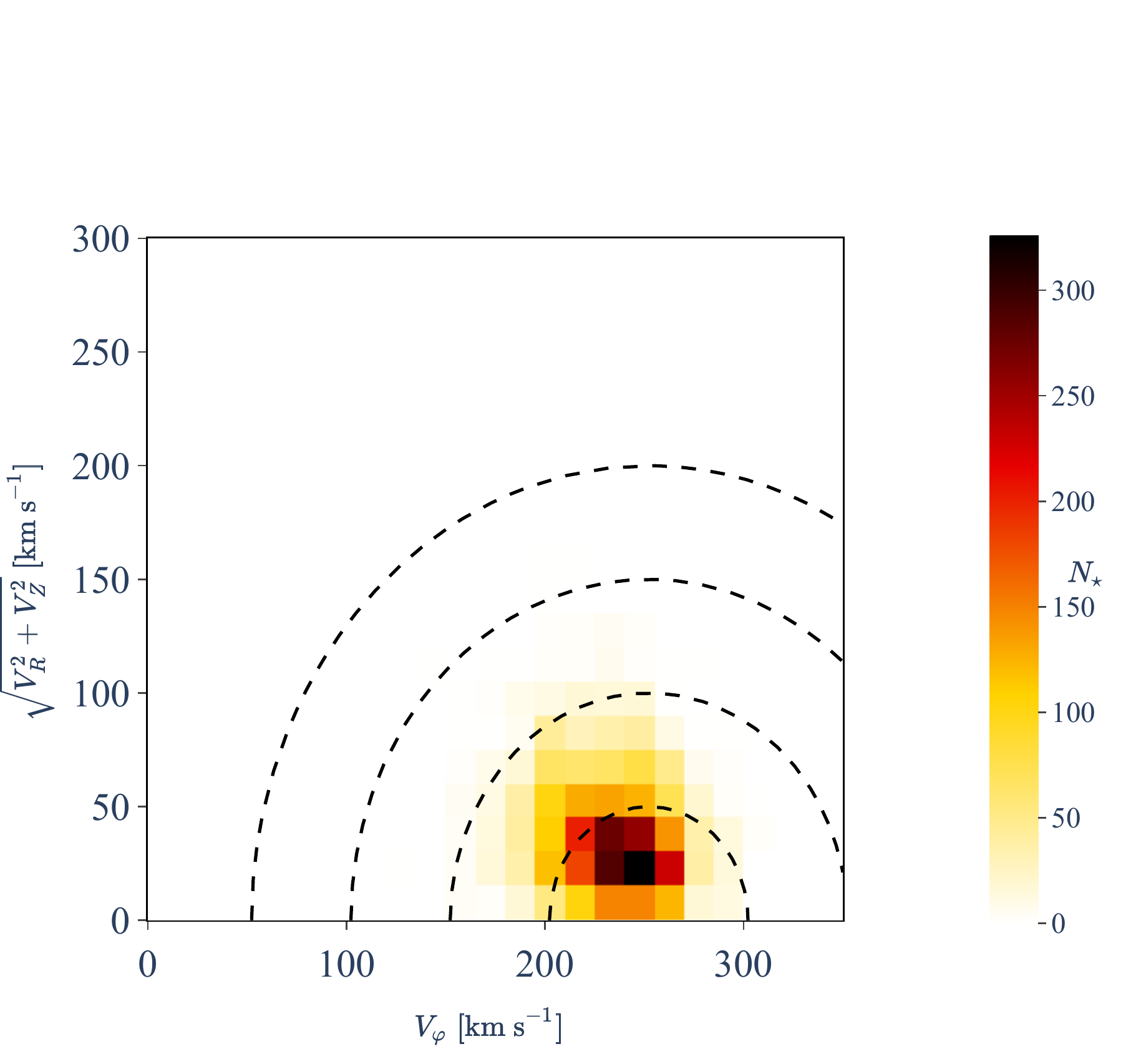}
     \includegraphics[width=0.33\hsize,clip=true,trim= 0cm 0cm 0cm 2cm]{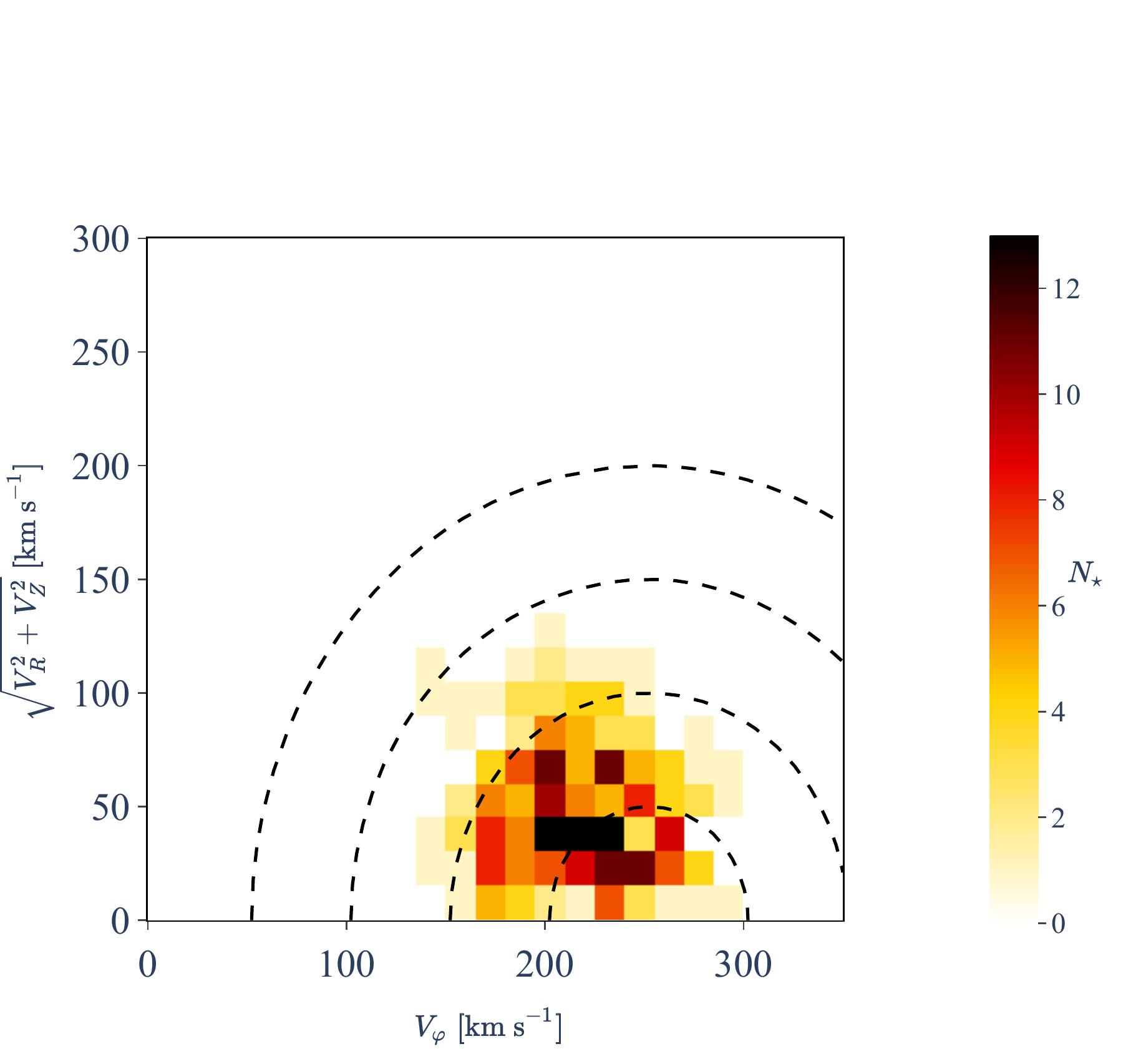}
     \includegraphics[width=0.33\hsize,clip=true,trim= 0cm 0cm 0cm 2cm]{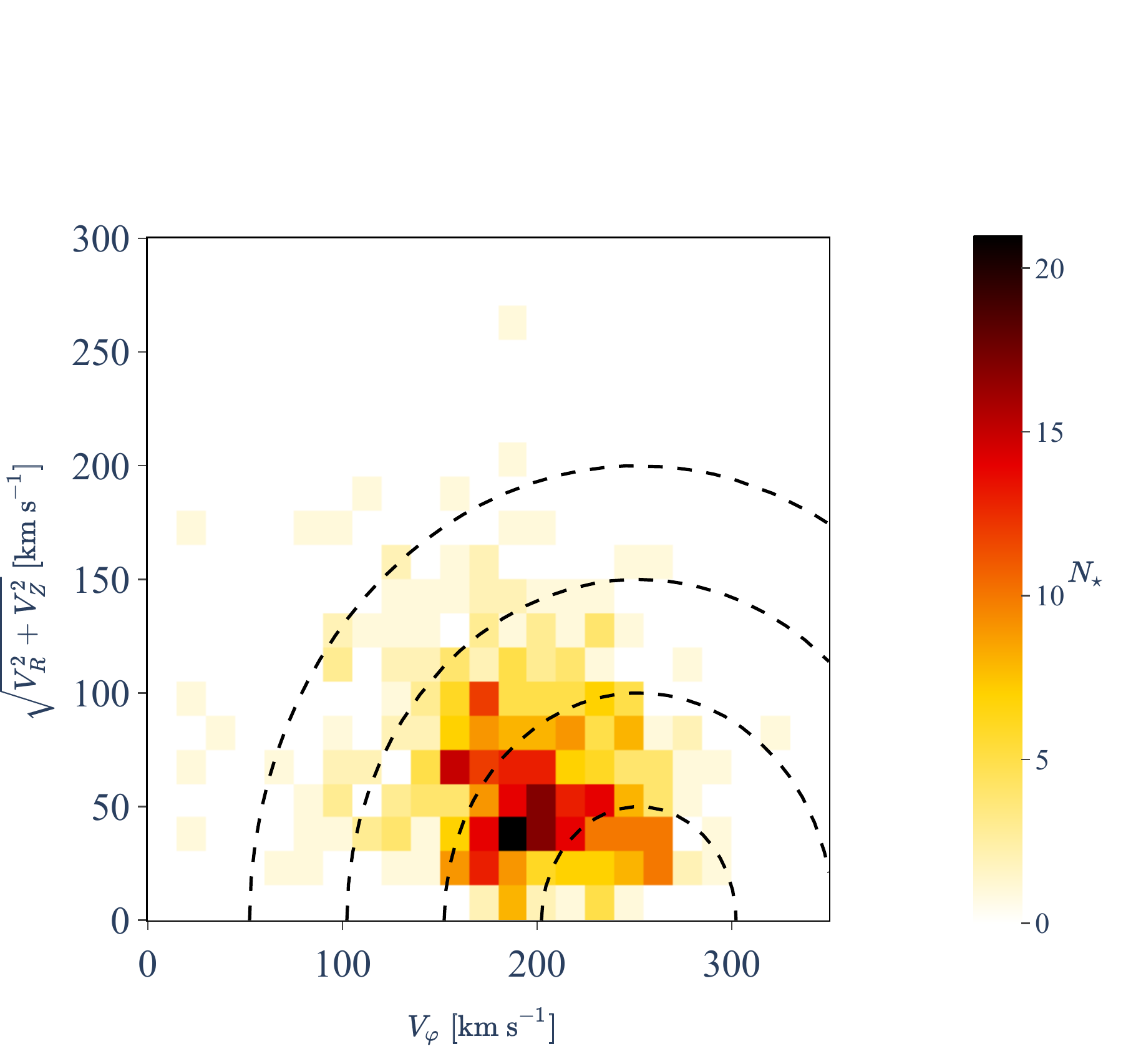}
   \caption{Toomre diagram for thin-disc (left panel), h$\alpha$mr thick-disc (middle panel), and h$\alpha$mp thick- disc (right panel) stars in our sample. The dashed curves indicate constant
space motion.}
   \label{toomre}
 \end{figure*}

 
\section{Kinematics}
 \label{Kinematics}
 
 The age--kinematics relations of stars provide crucial information about the formation, structure, and evolution of the Milky Way. The chemical composition of the stellar photosphere brings constraints on the chemical evolution history of the stellar population, while the stellar kinematics reveals the history of the stellar motion inside the Milky Way. \\
 
\begin{figure}
  \centering
    \includegraphics[width=0.49\hsize]{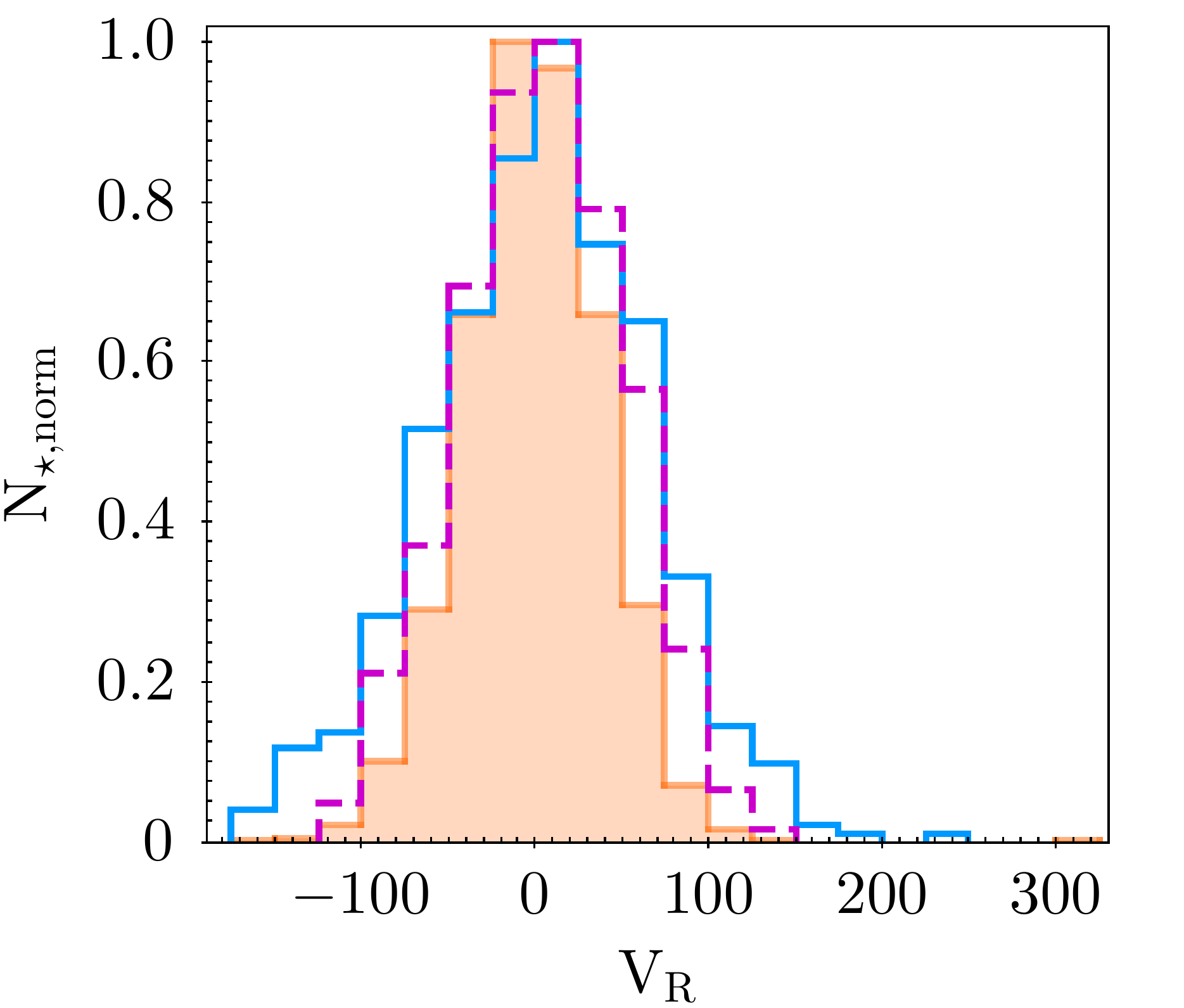}
     \includegraphics[width=0.49\hsize]{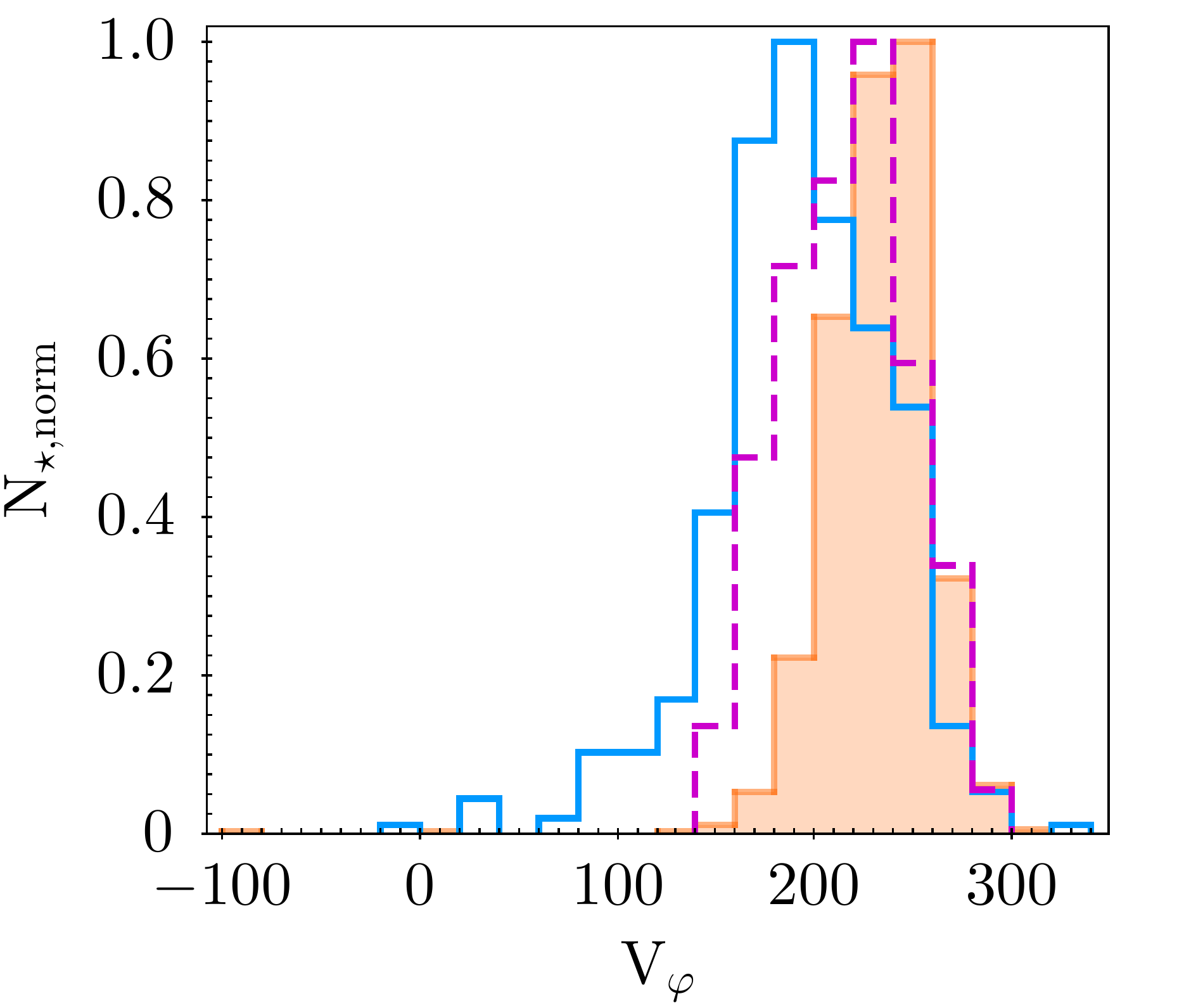}
     \includegraphics[width=0.49\hsize]{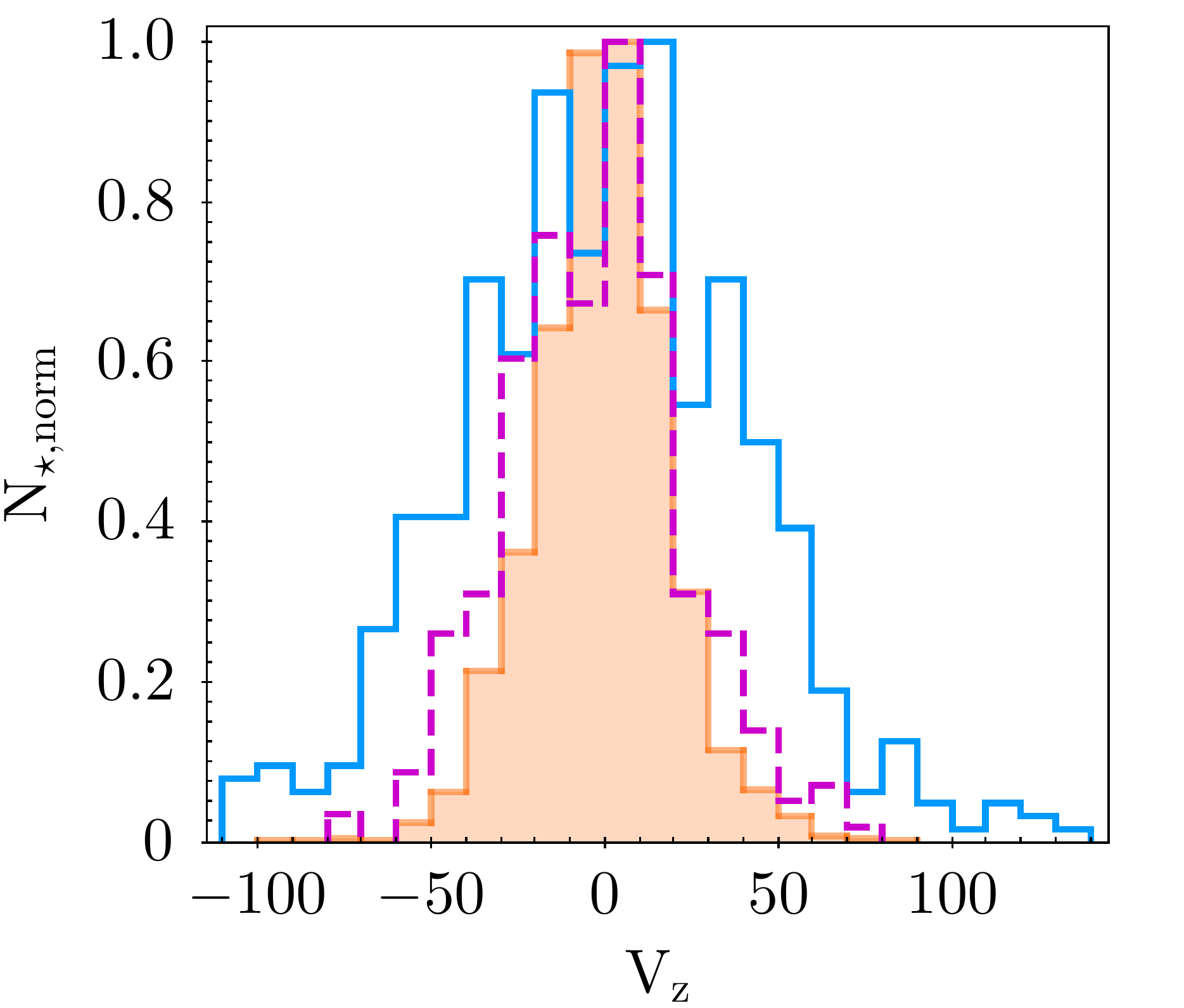}
   \caption{Normalised distribution of the velocities V$_R$, V$_\varphi$, and V$_Z$ (in km s$^{-1}$) for the thin-disc (orange solid line), h$\alpha$mp thick-disc (blue solid line), and the h$\alpha$mr thick-disc populations (magenta dashed line), with a clear tail showing the asymmetric drift in the $V_\varphi$ panel. These distributions are shown for the APOKASC sample}
   \label{distribution_V}
 \end{figure}

Table~\ref{meanvel} presents the mean value of the three velocity components for each stellar population. Figure \ref{toomre} displays the velocity components for the APOKASC sample in a Toomre diagram. This diagram is widely used to identify the major components of our Galaxy, such as thin and thick discs and the halo using only kinematics. We show the thin-disc, h$\alpha$mr, and h$\alpha$mp thick-disc populations ---identified by their chemical composition--- separately in the left, middle, and right panels, respectively. This figure clearly shows that the h$\alpha$mr thick-disc population has intermediate kinematics between the thin-disc and h$\alpha$mp thick-disc populations. The asymmetric drift can be seen from the thin- to the thick-disc diagram, indicating that the stars of the  h$\alpha$mp thick disc rotate slower than those of the thin disc. As shown in Fig.~\ref{distribution_V}, the velocity dispersions of the stars of the h$\alpha$mp thick disc are larger than those of the thin disc \citep[as also noted by][]{GiRe83, Soubiran03, Adibekyan13, Sharma14, Anders18}, while the velocity dispersions of the stars of the h$\alpha$mr thick disc are intermediate. The kinematics of the h$\alpha$mr thick disc appear to be more similar to those of the thin disc than to those of the h$\alpha$mp thick disc, showing a smaller distribution in V$_Z$ and a shift toward lower V$_\varphi$. 
As already discussed in Sect.~\ref{agedisp}, the h$\alpha$mp thick-disc stars are older than the thin-disc stars (see Table~\ref{meanvel}), and the h$\alpha$mr thick disc has a slightly greater contribution from younger stars than the h$\alpha$mp thick disc (see Fig.~\ref{distrib_age}). Considering this, Fig. \ref{distribution_V} shows the existing relations between velocity dispersion and stellar age.

 \begin{table*}[]
    \centering    
    \caption{ Mean galactocentric velocities and their dispersions (in km s$^{-1}$) for the three stellar populations.}
    \scalebox{0.85}{
    \begin{tabular}{ c c c c c c c c c c ||c c}
    Galactic component &$\overline{V_R}$  &$\overline{\sigma_R}$ &$\overline{V_\varphi}$ &$\overline{\sigma_\varphi}$ &$\overline{V_Z}$ &$\overline{\sigma_Z}$ &$\overline{[Fe/H]}$ & $\overline{[\alpha/Fe]}$ &$\overline{Age_{APOKASC}}$ &$\overline{Age_{APOKASCxM21}}$ & $\overline{Age_{M21}}$\\
    &&&&&&&&&\\
    \hline
    Thin disc &$-$0.72$\pm$0.6 & 31 &  +233$\pm$0.38 & 19 &$-$0.82$\pm$0.3 & 15 & 0.03  &  0.018 & 4.13& 4.00 & 4.81\\
    
    h$\alpha$mr thick disc & 4.54$\pm$2.7 & 38 & +216.7$\pm$1.81 & 25 & $-$2.23$\pm$1.46 & 20 & 0.06 &  0.09 &  7.36 & 8.55 & 9.57\\ 
            
    h$\alpha$mp thick disc & +3.02$\pm$1.85 & 47 & +194$\pm$1.85 & 32 &  +0.37$\pm$1.7  & 30 & $-$0.34 & 0.19 & 7.60 & 9.17 & 10.94\\
    \hline
    \end{tabular}}
   \tablefoot{The mean [Fe/H], [$\alpha$/Fe] and age are also indicated.}    \label{meanvel}
\end{table*}

  \subsection{Velocity relations with metallicity, [$\alpha$/Fe], and age}

Galactic archeology includes the study of the correlation between the kinematics and the chemical properties of the stars that compose our Galaxy. Indeed, the gradient of velocity with stellar metallicity and [$\alpha$/Fe] is important for the study of internal mechanisms such as radial migration \citep[e.g. ][]{SchBin09,Loebman11,Minchev16}, and could be a key observable with which to understand formation scenarios of the Galaxy itself. The velocity dispersions are crucial to constraining the heating mechanisms occurring in the stellar populations of the Milky Way \citep{Wielen75, Nordstrom04}. We discuss these points in the following sections. 

 \subsubsection{Velocities versus [Fe/H]}

\begin{figure}
  \centering
    \includegraphics[width=\hsize,clip=true,trim= 0.5cm 1cm 2cm 2.5cm]{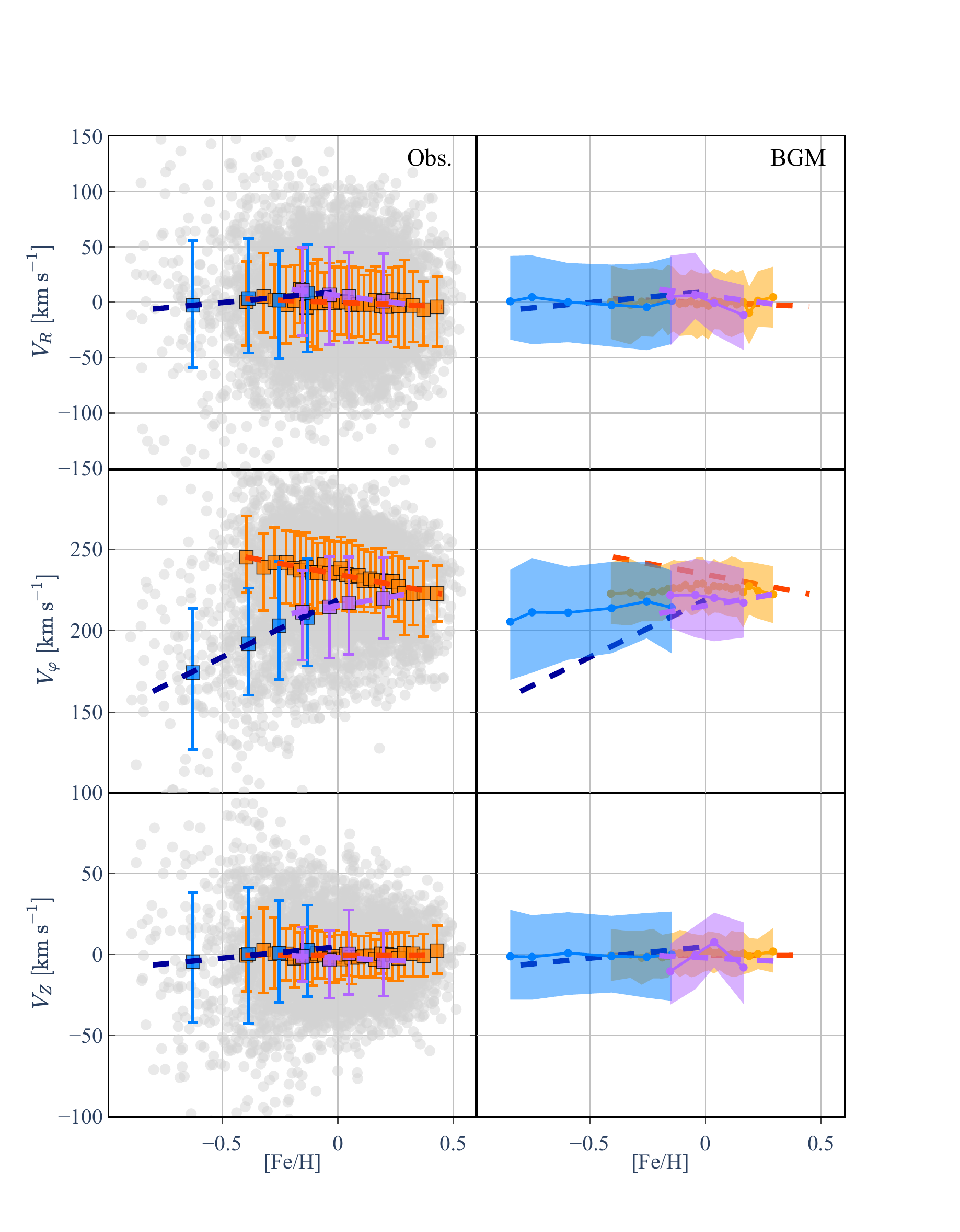}
       \caption{ Velocities in the Galactocentric cylindrical coordinate frame as a function of [Fe/H]. \textit{Left panels: }All sample stars are shown (grey dots). The mean value per [Fe/H] bin is shown for the thin- (orange), h$\alpha$mr- (purple), and h$\alpha$mp thick-disc (blue) populations. The mean value has been computed between the 5\% and 95\% quantiles of the sample to avoid extreme values. The $\pm$1$\sigma$ error bars are shown. The dashed lines are the linear fits. \textit{Right panels: }BGM predictions (orange, purple and blue solid lines) for the APOKASC sample as a function of [Fe/H] compared to the linear fits done using observations (dashed lines). The $\pm$1$\sigma$ error bars are shown with the shadow zones.  }
   \label{vel_feh}
 \end{figure}

Figure \ref{vel_feh} shows the mean velocities, $\overline{V_R}$, $\overline{V_\varphi}$, and $\overline{V_Z}$, as a function of [Fe/H] for the three stellar populations in the APOKASC sample. The mean values have been computed taking into account stars between the 5 and 95 percentiles to avoid extreme values (with a minimum of 30 stars per bin). 
The mean value of the rotational velocity for the thin and the h$\alpha$mp thick discs are in agreement with  the study of RAVE stars in the solar neighbourhood by \citet{Kordopatis13b}. There is no correlation between either the radial or vertical velocities and [Fe/H] for any of the stellar populations, but gradients can be seen for the rotational velocity $V_\varphi$. We use a simple linear fit to deduce the gradient of $\overline{V_\varphi}$ with  metallicity for stellar populations. Although these gradients depend on sample selection \citep{Minchev19}, we derive the gradient for each stellar population and compare them with each other. Quantitatively, we find: 
\begin{equation}
\frac{\partial V_\varphi}{\partial [Fe/H]}= 
\left\{
 \begin{array}{l}
 -26.76\pm 1.85 \hspace{0.2cm} \text{km s}^{-1}\text{dex}^{-1}\text{, for the thin disc,}\\23.8 \pm 2.3 \hspace{0.2cm}\text{km s}^{-1}\text{dex}^{-1}\text{, for the h$\alpha$mr thick disc,}\\
 70.2\pm  4.9 \hspace{0.2cm}\text{km s}^{-1}\text{dex}^{-1}\text{, for the h$\alpha$mp thick disc.}\\

 \end{array}
 \right.
 \label{vphieq}
\end{equation}

\begin{figure}
  \centering
 \includegraphics[width=0.98\hsize,clip=true,trim=0cm 5cm 4.5cm 2cm]{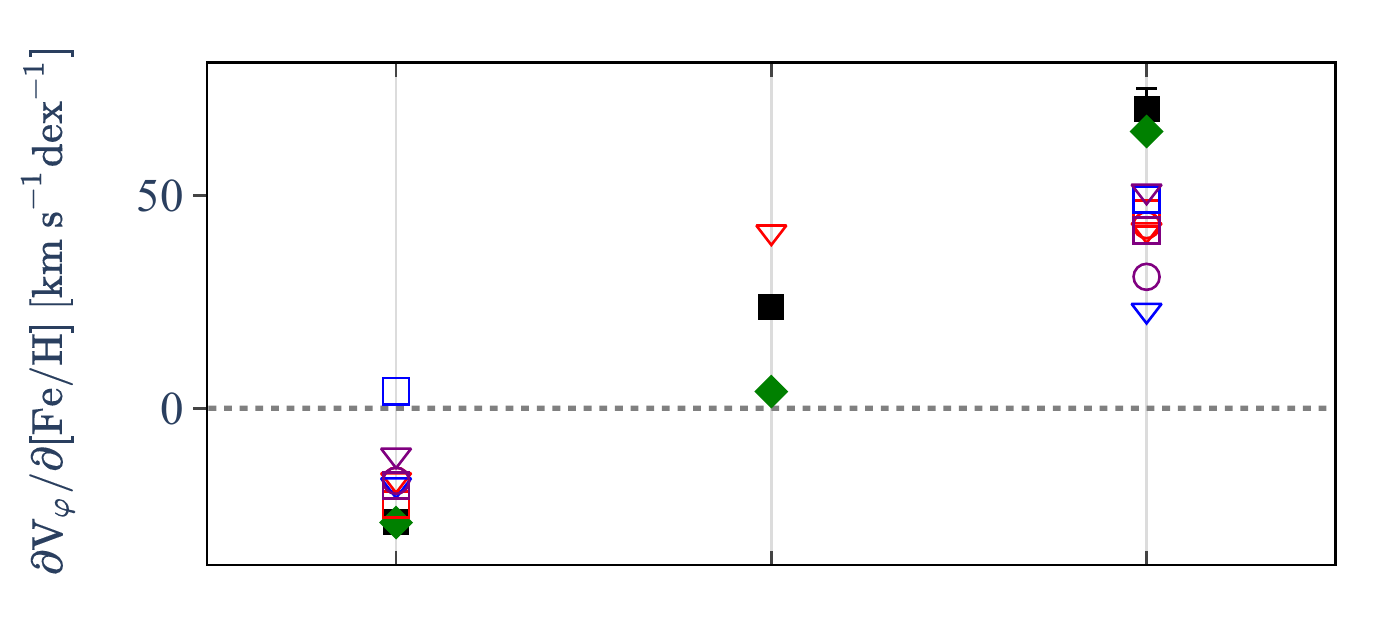}
 
    \includegraphics[width=0.98\hsize,clip=true,trim= 0cm 1.5cm 4.5cm 5.8cm]{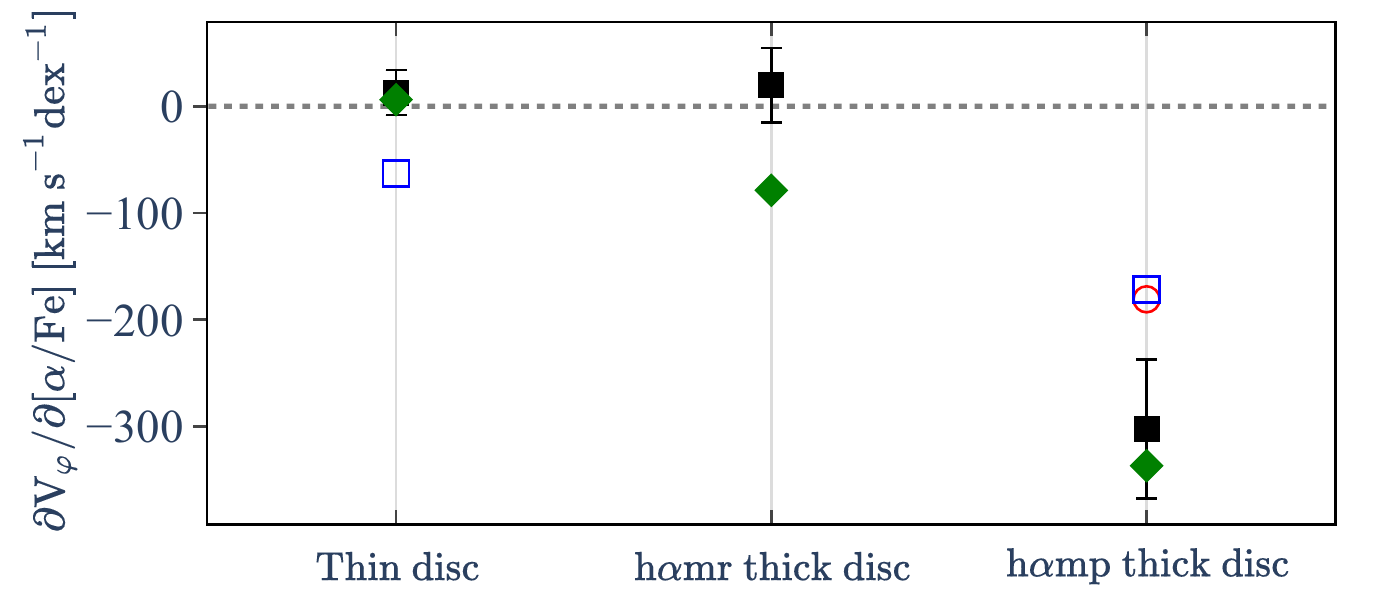}
    \caption{$\partial V_\varphi/\partial$[Fe/H] (top panel) and $\partial V_\varphi$/$\partial [\alpha$/Fe] (bottom panel) for the three stellar populations, using the APOKASC data (solid black square), the APOGEE-DR16  (green solid diamond). Previous results from literature have been added on the graph: from \citet{RecioBlanco14} (red open circle), \citet{Lee11b} (red open square), \citet{Adibekyan13} (red open triangle), \citet{Guiglion15} (blue open square), \citet{Prieto16} (blue open triangle), \citet{Peng18} (purple open square), and \citet{Yan19} (purple open circle).}
   \label{compa}
 \end{figure}

$V_\varphi$ increases with increasing [Fe/H] for the h$\alpha$mp thick-disc population, reaching values close to those of the stars of the thin discs. This trend has been underlined in the literature by, for example, \citet{Spagna10}. The strong gradient of V$_ {\varphi}$ in the h$\alpha$mp thick-disc population is partly due to the location of low-metallicity stars at larger Z. On the contrary, the thin-disc rotation velocity decreases with increasing [Fe/H]. This trend has already been shown in different models developed by \citet{SchBin09} and by \citet{Minchev13}. 
The correlation between V$_\varphi$ and their chemical properties has been extensively studied in the literature. The top panel of Fig.~\ref{compa} presents the rotational velocity gradient with [Fe/H] from the APOKASC catalogue compared to previous studies \citep{RecioBlanco14, Lee11b,Adibekyan13,Guiglion15,Wojno16,Prieto16,Peng18,Yan19}, showing good agreement with them.  Considering two stellar populations for the genuine thick disc, our determination for the h$\alpha$mp thick disc is slightly higher than those from most previous studies where the two populations are not separated \citep[except for ][]{Adibekyan13}.  In this context, it is interesting to note that the h$\alpha$mr thick-disc population shows an intermediate gradient between that of the h$\alpha$mp thick discs and the thin discs. Figure \ref{compa} also shows the gradient of the rotational velocity using [Fe/H] derived using DR16 of the APOGEE data (solid green diamond). The gradient between rotational velocity and metallicity is in agreement with the study done with DR14 for the thin and the h$\alpha$mp thick disc populations. \\ 
The relations between the mean velocity components and [Fe/H] from the BGM simulations are also shown in the right panels of Fig.~\ref{vel_feh}. Simulations and observations are in good agreement, showing no correlation between V$_R$ or V$_Z$ and [Fe/H] for any of the stellar populations. The BGM simulation shows a flat relation between V$_\varphi$ and [Fe/H] for the thin discs ($\partial V_\varphi /\partial$[Fe/H] = 3.16$\pm$2.35$ ~\text{km s}^{-1} \text{dex}^{-1}$) and a small increase for the h$\alpha$mp thick discs ($\partial V_\varphi /\partial$[Fe/H] = 12.8$\pm$4$~ \text{km s}^{-1} \text{dex}^{-1}$), while observations show a strong decrease and increase, respectively (see Eq. \ref{vphieq}). As underlined by \citet[][see their Fig.5]{Minchev19}, the gradient of V$_\varphi$ with the metallicity might offer clues as to the importance of radial migration. However, using RAVE data and a chemo-dynamical model \citep[see][]{Minchev13}, the same authors show that the positive gradient of V$_\varphi$ with metallicity in the high-[$\alpha$/Fe] stars already discussed in the literature can be the result of a certain distribution of age that can turn the originally negative gradient of mono-age populations (due to the asymmetric drift and the negative radial metallicity gradient) into a positive gradient.
Although the BGM takes into account the asymmetric drift in the computation of rotational velocity (see Sect.\ref{simulations}), the strong observed positive gradient is not reproduced by the BGM. This could be due to an underestimation of the velocity dispersions in the BGM and/or other processes such as radial migration, and should be explored further in a future study.

\subsubsection{Velocity versus [$\alpha$/Fe]}

\begin{figure}
  \centering
    \includegraphics[width=\hsize,clip=true,trim= 0.5cm 0.5cm 2cm 2.5cm]{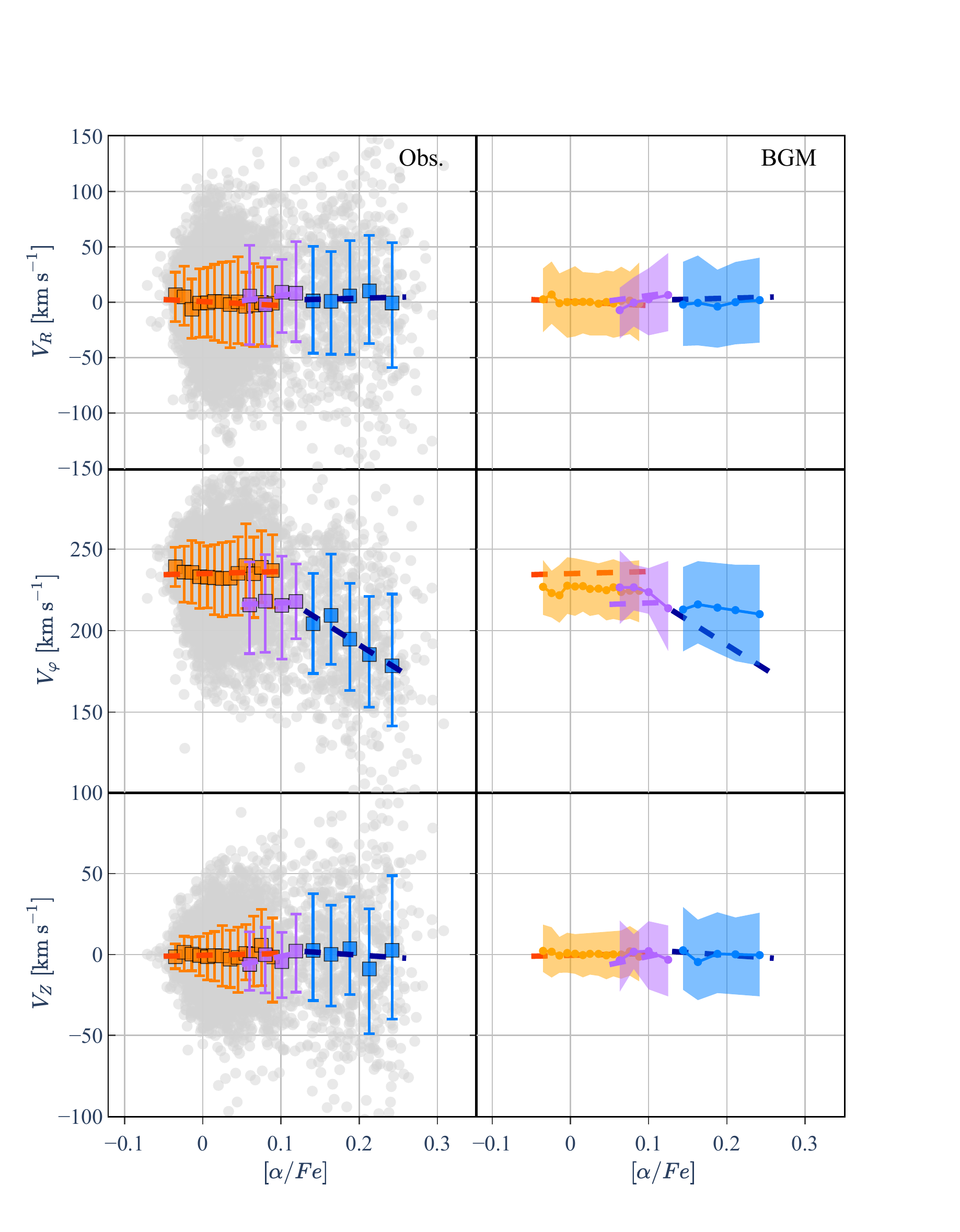}
       \caption{ Velocities in the Galactocentric cylindrical coordinate frame as a function of [$\alpha$/Fe]. The colours are the same as in Fig.\ref{vel_feh}.}   
        \label{vphi_alpha}
 \end{figure}

Figure \ref{vphi_alpha} presents the radial, rotational, and vertical velocities as a function of [$\alpha$/Fe] for the three stellar populations. The radial and vertical velocities show a flat behaviour with [$\alpha$/Fe] similar to that with [Fe/H] (see previous section). On the contrary, the rotational velocity shows a strong decrease with [$\alpha$/Fe] for the h$\alpha$mp thick-disc stars ($\partial$V$_\varphi$/$\partial$[$\alpha$/Fe]= $-$302 $\pm$ 65 km s$^{-1}$ dex$^{-1}$). These features are in good agreement with previous studies by \citet[][]{Haywood13,RecioBlanco14,Guiglion15}, and are also compatible with models including the radial migration effect \citep[e.g. ][]{SchBin09,Minchev14a}. In this context, it is interesting to note that the h$\alpha$mr thick-disc population shows almost the same flat gradient as the thin-disc population.\\

The relations between the mean velocities and the $\alpha$-abundance from the BGM simulations are also shown in the right panels of Fig.~\ref{vphi_alpha}. As already shown for [Fe/H], simulations and observations are in good agreement, showing no correlation between V$_R$ or V$_Z$ and [$\alpha$/Fe] for any of the stellar populations. Although the BGM simulations show a small decrease in V$_\varphi$ with [$\alpha$/Fe] for the h$\alpha$mp thick disc, the  gradient obtained with the BGM does not appear to reproduce that derived from observations. As the [$\alpha$/Fe] abundance is computed as a function of [Fe/H] in the BGM (see Sect.\ref{simulations}), this disagreement therefore reflects the behaviour of V$_\varphi$ with [Fe/H].

   \begin{figure*}
  \centering
  \hspace{0.8cm}\textbf{APOKASC} \hspace{7cm} \textbf{M21}\hspace{1cm}\\
    \includegraphics[width=0.98\hsize,clip=true,trim= 0.6cm 0.5cm 2cm 2.5cm]{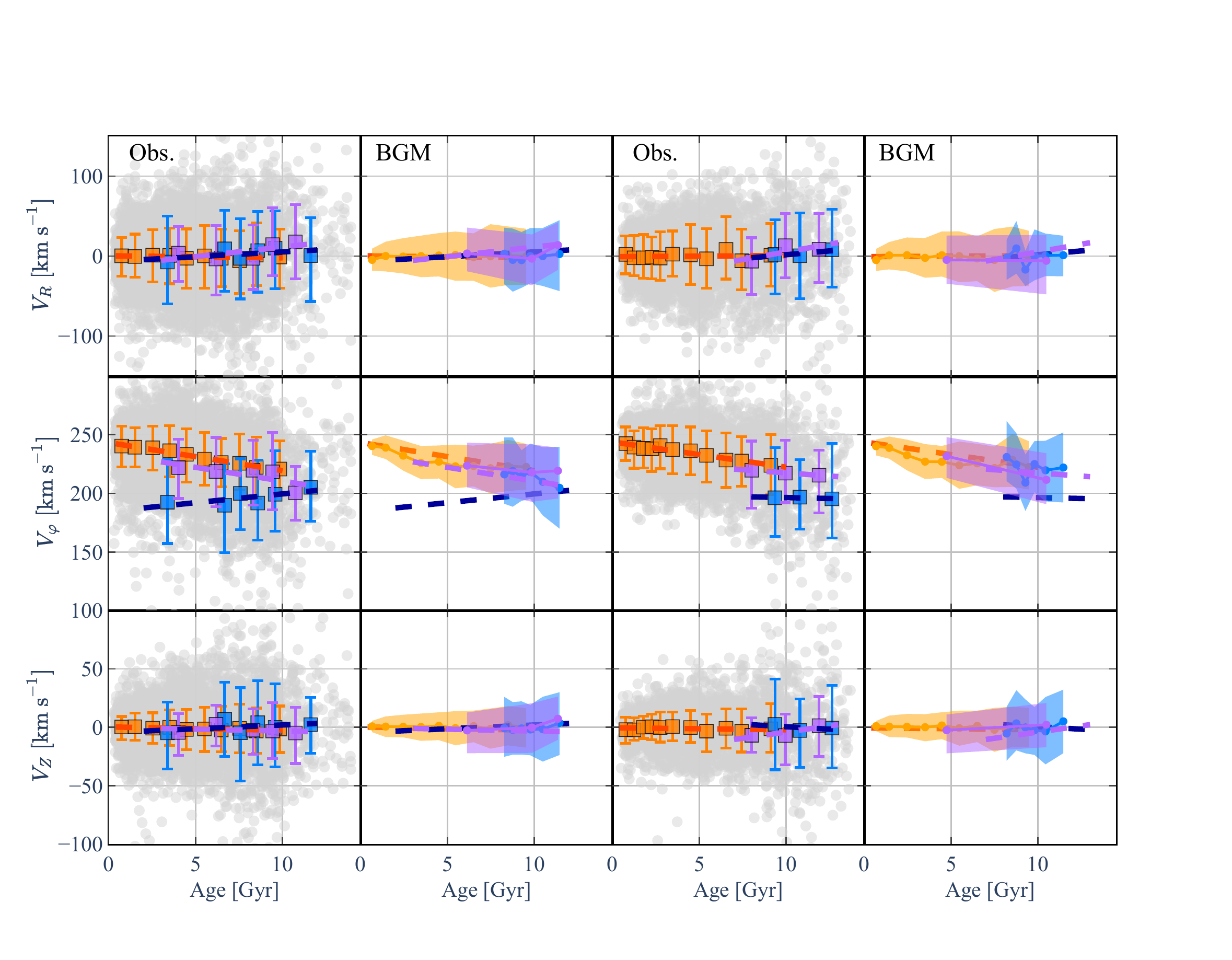}
   \caption{Velocities in the Galactocentric cylindrical coordinate frame as a function of stellar ages. Same as Fig.\ref{vel_feh}.} 
      \label{velocity_age}
 \end{figure*}

\subsubsection{Velocity versus stellar age}

In Galactic studies, the $\alpha$-over-iron enhancement is usually used as a proxy for stellar age. We investigate the velocity variations with stellar age, and compare the relations obtained with [Fe/H] and [$\alpha$/Fe]. Figure \ref{velocity_age} presents the radial, rotational, and vertical velocities as a function of stellar age for the three stellar populations and considering the two stellar age derivations (first and third columns). As shown for [Fe/H] and [$\alpha$/Fe], no correlations are identified between $V_R$ or $V_Z$ and stellar age. The gradient of V$_\varphi$ with stellar age is negative and is very similar for the thin disc considering either the APOKASC or the M21 sample ($-$2.38$\pm$0.14 and $-$2.14$\pm$0.18 km s$^{-1}$ Gyr$^{-1}$, respectively). Moreover, V$_\varphi$ decreases with [Fe/H] while it is almost constant with [$\alpha$/Fe]. These trends were found by \citet[][see their Fig.6]{Minchev13} using their chemo-dynamical model including the effects of radial migration and mergers on disc evolution. 
Unlike in other stellar populations, the behaviour of V$_\varphi$ with age for the h$\alpha$mp thick-disc stars is vastly different in the two samples: V$_\varphi$ increases with age by 1.49$\pm$0.75 km s$^{-1}$ Gyr$^{-1}$ in the APOKASC sample \footnote{In addition, in the APOKASC sample, V$_\varphi$ increases with age by 2.59$\pm$0.34 km s$^{-1}$ Gyr$^{-1}$ excluding overmassive stars (taking into account the selection criterion as M21; see their Sect 5.2) ; and by 2.21$\pm$0.08 km s$^{-1}$ Gyr$^{-1}$ considering more accurate stellar ages (relative error less than 25\%).} and decreases with age by $-$0.31$\pm$0.33 km s$^{-1}$ Gyr$^{-1}$ in the M21 sample. To investigate whether or not this feature is introduced by a possible selection bias, Figure \ref{velocity_age} presents the BGM simulations (second and fourth columns).  Simulations show a decrease in V$_\varphi$ with age by $-$3.99$\pm$1.23 km s$^{-1}$ Gyr$^{-1}$ and by $-$1.71$\pm$2.9 km s$^{-1}$ Gyr$^{-1}$ for the APOKASC and M21 samples, respectively. The different observational behaviour of V$_\varphi$ with age in the h$\alpha$mp thick-disc population is not due to sample selection but is probably induced by incorrect determination of the age of the stars with high mass loss (see Sect. 5.1 of M21). Indeed both samples are in agreement in the age range between 9 and 12 Gyr.  

 \subsection{Velocity dispersions}

 \begin{figure*}
  \centering 
  \includegraphics[width=0.33\hsize,clip=true,trim= 0cm 0cm 1.5cm 2cm]{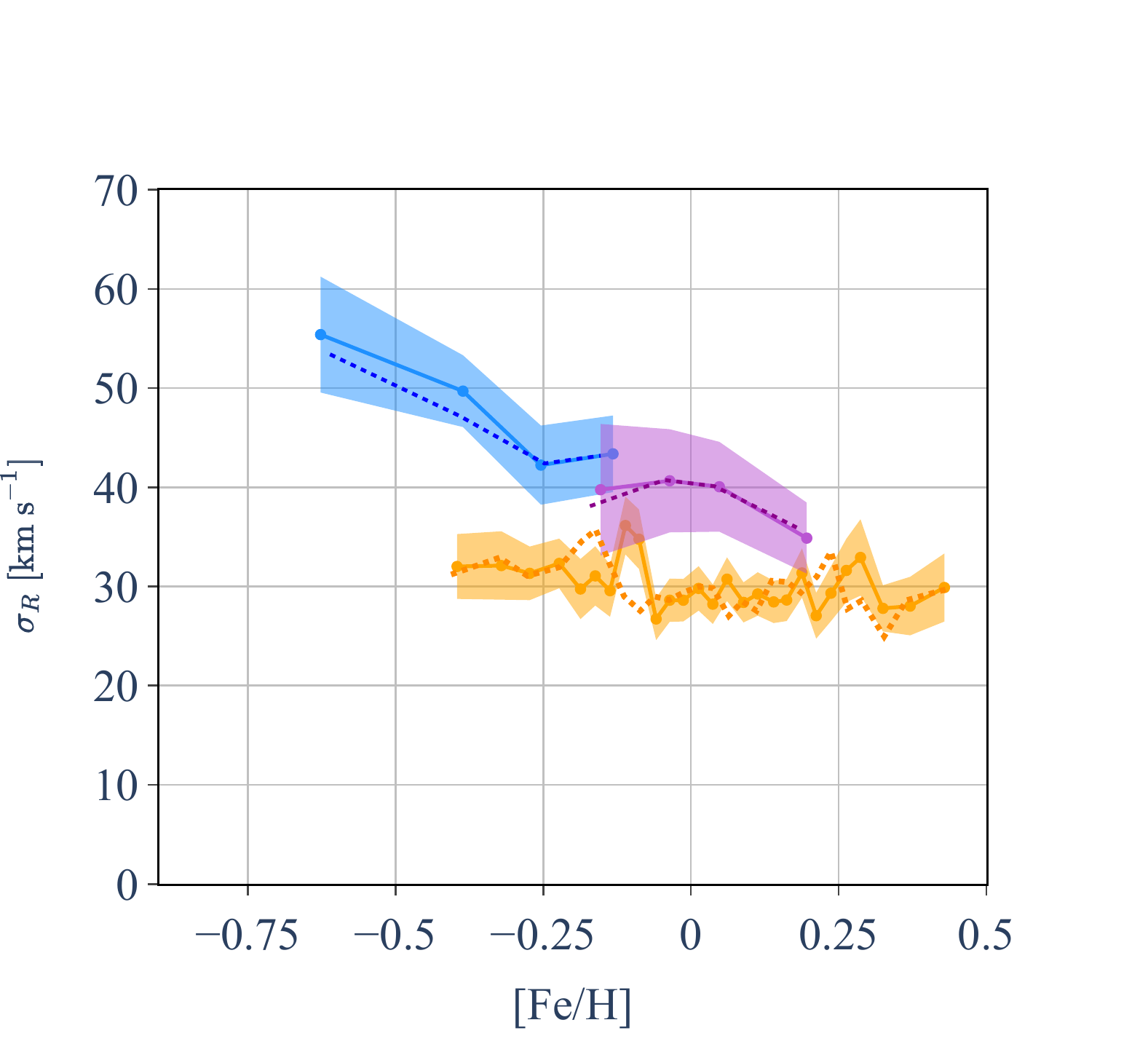} 
   \includegraphics[width=0.33\hsize,clip=true,trim= 0cm 0cm 1.5cm 2cm]{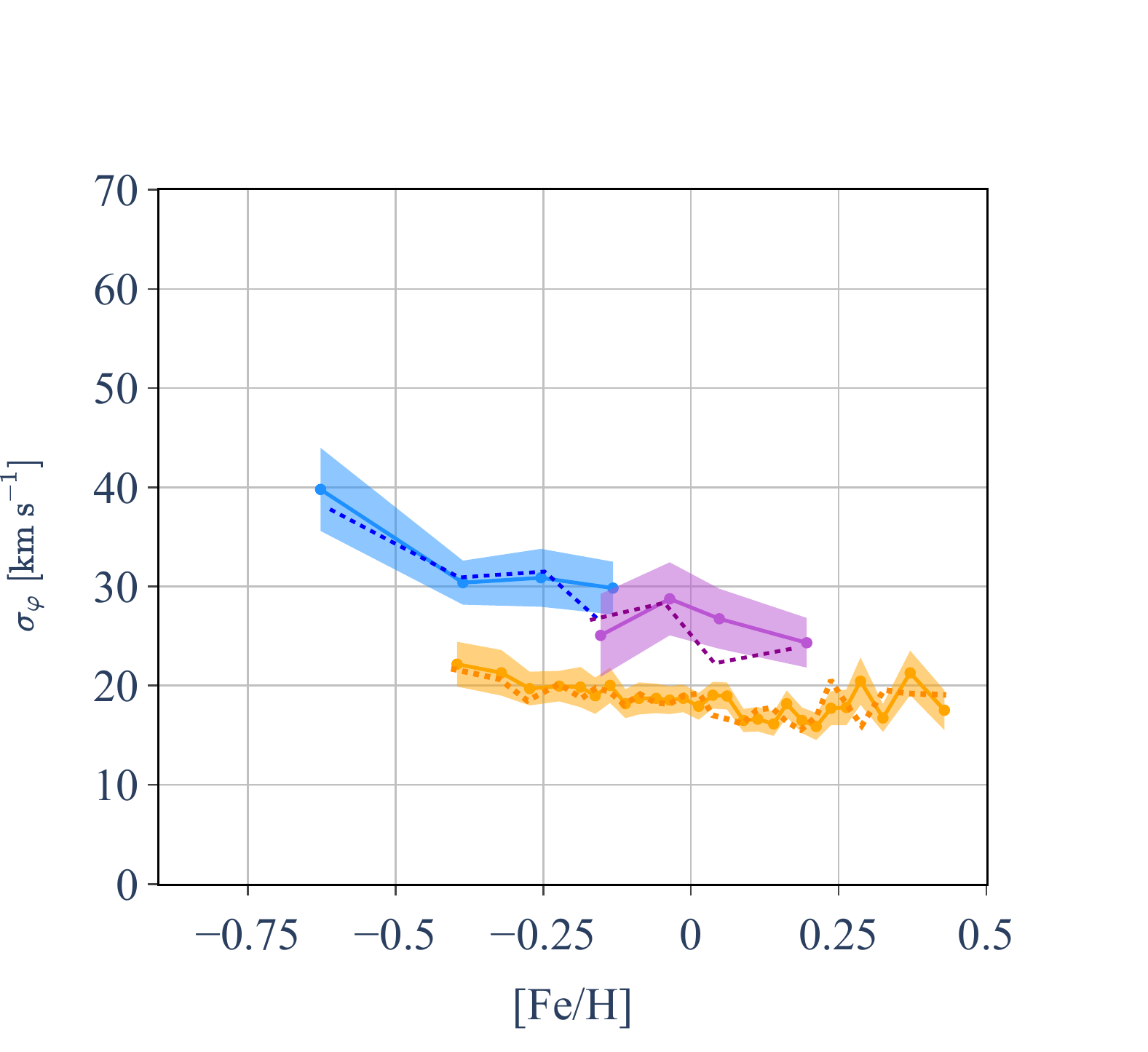}
    \includegraphics[width=0.33\hsize,clip=true,trim= 0cm 0cm 1.5cm 2cm]{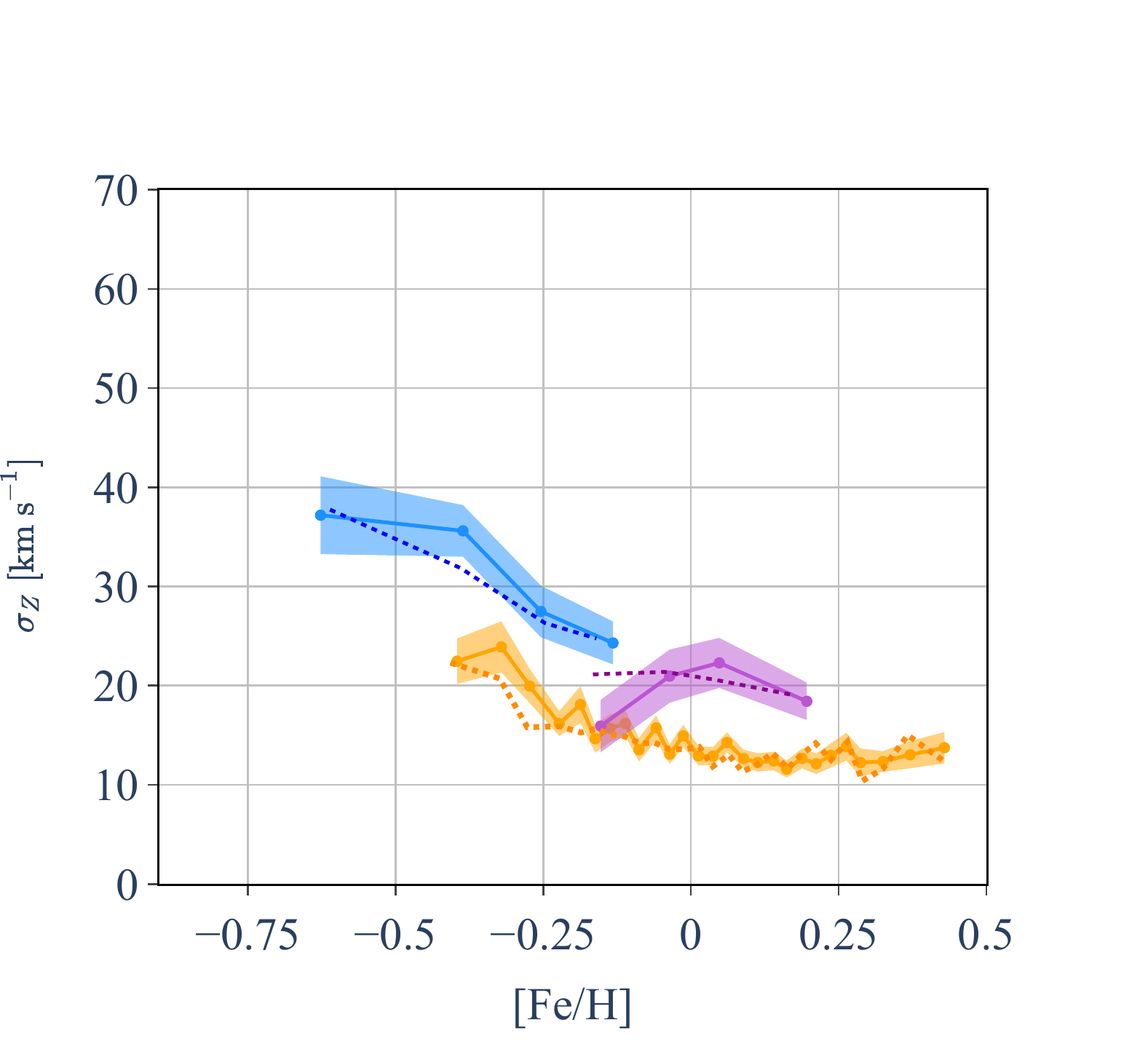}
    
  \includegraphics[width=0.33\hsize,clip=true,trim= 0cm 0cm 1.5cm 2cm]{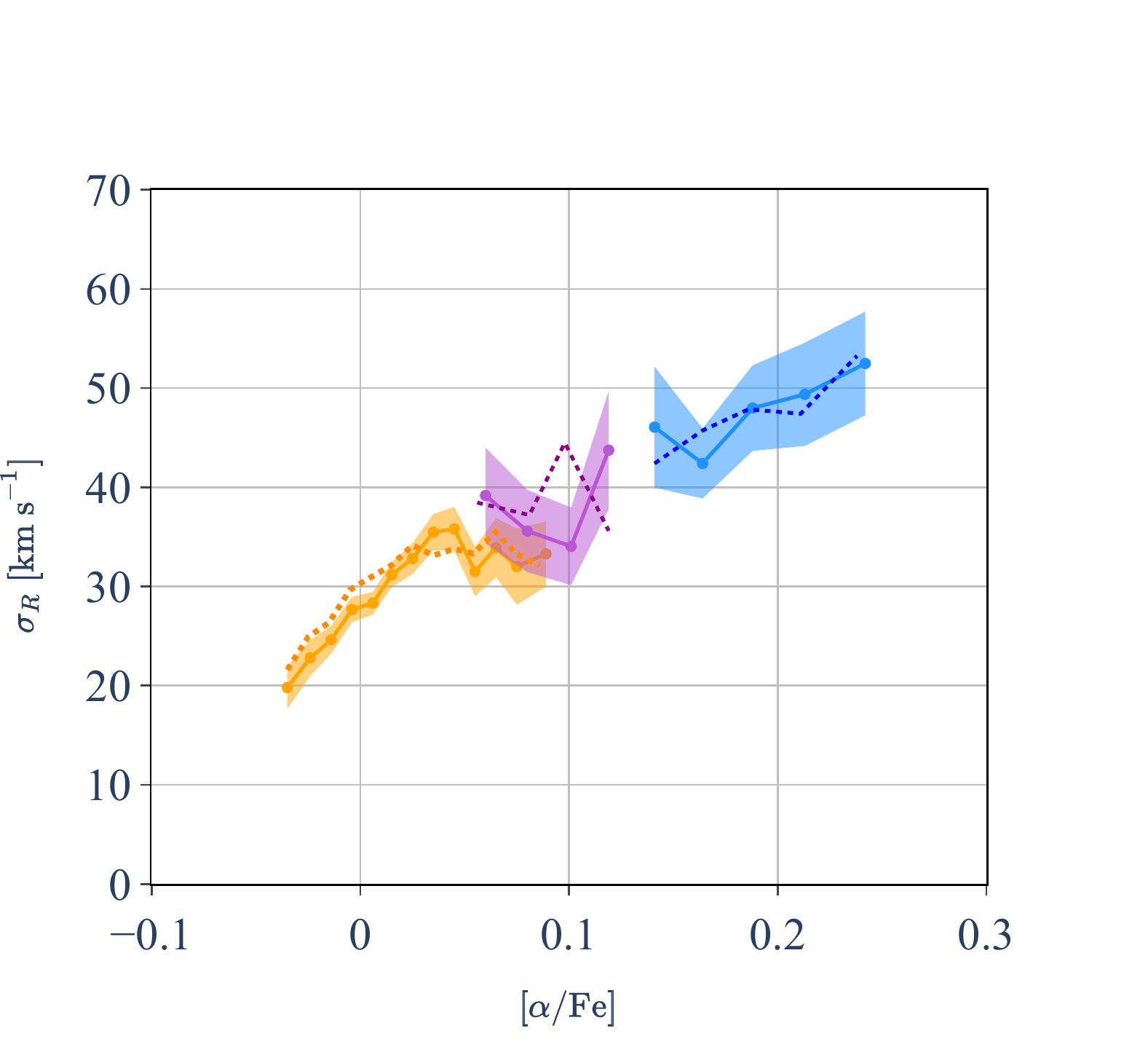} 
   \includegraphics[width=0.33\hsize,clip=true,trim= 0cm 0cm 1.5cm 2cm]{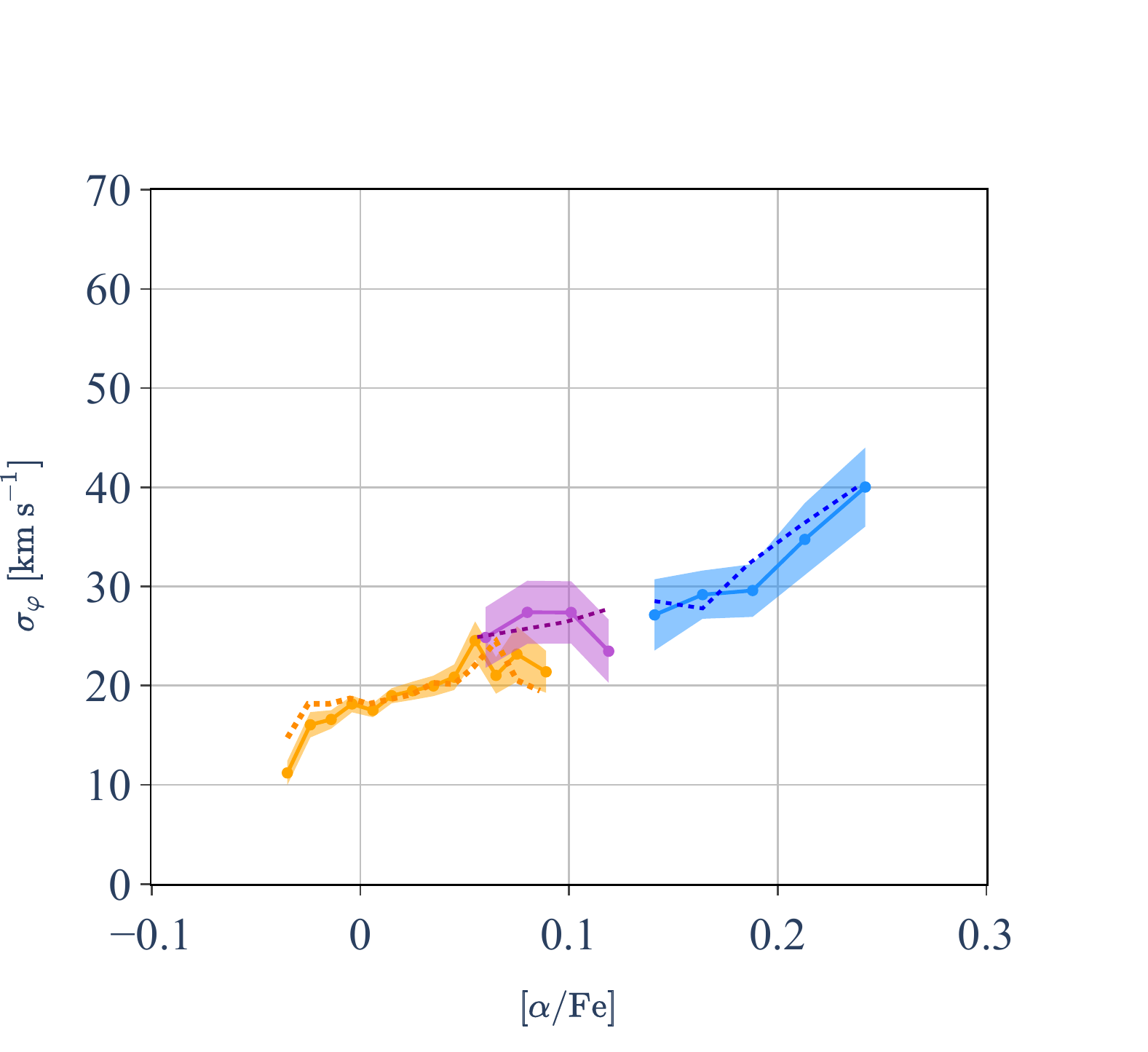}
    \includegraphics[width=0.33\hsize,clip=true,trim= 0cm 0cm 1.5cm 2cm]{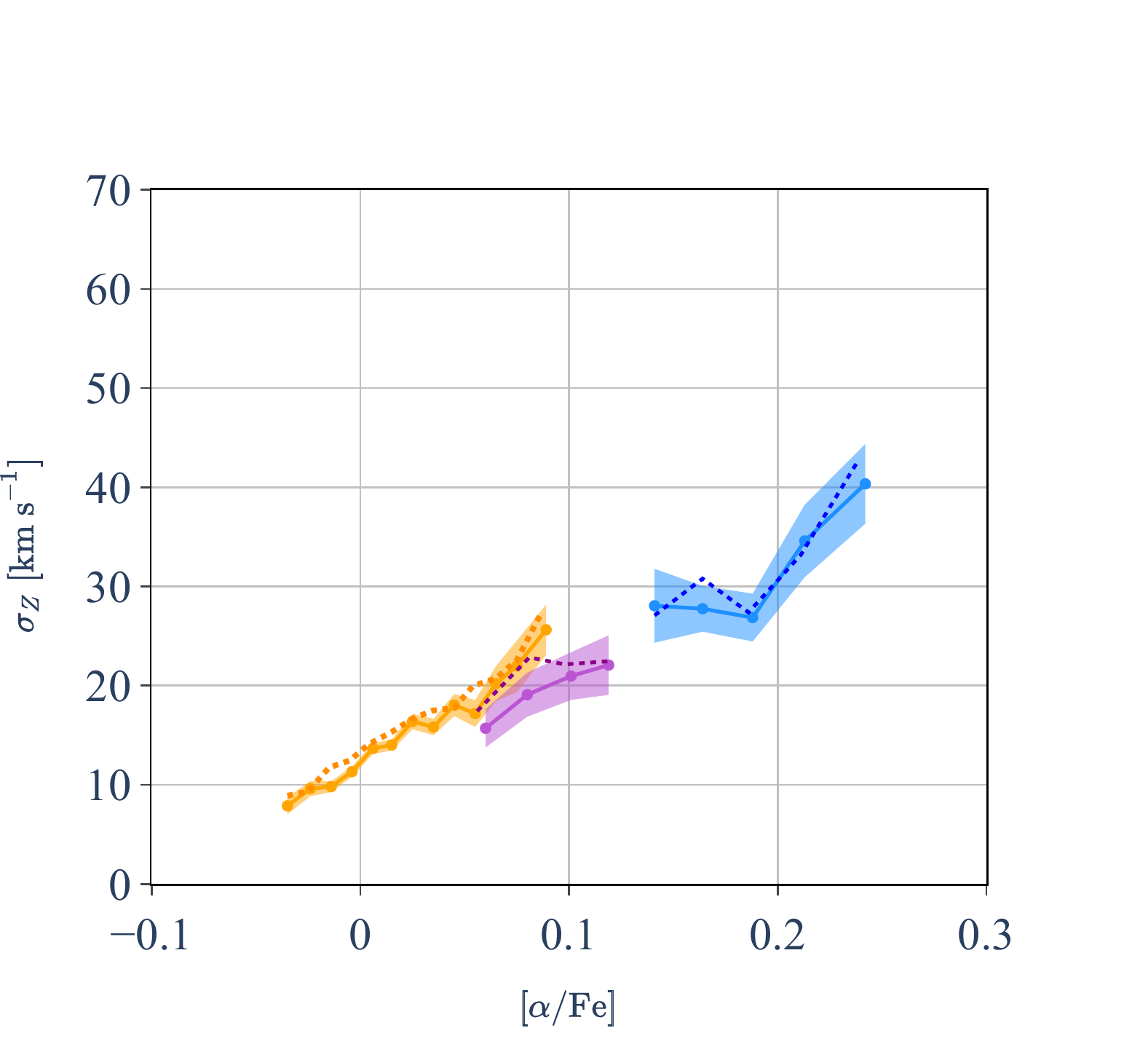}
    
  \includegraphics[width=0.33\hsize,clip=true,trim= 0cm 0cm 1.5cm 2cm]{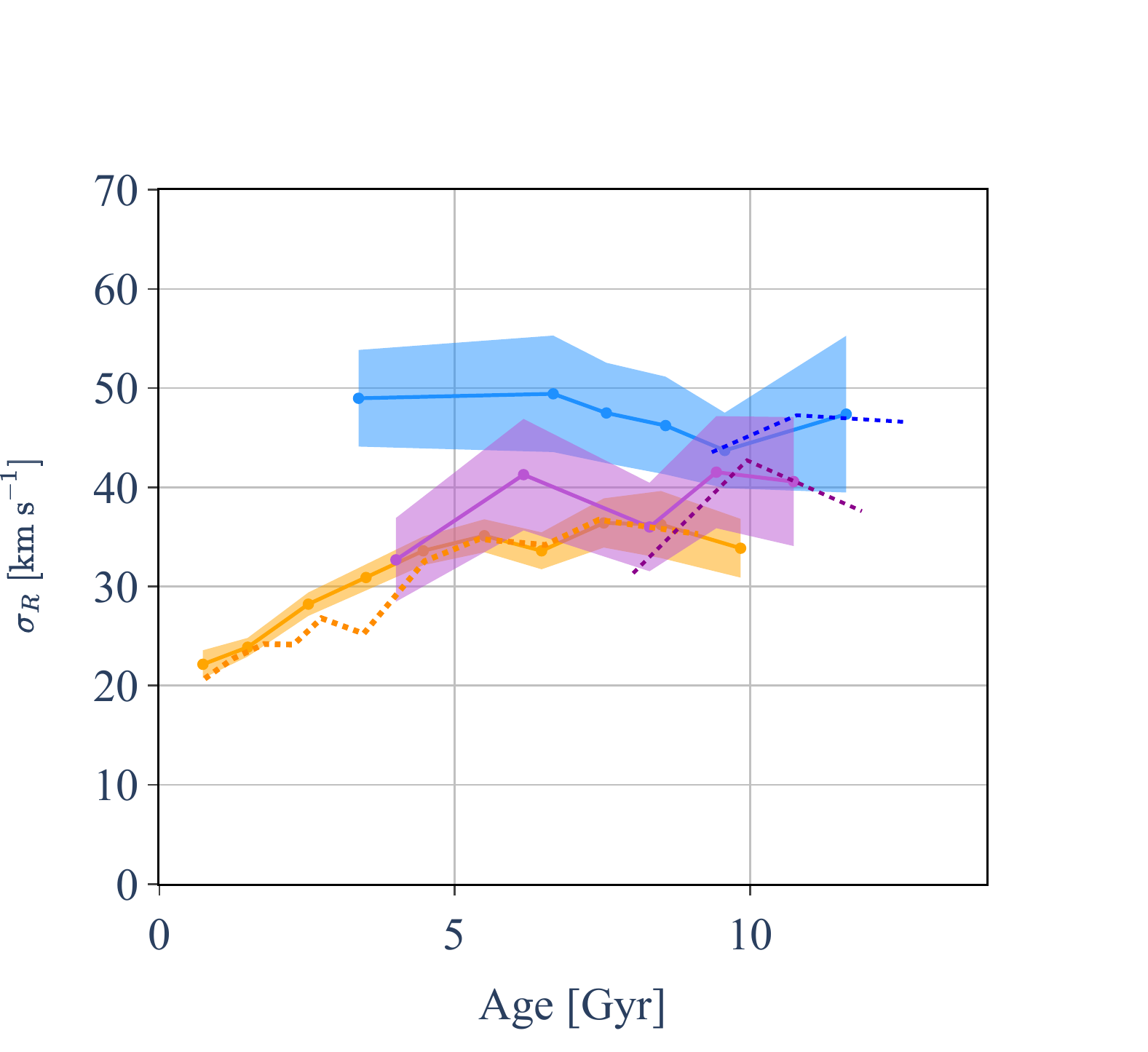} 
   \includegraphics[width=0.33\hsize,clip=true,trim= 0cm 0cm 1.5cm 2cm]{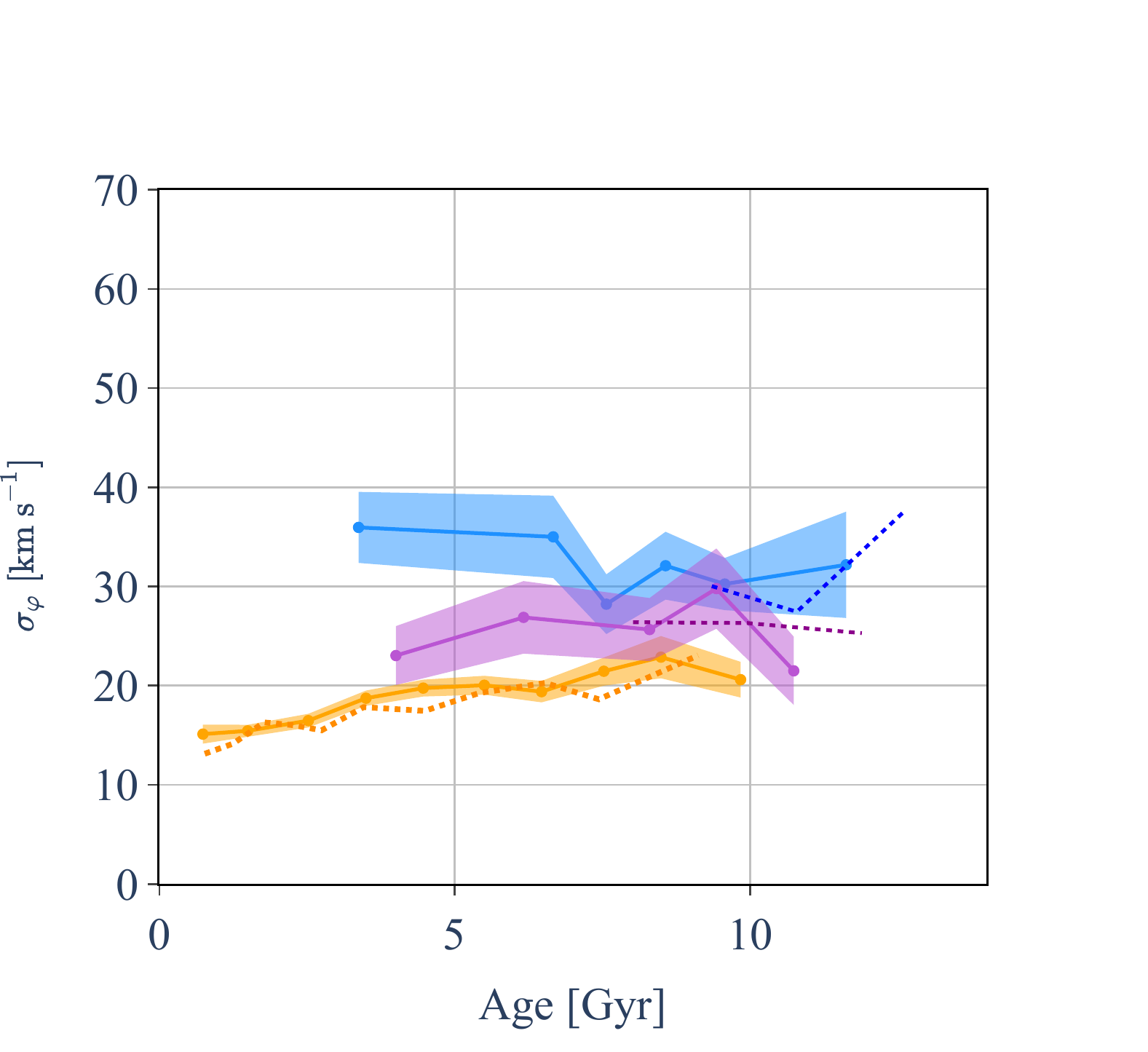}
    \includegraphics[width=0.33\hsize,clip=true,trim= 0cm 0cm 1.5cm 2cm]{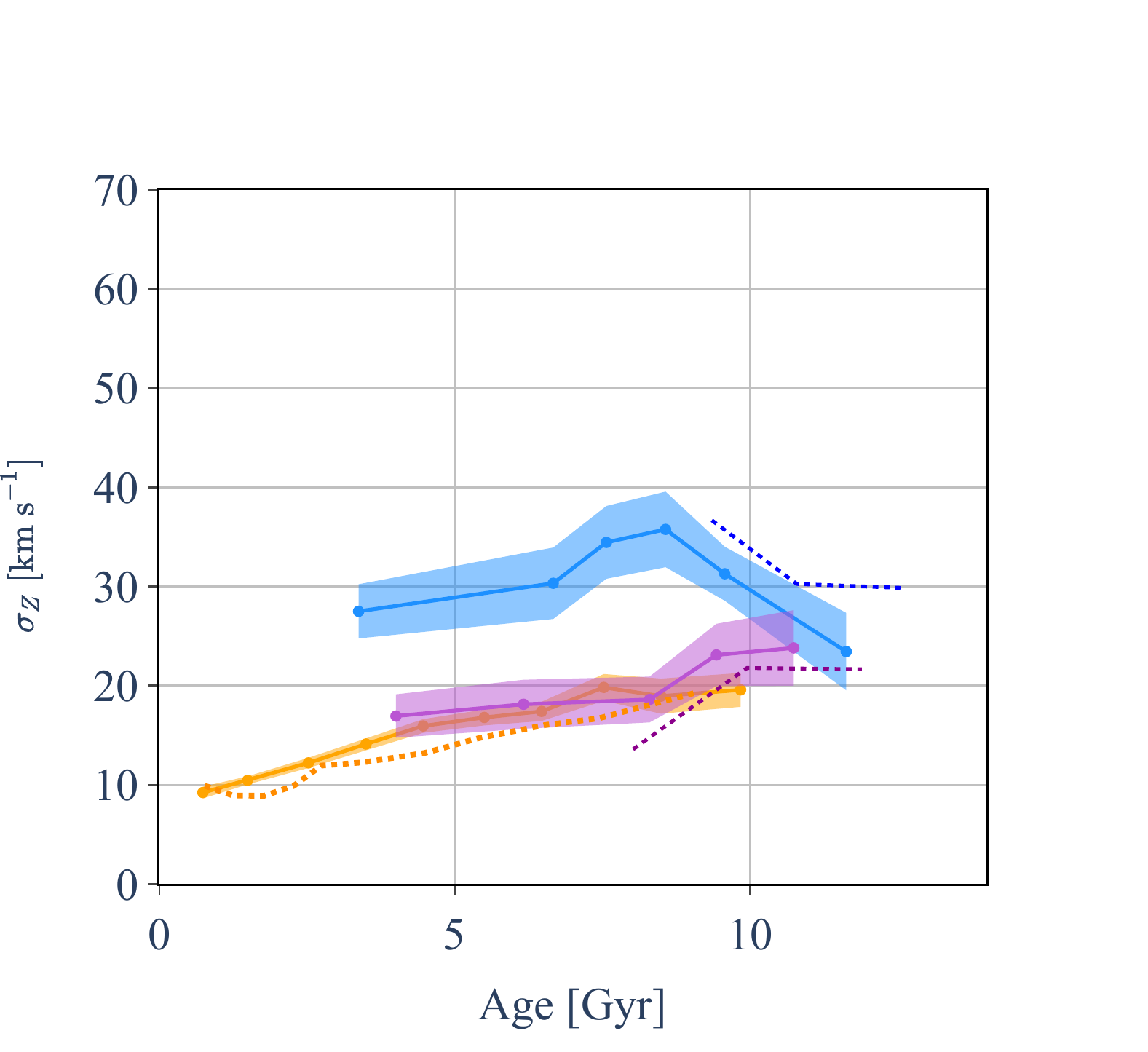}
  \caption{Dispersion of the velocities V$_R$ (left panels), V$_\varphi$ (middle panels), V$_Z$ (right panels) as a function of [Fe/H] (top panels), [$\alpha$/Fe] (middle panels), and age (bottom panels) for the thin-disc, h$\alpha$mp, and h$\alpha$mr thick-disc populations (orange, blue, and magenta symbols, respectively). Solid lines represent velocity dispersions computed using chemistry and stellar ages from the APOKASC catalogue, while dotted lines represent velocity dispersions computed using abundances derived from DR16 APOGEE data (top and middle panels) and using stellar ages derived by M21 (bottom panels). Error bars on the velocity dispersions are also shown. }
   \label{dispV}
 \end{figure*}

We also investigate the Galactic velocity dispersions for the chemically separated populations in our sample.  Figure \ref{dispV} presents these dispersions as a function of [Fe/H], [$\alpha$/Fe] and stellar ages for the three stellar populations. We compute the dispersion of $V_R$, $V_\varphi$, and $V_Z$ by taking the standard deviation for each bin, assuring a minimum of 30 stars per bin. The error bars come from the standard error of the standard deviations. The velocity dispersions are shown using chemistry and stellar ages from the APOKASC catalogue (solid line), DR16 abundances (dotted lines for the top and middle panels), and stellar ages from M21 (dotted line on bottom panels). \\
The average values of the velocity dispersions are in Table~\ref{meanvel}. They are in close agreement with the relation $\overline{\sigma_Z} $< $\overline{\sigma_\varphi} $ < $\overline{\sigma_R}$ for all velocity components and $\overline{\sigma_{Z}}\simeq $ 0.5 $ \overline{\sigma_{R}}$ for the thin and the h$\alpha$mr thick discs  \citep{QuiGar01,Holmberg07, Yu18, Mackereth19b}. As published in previous studies \citep{Wojno16, Bensby05}, we find a constant offset of $\sim$16 km s$^{-1}$ between the average dispersions of h$\alpha$mp thick-disc and thin-disc sequences. We also note that a constant offset of $\sim$6 km s$^{-1}$ is found between average dispersions of h$\alpha$mr thick-disc and thin-disc populations. \\

As expected, velocity dispersions are higher for the h$\alpha$mp thick-disc population than for the thin-disc population \citep[e.g. ][]{RecioBlanco14, Grieves18}, with the h$\alpha$mr thick-disc population lying in between (see Fig. \ref{dispV} and Table~\ref{meanvel}). On the one hand, studying Gaia-ESO data, \citet{RecioBlanco14} found no significant dependence of $\sigma_Z$ or  $\sigma_\varphi$ on [Fe/H] (see their Fig. 20). On the other hand, \citet{Grieves18} found a decrease in velocity dispersion for the thin-disc stars and in $\sigma_\varphi$ only for the thick-disc stars with increasing [Fe/H]. We find a slight decrease in $\sigma_R$, $\sigma_\varphi$ , and $\sigma_Z$ with increasing [Fe/H] for the thin-disc population, which is in agreement with the results of \citet{Grieves18}, and  the same for the h$\alpha$mp thick-disc population. This may be due to the fact that our sample is in a very local volume (see Fig. \ref{ZRfig}) as in the sample of \citet{Grieves18} and contrary to that of \citet{RecioBlanco14} who considered a larger volume where more stars from other radii are included. In addition, the h$\alpha$mr thick-disc population shows a small increase in velocity dispersions up to solar metallicity and a decrease for higher metallicities.
Additionally, as suggested by \citet{RecioBlanco14}, $\sigma_\varphi$ and $\sigma_Z$ are very similar within the error bars for the thick disc, and do not suggest a smaller $\sigma_\varphi$ than $\sigma_Z$, as discussed by \citet{Binney12}.\\

The middle panels of Fig.\ref{dispV} present the velocity dispersions with [$\alpha$/Fe]. For the thin disc, the velocity dispersions $\sigma_R$ and $\sigma_\varphi$ increase with [$\alpha$/Fe], but only up to $\sim$0.05 dex, while the trend of $\sigma_Z$ stays positive. As already shown by \citet{Lee11b} and \citet[][]{Guiglion15} for more metal-rich stars, the velocity dispersions also increase as a function of [$\alpha$/Fe] for the h$\alpha$mp thick-disc population. Although the [$\alpha$/Fe] abundances are usually used as a proxy for age, the behaviours of the velocity dispersion of the h$\alpha$mp thick-disc stars are completely different as a function of age and [$\alpha$/Fe].  The trends are the same when DR14 or DR16 APOGEE data are considered.\\

Bottom panels of Fig.\ref{dispV} present the velocity dispersions with stellar age derived from APOKASC (solid lines) and M21 (dashed lines) catalogues. It is important to recall that the M21 sample excludes the low-mass red clump stars in order to avoid the effect of mass loss on the age determination. Although the mass--age relations are quite robust for giants, stars can gain and lose mass in different events (e.g. binarity, mass-loss, accretion), inducing a significant age bias: stars appear younger or older than they are. Also, with the more robust ages of M21, the  ages of h$\alpha$mp thick-disc stars do not extend to younger ages \citep[see discussion about overmassive stars in ][]{Miglio21}. The figure shows a relationship between the kinematic dispersion of stars and their age (at least in the thin disc) in the sense that the older the thin-disc population, the higher its velocity dispersion. This increase in velocity dispersion with age has been attributed to secular evolution effects in the disc \citep[e.g.][]{SpiSch51, Sellwood14}. We estimate that the oldest thin-disc stars (age$\sim$10 Gyr) have radial, rotational, and vertical dispersion  equal to $\sigma_R$ $\sim$34 km s$^{-1}$, $\sigma_\varphi$ $\sim$20 km s$^{-1}$, and $\sigma_Z$ $\sim$20 km s$^{-1}$, respectively, which is significantly higher than the youngest objects, which reach only $\sigma_R$ $\sim$22 km s$^{-1}$, $\sigma_\varphi$ $\sim$15 km s$^{-1}$, and $\sigma_Z$ $\sim$10 km s$^{-1}$. This result is in agreement with previous studies \citep[e.g.][]{Haywood13}. This increase in $\sigma_\varphi$ is accompanied by a decrease in the rotational velocity V$_\varphi$ of thin-disc stars with age (see Fig.~\ref{velocity_age}). This shows that the relative rotational velocity dispersion, V$_\varphi$/$\sigma_\varphi$, is higher in older stars than in younger stars, which is indicative of a dynamical heating mechanism in the thin disc.
Furthermore, this figure shows a significant difference in velocity dispersion between the thin-disc and the h$\alpha$mp thick-disc populations, indicating that the chemically defined stellar populations have a different history imprinted in their kinematics (see M21 for a discussion).  \\ 

\begin{figure*}
  \centering 
      \includegraphics[width=0.4\hsize,clip=true,trim= 0cm 0cm 2cm 2cm]{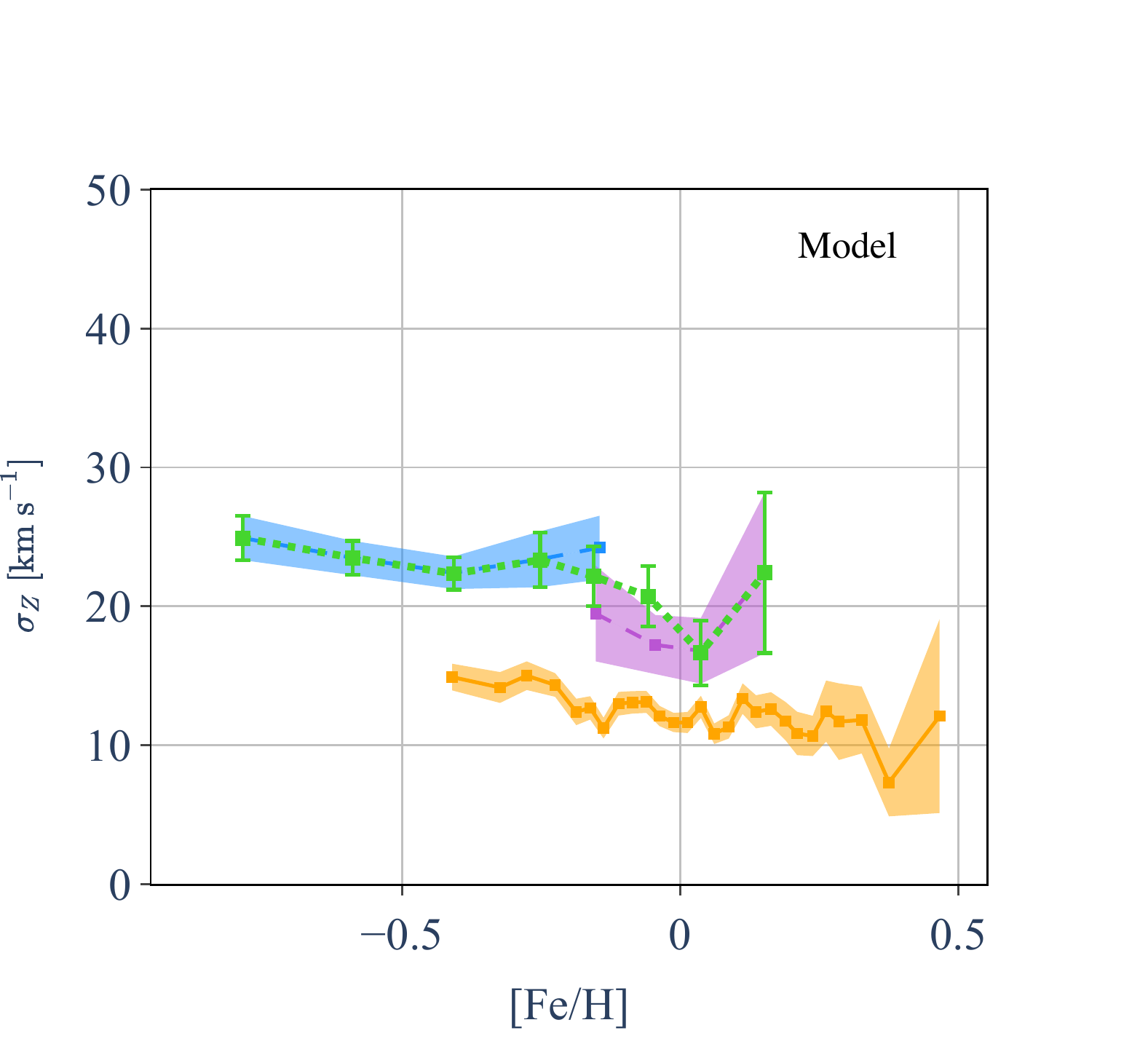} 
      \includegraphics[width=0.4\hsize,clip=true,trim= 0cm 0cm 2cm 2cm]{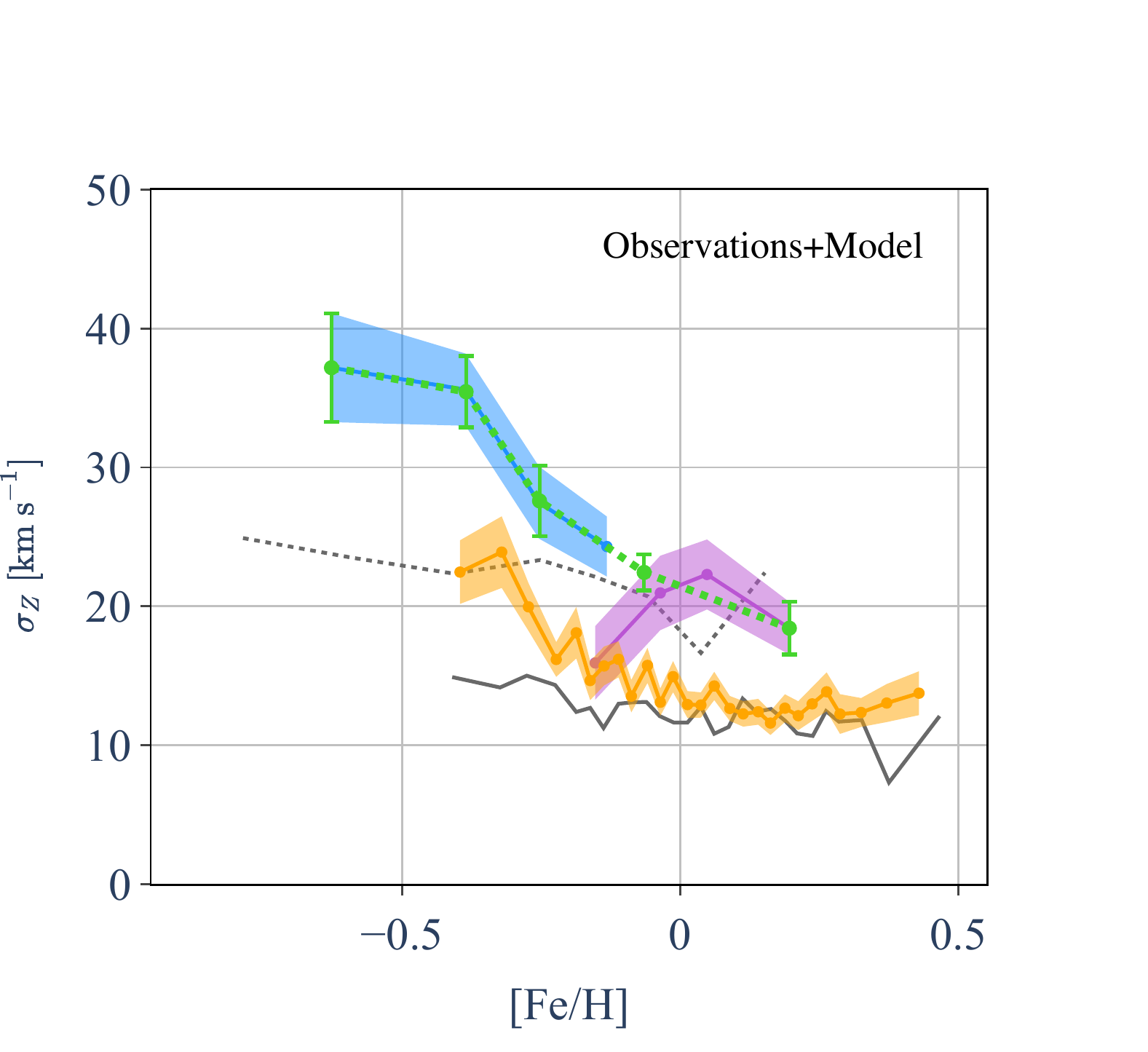}
      
      \includegraphics[width=0.4\hsize,clip=true,trim= 0cm 0cm 1.8cm 2cm]{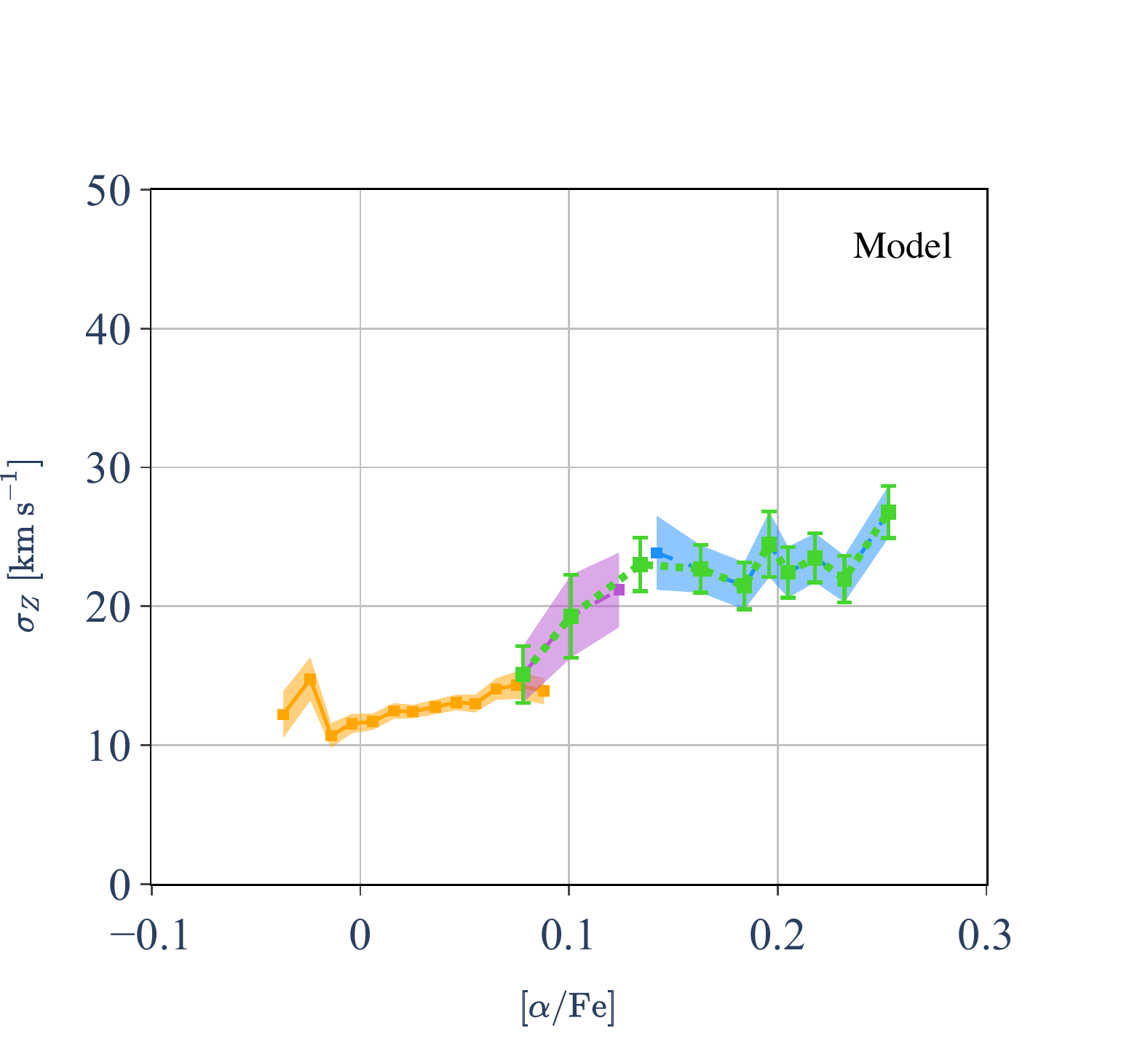} 
      \includegraphics[width=0.4\hsize,clip=true,trim= 0cm 0cm 1.8cm 2cm]{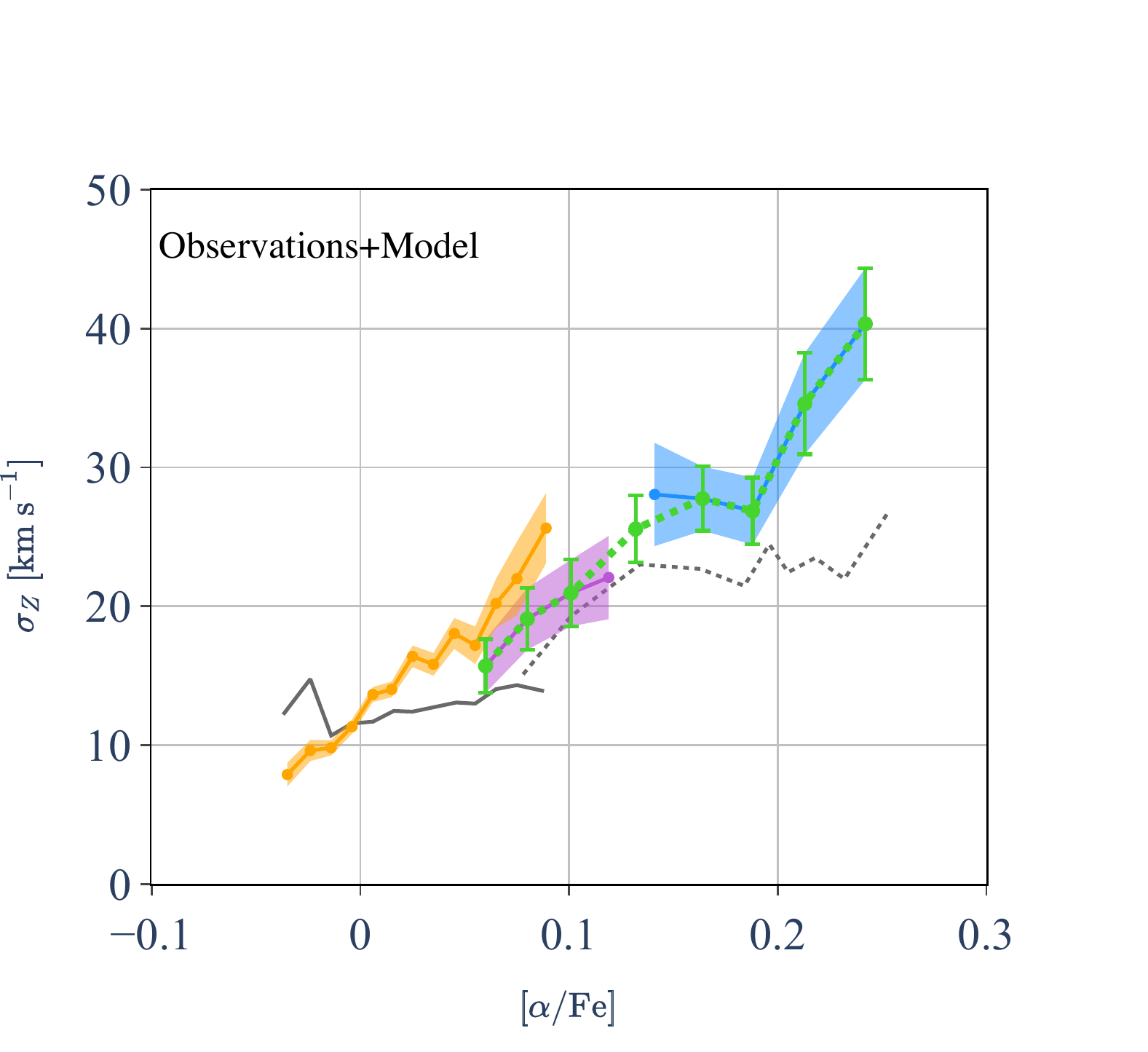}
      
      \includegraphics[width=0.4\hsize,clip=true,trim= 0cm 0cm 2cm 2cm]{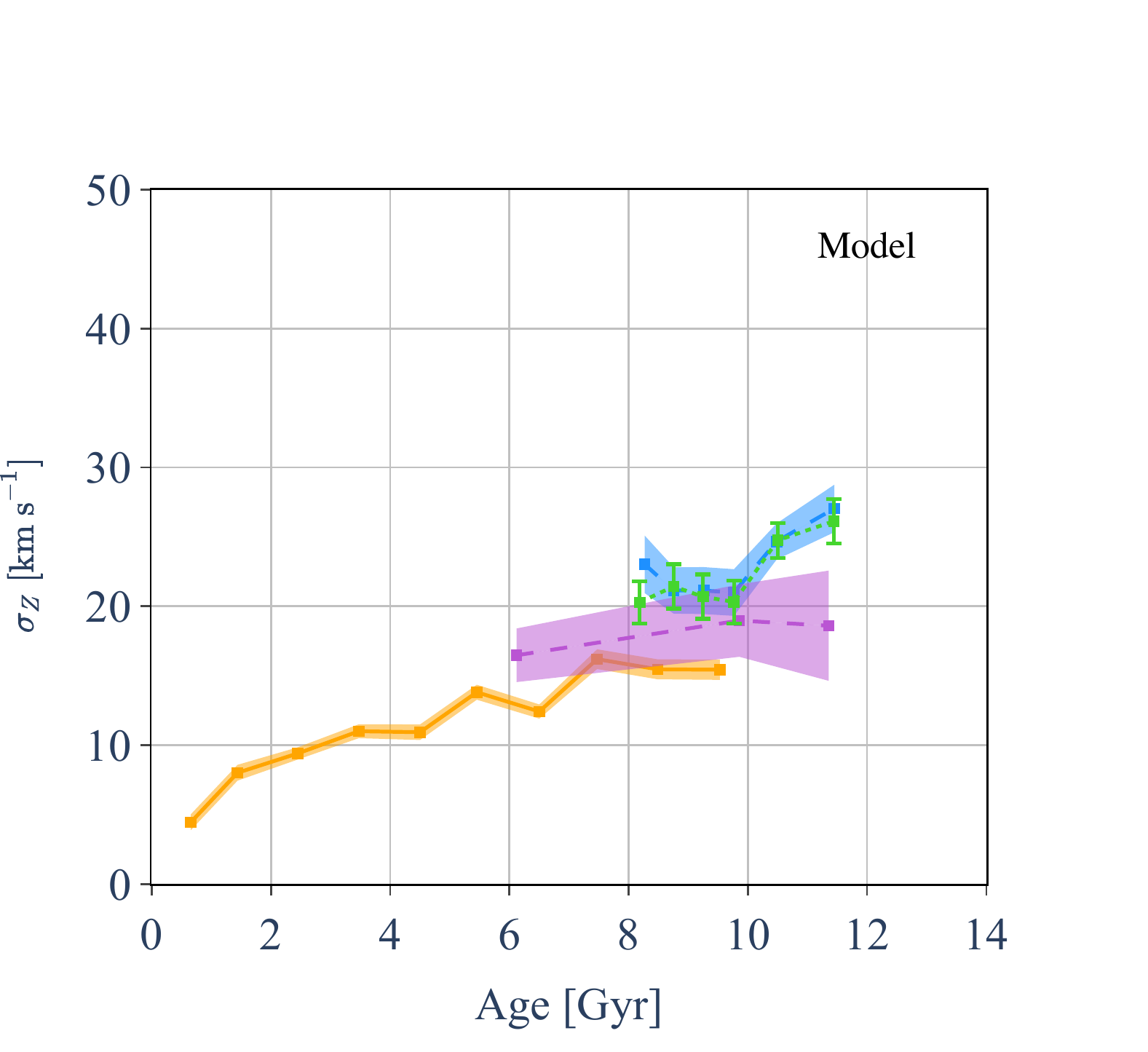} 
      \includegraphics[width=0.4\hsize,clip=true,trim= 0cm 0cm 2cm 2cm]{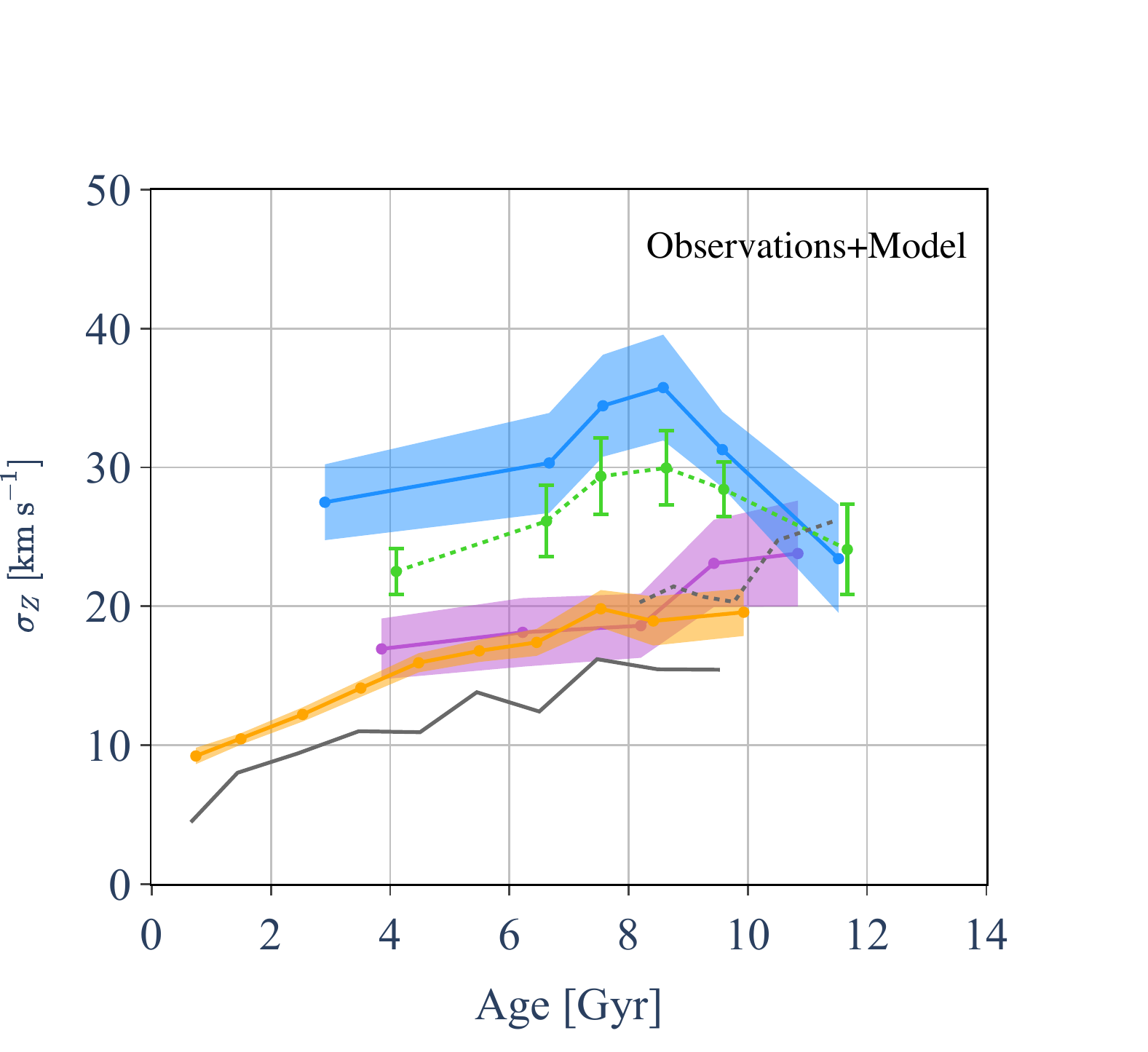}
   \caption{Dispersion of the vertical velocity as a function of metallicity (top panels), [$\alpha$/Fe] (middle panels), and stellar age (bottom panels) for the three stellar populations:  thin disc (orange), h$\alpha$mp thick (blue), and h$\alpha$mr thick disc (magenta). The whole thick disc (i.e. h$\alpha$mp + h$\alpha$mr thick discs) is also shown with the green dashed line. $\sigma_Z$ are computed from BGM (left panels) and from observations using the APOKASC sample (right panels). BGM predictions for the thin disc and the `whole' thick disc are overplotted on the right panels with grey solid and dashed lines, respectively.}
   \label{dispVz_model}
 \end{figure*}

Figure \ref{dispVz_model} shows the dispersion of the vertical velocity simulated by the BGM as a function of [Fe/H], [$\alpha$/Fe], and stellar age for the three stellar populations. We also show the whole thick disc putting together h$\alpha$mr and h$\alpha$mp thick-disc stars. The left panels of Fig.\ref{dispVz_model} show the BGM simulations for all stellar populations, while in the right panels we compare the observed and simulated $\sigma_Z$. 
The BGM predicts a lower dispersion of the vertical velocity than the observations with increasing metallicity, [$\alpha$/Fe], and stellar age, for all stellar populations. At a given age, the observed $\sigma_Z$ is $\sim$30\% higher than the predicted one. Moreover, at a given age, the difference between the $\sigma_Z$ of h$\alpha$mp thick-disc stars and and that of  the thin-disc stars is higher in the observations than in the model. For all stellar populations, the simulated $\sigma_Z$ from the BGM is flatter than $\sigma_Z$ deduced from observations. At a given age, the observed $\sigma_Z$ changes significantly between the h$\alpha$mr and h$\alpha$mp thick-disc stars, such that $\sigma_Z$(8.5 Gyr)$\sim$18.6$\pm$2.3 km s$^{-1}$ and $\sigma_Z$(8.5 Gyr)$\sim$35.7$\pm$3.8 km s$^{-1}$, respectively. Moreover, the age--velocity relations derived by \citet{Miglio21} give $\sigma_Z$(8.5 Gyr)$\sim$ 23 km s$^{-1}$ and 34 km s$^{-1}$ for the low- and high-[Mg/Fe] stellar populations, respectively. These values are in good agreement with our results for the h$\alpha$mp thick-disc and the thin-disc stars deduced from the APOKASC sample ($\sigma_Z$(8.5 Gyr)$\sim$18.9$\pm$1.8 km s$^{-1}$). This difference between the two stellar populations in the $\sigma_Z$ is an important observational constraint: it suggests that the thick disc was not formed by secular processes but either because of merger events or strong gas accretion \citep[see e.g.][]{Brook04, Minchev13, Martig14, Miglio21}.

Moreover, the observations show a maximum in the $\sigma_Z$ behavior with age at $\sim$8 Gyr for the  h$\alpha$mp thick-disc stars. This observational feature in $\sigma_Z$ is not seen in other velocity components V$_\varphi$ and V$_Z$.
These features are not visible in BGM simulation, because it uses simple assumptions that do not take into account radial migration or mergers in the input kinematics (see following section). Therefore, we believe that these behaviours are not induced by sample selection but are real features.

\subsection{Implications for Galactic disc formation}
 
Here we discuss our results in the context of different Galactic disc formation scenarios. Different formation scenarios have been proposed in the literature to explain different chemical and kinematic properties of stars of the thin and the thick discs. Multiple studies have proposed vertical heating of an already existing disc by different kinds of mergers \citep[e.g. ][]{Quinn93, VilHel08}. \citet{Abadi03} and \citet{Brook04,Brook07} suggested that the formation of the thick disc is due to stellar accretion from mergers of dwarf galaxies or the coalescence of a gas-rich satellite, respectively. On the other hand, radial migration \citep{SelBin02} has been proposed to explain the emergence of the thick disc. Two different radial migration processes have been distinguished: (1) the radial migration driven by corotation and (2) the Lindblad resonances of the spiral arms. Nevertheless, different studies using numerical models show that the radial migration driven by corotation cannot significantly contribute to thickening of the disc  \citep[e.g.][]{Minchev12, Martig14}. \citet{Bournaud09} proposed that the thick disc formation could be explained by stellar scattering induced by massive clumps. Alternatively, chemical analyses suggested early on that the thick disc may be the result of fast gas accretion \citep[e.g.][]{Chiappini97, Haywood13, Spitoni20,KawChi16}.

The highest velocity dispersions of the h$\alpha$mp thick disc are in agreement with different scenarios such as perturbations by mergers at the beginning of the disc formation \citep{Wyse01, Quinn93}, by clumpy disc instabilities \citep{Bournaud09}, by stars born hot at high redshift \citep[i.e. z$\sim$1 to 2][]{Brook05, Brook12}, and by accreted satellite or stellar populations \citep{Abadi03, Meza05}. 

\citet{Minchev14a} underline a strong inversion of the velocity dispersions at [Mg/Fe]>0.4 dex using a sample of giant stars in the RAVE survey. They also studied the SEGUE G-dwarf stars and found the same inversion at [$\alpha$/Fe]$\sim$ 0.35-0.4 dex. For all velocity components, these authors found that the most metal-poor and [Mg/Fe]-rich stars have velocity dispersions comparable to those of the most metal-rich stars. Using a chemo-dynamical model \citep{Minchev13,Minchev14b}, they explained this feature by the stronger effect of mergers on the outer part of the discs and by the radial migration of older stellar populations with cooler kinematics born in the inner disc. 
Our sample has a restricted metallicity range and the most [$\alpha$/Fe]-rich stars are around 0.3 dex. This explains why we cannot see the same inversion of the velocity dispersion as \citet[][see their Fig. 1 and 2]{Minchev14a}. Contrary to this latter study, we have access to the stellar ages derived by asteroseismology. Figures \ref{dispV} and \ref{dispVz_model} show a similar inversion of  $\sigma_Z$ for the h$\alpha$mp thick-disc population around 8 Gyr. This typical inversion is not predicted by the BGM simulation and therefore strongly suggests merger effects \citep[e.g.][]{Belokurov20} and/or radial migration \citep[][and Chiappini et al. in prep.]{Miglio21}.

Moreover, the h$\alpha$mr thick-disc population has a very different vertical velocity dispersion with age from that of the h$\alpha$mp thick-disc population ---but it is very similar to that of the thin disc--- and these differences are not predicted by the BGM simulation. This could be due to a different formation scenario for the h$\alpha$mr thick-disc stars compared to the h$\alpha$mp thick-disc population, and probably closer to the thin disc. \citet{Anders18} suggested that stars of the h$\alpha$mr thick disc had a different origin and history to the thick- and thin-disc stars migrating from the inner disc, while \citet{Ciucua20} proposed this population as a transition region connecting the old thick-disc and the young thin-disc populations. The results shown constitute an important constraint to future simulations of the formation of the Milky Way.

\section{Conclusions}
\label{conclu}

Understanding the structure and evolution of the Milky Way requires a sound understanding of the relations between stellar age, chemical properties, and kinematics. In the present paper, we study the APOKASC-\gaia sample, combining all crucial constraints from different kinds of surveys. We also use stellar ages computed by \citet{Miglio21} for a subsample of the APOKASC catalogue. We complement the above information with that of the ESA satellite \gaia which provides proper motions, radial velocities,  accurate parallaxes, and StarHorse distances \citet{Queiroz20} to derive kinematic parameters for a large number of stars in the Milky Way. Moreover, to avoid selection effects on the interpretation of the studied samples, we built mock catalogues using the Besan\c con Galaxy model \citep[e.g.][]{Lagarde17,Lagarde19} and taking into account selections made in APOKASC and M21 samples.  
We study the properties of three chemically defined populations: the thin, high-$\alpha$ metal-rich (h$\alpha$mr), thick, and high-$\alpha$ metal-poor (h$\alpha$mp) thick discs. Our main results can be summarised as follows:\\

\textit{Age distributions}\\

By comparing the age distributions we find that the APOKASC sample shows a peak around 1-2 Gyr that is not seen in the M21 distribution, and is not simulated by the BGM. We suggest that this peak is created by the strong dependency of the age determination for red clump stars on the physics included in stellar models (e.g. hydrodynamical processes, mass loss or accretion, and binarity). On the other hand, we find good agreement between the age distribution derived using M21 and that suggested by the BGM simulation within small statistical fluctuations. A small hint of SFR enhancement between 2 and 5.5 Gyr in both samples remains to be confirmed with larger seismic samples. We find that the age distribution of the thin disc stars is mostly between 0 and 10 Gyr and in good agreement between APOKASC and M21. Moreover, we underlined a significant difference between the mean ages derived for the h$\alpha$mr and h$\alpha$mp thick-disc stars from  APOKASC and M21: ($\overline{Age_{M21}}$- $\overline{Age_{APOKASCxM21}}$$\sim$1.02 and 1.77 Gyr, respectively).\\ 

\textit{Age versus metallicity and [$\alpha$/Fe] relations}\\

We show that both samples show a flat age--metallicity relation for the thin disc and an increase in [$\alpha$/Fe] with the stellar age. No correlation between stellar age and metallicity is found for the stars of the h$\alpha$mr thick disc. Concerning the h$\alpha$mp thick-disc population, our conclusion is different. While both relations are flat for the APOKASC sample, the M21 sample shows negative and positive gradients for the age--metallicity and age--[$\alpha$/Fe] relations, respectively. The comparison with the BGM simulations cannot explain these differences. This suggests that important processes such as radial migration and accretion due to mergers play an important role in shaping our Galaxy.\\

\textit{Kinematics properties}\\

As the metallicity and $\alpha$-abundances are usually used as proxies for stellar age, we investigated the relations between the kinematics properties and metallicity, [$\alpha$/Fe], and age. We confirm no correlation between radial or vertical velocities and [Fe/H] or [$\alpha$/Fe] for either of the three stellar populations. We also confirm that there is not correlation  between radial or vertical velocities and stellar age. For the thin disc, V$_\varphi$ decreases with metallicity and age while it is almost constant with [$\alpha$/Fe]. By looking at V$_\varphi$/$\sigma_\varphi$, we show evidence in support of a dynamical heating mechanism in the thin disc. 
Indeed the APOKASC and M21 samples show that V$_\varphi$ decreases with age for the thin disc with a very similar gradient ($\partial V_\varphi /\partial$age = $-$2.38$\pm$0.14 and $-$2.14$\pm$0.18 km s$^{-1}$Gyr$^{-1}$, respectively). 
On the other hand, in the h$\alpha$mp thick-disc population, the azimuthal velocity behaviours with age show opposite gradients in the two samples. This difference is not due to the sample selections as shown by the BGM simulations, but is probably induced by incorrect determination of the age of stars with high mass loss in the APOKASC sample.\\

\textit{Characterisation of the h$\alpha$mr thick disc }\\

The age distribution in the h$\alpha$mr thick-disc population mimics that of the h$\alpha$mp thick-disc population, with a similar mean age. The [$\alpha$/Fe] abundances of this population also differ from those of the other populations, while its [Fe/H] is very similar to that of the thin disc (see Sect.\ref{chemicalpop}). The kinematics of the h$\alpha$mr thick-disc population seems  to follow  that of the thin-disc population  more closely than that of the h$\alpha$mp thick disc  (see Sect.\ref{Kinematics}). The $\sigma_Z$ with age is clearly lower for the h$\alpha$mr thick disc than for the h$\alpha$mp thick disc and is almost the same as that for the thin disc, representing a link between the thin and the h$\alpha$mp thick discs. These behaviours are not simulated by the BGM, suggesting a different formation scenario for these stars that is not included in the model. These properties might suggest a different origin and history for these stars, by migration from the inner disc as proposed by \citet{Anders18} or as a transition region between the old thick disc and the young thin disc as proposed by \citet{Ciucua20}. This feature could be investigated in more detail using a chemo-dynamical model. \\

\textit{Galactic disc formation}\\

We highlight the different behaviours of the dispersion of  vertical velocity with age in the BGM simulations and in the observations. The simple assumptions in the model do not explain the inversion in the relation between $\sigma_Z$ and age for the h$\alpha$mp thick-disc stars around 8 Gyr that we observe. These comparisons underline the need for a more complex chemo-dynamical scheme to explain the data, perhaps including mergers and radial migration effects as discussed previously by \citet{Minchev14b}, \citet{Belokurov20}, and \citet{Miglio21}. \\

Current asteroseismic missions are limited in either Galactic volume coverage or duration of observations, which in turn limits the precision achievable for inferred stellar properties such as stellar age. The future space mission PLATO \citep{PLATO} will provide stellar ages in different directions in the Milky Way for a large number of stars and with sufficient data quality to reach 10\% precision for age \citep[for more details see][]{Miglio17, Montalban21}. Furthermore, future spectroscopic surveys will provide a complementary chemical vision of our Galaxy, allowing investigation of the age--metallicity relation over a large range of $R$ and $Z$, probing all stellar populations in the Milky Way.

\begin{acknowledgements}
   
N.L. and C.R. acknowledges financial support from "Programme National de Physique Stellaire" (PNPS). N.L., C.R., A.R. and O.B. acknowledges financial support from the "Programme National Cosmology et Galaxies (PNCG)" of CNRS/INSU, France. Simulations have been executed on computers from the Utinam Institute of the Universit\'e de Franche-Comt\'e, supported by the R\'egion de Franche-Comt\'e and Institut des Sciences de l'Univers (INSU). This work has made use of data from the European Space Agency (ESA) mission {\it Gaia} (\url{https://www.cosmos.esa.int/gaia}), processed by the {\it Gaia}
Data Processing and Analysis Consortium (DPAC, \url{https://www.cosmos.esa.int/web/gaia/dpac/consortium}). Funding for the DPAC
has been provided by national institutions, in particular the institutions participating in the {\it Gaia} Multilateral Agreement.
F.F., A.F., R.M., M.R.,T.A. acknowledge supported by the Spanish Ministry of Science, Innovation and University (MICIU/FEDER, UE) through grant RTI2018-095076-B-C21, the Institute of Cosmos Sciences University of Barcelona (ICCUB, Unidad de Excelencia ``Mar\'{\i}a de Maeztu'') through grant CEX2019-000918-M, the Ramon y Cajal Fellowship RYC2018-025968-I. This project has received funding from the European Union's Horizon 2020 research and innovation programme under the Marie Sklodowska-Curie grant agreement  No. 800502. 
AM acknowledges funding from the European Research Council (ERC) under the European Union’s Horizon 2020 research and innovation programme (grant agreement No. 772293  - project ASTEROCHRONOMETRY, \url{https://www.asterochronometry.eu}. We also acknowledge the International Space Science Institute, Bern, Switzerland for providing financial support and meeting facilities.)

 \end{acknowledgements}

\bibliographystyle{aa}
\bibliography{Reference}
\end{document}